\documentclass[aps,prb,twocolumn,superscriptaddress,longbibliography,showpacs]{revtex4-1}
\usepackage{amsmath,amssymb,wasysym,graphicx}
\usepackage[utf8]{inputenc}
\usepackage[T1]{fontenc}
\usepackage{xcolor}

\IfFileExists{newtxtext.sty}
{\usepackage{newtxtext,newtxmath}}
{\IfFileExists{stix.sty}
	{\usepackage{stix}}
	{\IfFileExists{mathptmx.sty}
		{\usepackage{mathptmx}}{} } }

\usepackage{textcomp}

\usepackage{bm}

\IfFileExists{siunitx.sty}{\usepackage{booktabs,siunitx}}{}

\usepackage{color}
\definecolor{LinkColor}{rgb}{0.256,0.439,0.588}
\usepackage{hyperref}
\hypersetup{
	pdfauthor={good guys},
	pdftitle={good title},
	colorlinks=true,
	citecolor=LinkColor,
	linkcolor=LinkColor,
	urlcolor=LinkColor
}

\newcommand{\beq} {\begin{equation}}
\newcommand{\eeq} {\end{equation}}
\newcommand{\bea} {\begin{eqnarray}}
\newcommand{\eea} {\end{eqnarray}}
\newcommand{\be} {\begin{equation}}
\newcommand{\ee} {\end{equation}}

\begin{document}

\title{Emergent self-duality in long range critical spin chain: from deconfined criticality to first order transition}

\author{Sheng Yang}
\altaffiliation{The first two authors contributed equally.}
\affiliation{Zhejiang Institute of Modern Physics and School of Physics, Zhejiang University, Hangzhou 310027, China}
\author{Zhiming Pan}
\altaffiliation{The first two authors contributed equally.}
\affiliation{Institute for Theoretical Sciences, WestLake University, Hangzhou 310024, China}
\author{Da-Chuan Lu}
\email{dclu137@gmail.com}
\affiliation{Department of Physics, University of California, San Diego, CA 92093, USA}
\author{Xue-Jia Yu}
\email{xuejiayu@fzu.edu.cn}
\affiliation{Department of Physics, Fuzhou University, Fuzhou 350116, Fujian, China}
\affiliation{Fujian Key Laboratory of Quantum Information and Quantum Optics,College of Physics and Information Engineering,Fuzhou University, Fuzhou, Fujian 350108, China}

%-------------------------------------
\begin{abstract}
{Over the past few decades, tremendous efforts have been devoted to understanding self-duality at the quantum critical point, 
which enlarges the global symmetry and constrains the dynamics. 
One-dimensional spin chain is an ideal platform for the theoretical investigation of these extoic phenomena, due to the powerful simulation method like density matrix renormalization group. Deconfined quantum criticality with self-duality at the critical point has been found in an extended short-range spin chain.
In this work, we employ large-scale density matrix renormalization group simulations to investigate a critical spin chain with long-range power-law interaction $V(r) \sim 1/r^{\alpha}$. 
Remarkably, we reveal that the long-range interaction drives the original deconfined criticality towards a first-order phase transition as $\alpha$ decreases. 
More strikingly, the emergent self-duality leads to an enlarged symmetry and manifests at these first-order critical points. 
This discovery is reminiscent of self-duality protected multicritical points and provides the example of the critical line with generalized symmetry.
Our work has far-reaching implications for ongoing experimental efforts in Rydberg atom quantum simulators.}
\end{abstract}

%-------------------------------------
\date{\today}
\maketitle

%%%%% Main Text %%%%%%%%%
%-------------------------------------
%-------------------------------------
\section{Introduction} \label{sec:introduction}
Quantum critical points (QCPs) and associated emergent phenomena in strongly correlated many-body systems stand as central topics within realms of both condensed matter and high-energy communities~\cite{sachdev2023quantum,sachdev_2011,wilson1983rmp}. A special property observed in certain QCPs is self-duality. 
Its origin can trace back to the Kramers-Wannier duality of the two-dimensional (2D) classical Ising model~\cite{Kramers1941pr,kadanoff1971prb} and has subsequently been established in a series of models featuring deconfined quantum critical points (DQCP)~\cite{xu2012unconventional,senthil2004prb,Senthil_2005,levin2004prb,senthil2005,swingle2012prb,wang2017prx,bi2019prx,bi2020prr,jiang2019_1ddqcp,senthil2023deconfined,aksoy2023liebschultzmattis}. 
More specifically, DQCP exhibits anomalies~\cite{max2018prb}, fractionalization~\cite{ma2018prb}, self-duality~\cite{wang2017prx,qin2017prx}, and emergent symmetry~\cite{sandvik2007prl,nahum2015prl,ma2019prl}. 
Despite extensive theoretical~\cite{wang2017prx,senthil2023deconfined,ma2020prb,nahum2020prb,lu2022nonlinear,ji2022boundary,prembabu2022boundary,zhang2023prl,shyta2022prl,Joshi2020prx,lee2019prx,you2018prx,jian2018prb,jian2017emergent,zou2021prx,song2019unifying,song2020prx,lukas2017prb,jian2017emergent}, numerical~\cite{sandvik2010prl,kaul2012prl,Sandvik_2020,shao2016quantum,block2013prl,kaul2008prb,kaul2012prl,yu2016prl,he2021scipost,shu2023equilibration,li2019deconfined,liu2023does,zhang2018prl,chen2023phases,song2023deconfined,liu2023prl,liao2022prb,zhao2022prl,wang2022scipost,daliao2023teaching,liu2022emergent,you2022deconfined,Xi_2022,shu2022prb,liu2022prx,liu2022prl,shu2022prl,huang2020prr,zeng2020prb,sun2019prb,robert2019prb,nahum2015px,zhou2023mathrmso5,liu2019superconductivity,demidio2023leeyang,jonathan2017prl,chen2013prl}, and experimental explorations~\cite{guo2020prl,cui2023proximate,song2023unconventional}, there is still ongoing efforts surrounding how exactly they have been implemented in lattice models and experimental setups.

Recently, quantum simulators like Rydberg atoms and ion traps have emerged as powerful tools for simulating exotic quantum phases and phase transitions~\cite{Saffman2010rmp,deng2005pra,Lahaye_2009,ritsch2013rmp,Carr_2009,blatt2012quantum,britton2012engineered,islam2013emergence,richerme2014non,jurcevic2014quasiparticle,yu2022prb}. 
These systems offer intriguing opportunities for exploring long-range interactions in many-body systems, an area extensively investigated in condensed matter and ultracold atom physics~\cite{defenu2018prl,defenu2021longrange,defenu2023outofequilibrium,syed2021prb,defenu2016prb,gong2016prb,gong2016prb2,shen2023long}. 
The critical behavior of long-range interacting systems has been widely studied in both quantum spin models~\cite{fisher1972prl,defenu2017prb,Defenu_2020,giachetti2021prl,giachetti2022prb,codello2015prd,song2023dynamical,song2023quantum,zhao2023finitetemperature,defenu2015pre,maghrebi2016prb,longrangeboson2017,yang2020deconfined,sandvik2010prl2,yu2023pre} and interacting fermion models~\cite{vu2022prl,leaw2019universal,ghazaryan2015prb,vodola2014prl}. 
The presence of long-range interactions can effectively modify the system's dimensionality~\cite{defenu2015pre,defenu2016prb,defenu2017prb,Defenu_2020,defenu2021longrange}, leading to effects such as the breakdown of quantum-classical correspondence and the Mermin-Wagner theorem~\cite{longrangeboson2017,song2023quantum,zhao2023finitetemperature,yang2020deconfined}, thus can dramatically alter the phases and phase transitions. 
For instance, even in the case of conventional QCPs, the influence of long-range interactions generally gives rise to three distinct universality classes~\cite{defenu2021longrange}: the mean-field universality class, the long-range ``non-classical'' universality class, and the short-range universality class.
Certainly, a crucial question arises: How do long-range interactions influence unconventional QCPs, like the DQCP? Does this interaction lead to new physical phenomena? Currently, these questions remain unanswered.

In this work, we address the fate of a 1D DQCP with long-range power-law decay interaction $V(r) \sim \frac{1}{r^{\alpha}}$, using both lattice simulations of a frustrated quantum spin model and renormalization group (RG) calculations of a proposed Luttinger-liquid field theory. 
Remarkably, we find there still exist DQCP and emergent symmetry for fast decay of the long-range interaction above a critical power $\alpha_{\rm c}\approx 1.95$. 
This result is consistent with the predictions from bosonization and RG analyses. 
The most intriguing observation is that as the long-range interaction decays slower (i.e., $\alpha < \alpha_{\rm c}$), the DQCP turns into a first-order phase transition with enlarged symmetry preserved by the emergent self-duality. 
This discovery goes beyond traditional understandings, particularly as emergent symmetry is generally associated with continuous QCPs. Our work clarifies and significantly extends the discussion of the DQCP phenomena in long-range interacting systems, and its signatures should be detectable in existing Rydberg atom experiments.

%-------------------------------------

The rest of this paper is organized as follows: Sec.\ref{sec:model} presents the lattice model of the 1D DQCP with long-range power-law interaction and outlines the employed numerical method. Sections\ref{sec:numerics} and \ref{sec:theory} depict the phase diagram of the 1D DQCP with long-range interaction and the finite-size scaling of the critical behavior, along with an effective field theory to explain the aforementioned numerical results. The discussion and conclusion are presented in Sec.~\ref{sec:con}. Additional data for our analytical and numerical calculations are provided in the Appendices.

\section{Model and Method}
\label{sec:model}

\begin{figure}[tb]
\includegraphics[width=\linewidth]{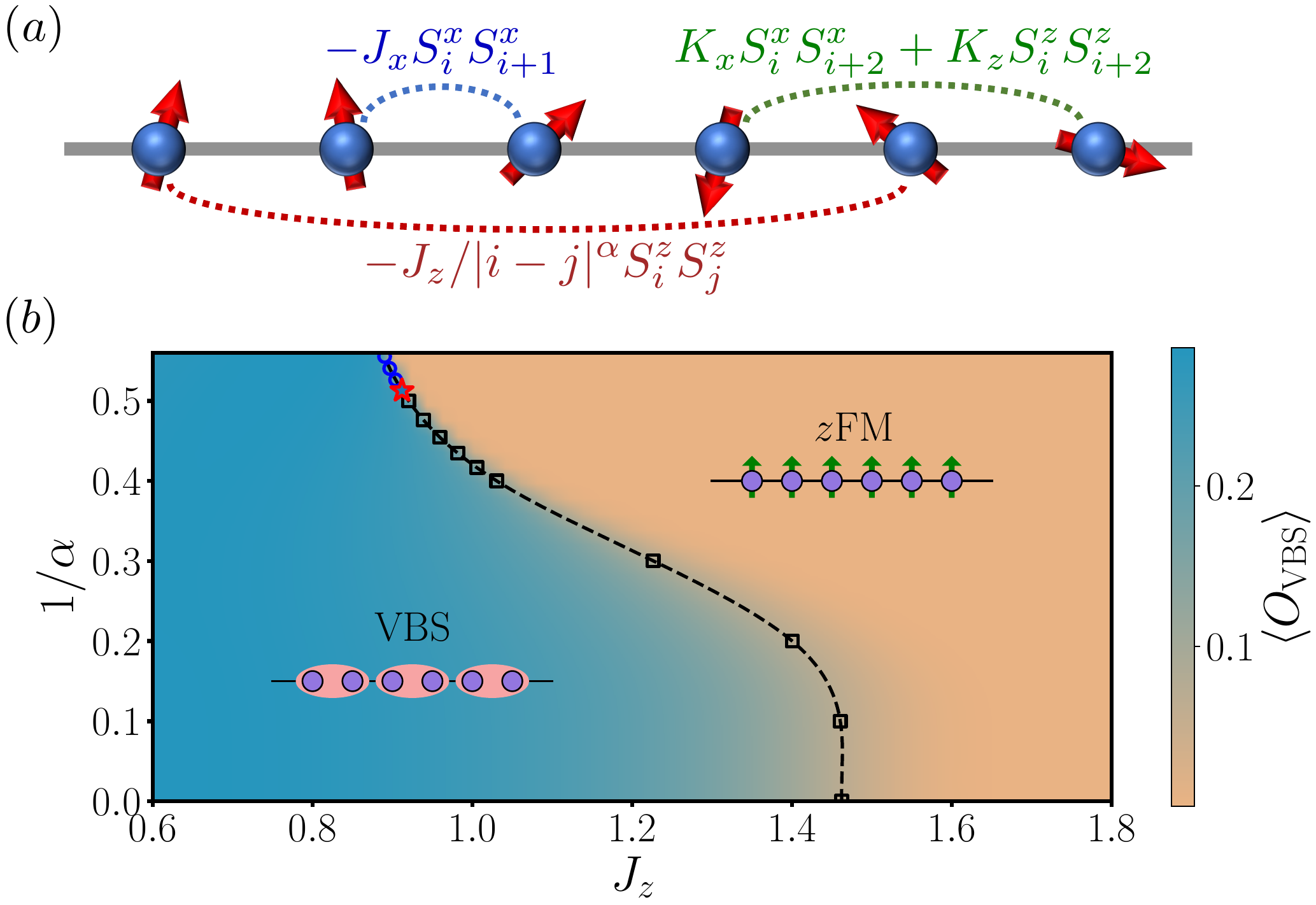}
\caption{(a) Schematic representation of the long-range JM model. (b) Phase diagram of the 1D spin Hamiltonian~\eqref{E1}, mapped out by the VBS order parameter with $L=128$. The markers obtained from the extrapolation of $U_{z\rm FM}$ crossing points demarcate the boundaries between the $z{\rm FM}$ and VBS ordered phases and the dashed line is a guide to the eye (see Figs.~\ref{fig:Binder_all} and~\ref{fig:cp_all} for the details). The black squares indicate the continuous phase transition, while the blue circles indicate the first-order phase transition ($\alpha < \alpha_{\rm c}$ with the estimated multicritical point $\alpha_{\rm c}\approx 1.95$ marked by the red star).
}
\label{fig:phase_diagram}
\end{figure}

The system under study is a frustrated quantum spin chain proposed by Jiang and Motrunich (JM model)~\cite{jiang2019_1ddqcp}, with additional long-range power-law interactions, depicted in Fig.~\ref{fig:phase_diagram}(a). 
The model is defined by the following Hamiltonian: 
\begin{equation}
\begin{aligned}
\label{E1}
H_{\rm LRJM} = & \sum_{i}(-J_{x}S^{x}_{i}S^{x}_{i+1}+K_{x}S^{x}_{i}S^{x}_{i+2}+K_{z}S^{z}_{i}S^{z}_{i+2}) \\
& -\frac{J_{z}}{N(\alpha)}\sum_{i<j}\frac{S^{z}_{i}S^{z}_{j}}{|i-j|^{\alpha}}\,,
\end{aligned}
\end{equation}
where $\bm{S}_{i}=(S^{x}_{i},S^{y}_{i},S^{z}_{i})$ represents the spin-$1/2$ operator on each site $i$. 
$J_{\gamma}$/$K_{\gamma}$ ($\gamma=x,z$) corresponds to the nearest/next-nearest neighbor ferromagnetic(FM)/antiferromagnetic(AFM) couplings. 
For simplicity, we set $K_{x} = K_{z} = 0.5$ and $J_{x} = 1.0$ as the energy unit below. 
The parameter $\alpha$ tunes the power of long-range $S^{z}-S^{z}$ interactions, which tends to the nearest-neighbor short-range JM model in the limit $\alpha \rightarrow \infty$. 
The Kac factor $N(\alpha)(= \frac{1}{L-1}\sum_{i<j}\frac{1}{|i-j|^{\alpha}}$) is included to keep the Hamiltonian extensive. 

When $\alpha \rightarrow \infty$, the original JM model exhibits a 1+1D analogy of DQCP~\cite{jiang2019_1ddqcp,roberts2019dqcp1d,mudry2019_1dqdcp,huang2019_1ddqcp}.
By tuning $J_z$, the system undergoes a continuous quantum phase transition between a valence-bond-solid (VBS) phase ($J_{z}\approx1$) and a spin-ordered ferromagnetic (called $z$FM) phase ($J_{z}\gg1$), which corresponds to the horizontal line $1/\alpha=0$ in Fig.~\ref{fig:phase_diagram}(b). The phase transition is analogous to the 2+1D DQCP \cite{Senthil_2005} as it represents a direct continuous transition between two incompatible spontaneous symmetry breaking phases\cite{jiang2019_1ddqcp,roberts2019dqcp1d}.
Moreover, it can be analytically described by a Luttinger-like field theory with central charge $c=1$, featuring an emergent O(2)$\times$O(2) symmetry at the deconfined critical point~\cite{huang2019_1ddqcp}.

To obtain the ground-state properties of the Hamiltonian $H_{\rm LRJM}$, we adopt the density matrix renormalization group (DMRG)~\cite{white1992prl,schollwock2005rmp,SCHOLLWOCK201196} based on the matrix product state (MPS) technique~\cite{SCHOLLWOCK201196,verstraete2004prl}, which has established itself as one of the best numerical approaches nowadays for one dimensional strongly-correlated systems. 
Our focus lies in exploring the resulting critical behaviors arising from the interplay between the DQCP and long-range interactions.
For most of the calculations, we consider system sizes $L=32$--$256$, while for reliable finite-size scaling analyses, we simulate systems of size $L=192$--$384$. 
To guarantee the numerical accuracy and efficiency in practical calculations, we perform at most $50$ DMRG sweeps with a gradually increased MPS bond dimension, $\chi \leq \chi_{\rm max} = 2048$, under the open boundary condition. Once the MPS energy has converged up to the order $10^{-10}$, the sweeping route would be stopped and the final MPS is believed to be a faithful representation of the true ground state.

\section{Numerical Results}
\label{sec:numerics}

\subsection{Quantum phase diagram: an overview}
Before the illustration of the numerical results, we first summarize the main findings about the long-range JM model in Eq.~\eqref{E1}. An accurate ground-state phase diagram expanded by the axes of $1/\alpha$ and $J_{z}$ is displayed in Fig.~\ref{fig:phase_diagram}(b). It is found that the long-range interaction physics can be classified into two distinct regimes separated by a critical power $\alpha_{\rm c}$. For $\alpha>\alpha_{\rm c}$, the long-range power-law interaction decays very fast such that the interaction tail does not bring any essential change to the DQCP compared with the original model with nearest interaction. The VBS-to-$z$FM transition remains a direct continuous transition characterized by DQCP properties. This large $\alpha$ regime is in some sense roughly consistent with the classification given in Ref.~\cite{defenu2021longrange}, which asserts that if $\alpha$ is larger than a certain critical value, the critical behavior should be indistinguishable from its short-range limit. However, the long-range interaction does extend the $z$FM phase region and the critical point shifts gradually towards smaller $J_z$ with decreasing $\alpha$ as expected. Remarkably, what makes our results fundamentally different from the previous literature (see Ref.~\cite{defenu2021longrange} and references therein) is the small $\alpha$ regime. For the case of $\alpha<\alpha_{\rm c}$, the phase transition is no longer continuous but is now driven into a first-order one by the sufficiently strong long-range interaction. More strikingly, our numerical calculations and low-energy field theory analysis consistently support that the self-duality emerged at the 1+1D DQCP is survived from the strong long-range interaction, giving rise to a first-order phase transition with an enlarged O(2)$\times$O(2) symmetry, which is not well-studied in previous works. 

\subsection{Continuous phase transition at $\alpha>\alpha_{\rm c}$} \label{subsec:continuous}
Similar to the case of conventional QCPs, it is found that the 1+1D analogy of the DQCP hosted in the original short-range JM model is robust against the long-range interaction when the power $\alpha$ is large enough. 

To unveil the continuous nature of the transition, we calculate the associated order parameters, respectively, given by
\begin{equation}
\label{eq:order}
O_{z\rm FM} = \frac{1}{L'} \sum_{i} S_{i}^{z}\ \text{ and }\ O_{\rm VBS} = \frac{1}{L'} \sum_{i} (-1)^{i} \bm{S}_{i}\cdot\bm{S}_{i+1}\,, 
\end{equation}
where the summation is restricted within the middle $L' = L/2$ subsystem to reduce the boundary effect, and resort to standard finite-size scaling analyses according to the scaling form~\cite{Nigel1992}
\begin{equation}
\label{eq:fss}
\langle{O_{z\rm FM/VBS}}\rangle = L^{-\beta/\nu} f[L^{1/\nu}(J_{z}-J_{z}^{\rm c})]\,,
\end{equation}
where $\beta$ and $\nu$ are critical exponents of order parameter and correlation length, respectively. If the phase transition is continuous, the critical exponents extracted independently from the data collapses of $O_{z\rm FM}$ and $O_{\rm VBS}$ should be identical.

\begin{figure}[tb]
\includegraphics[width=\linewidth]{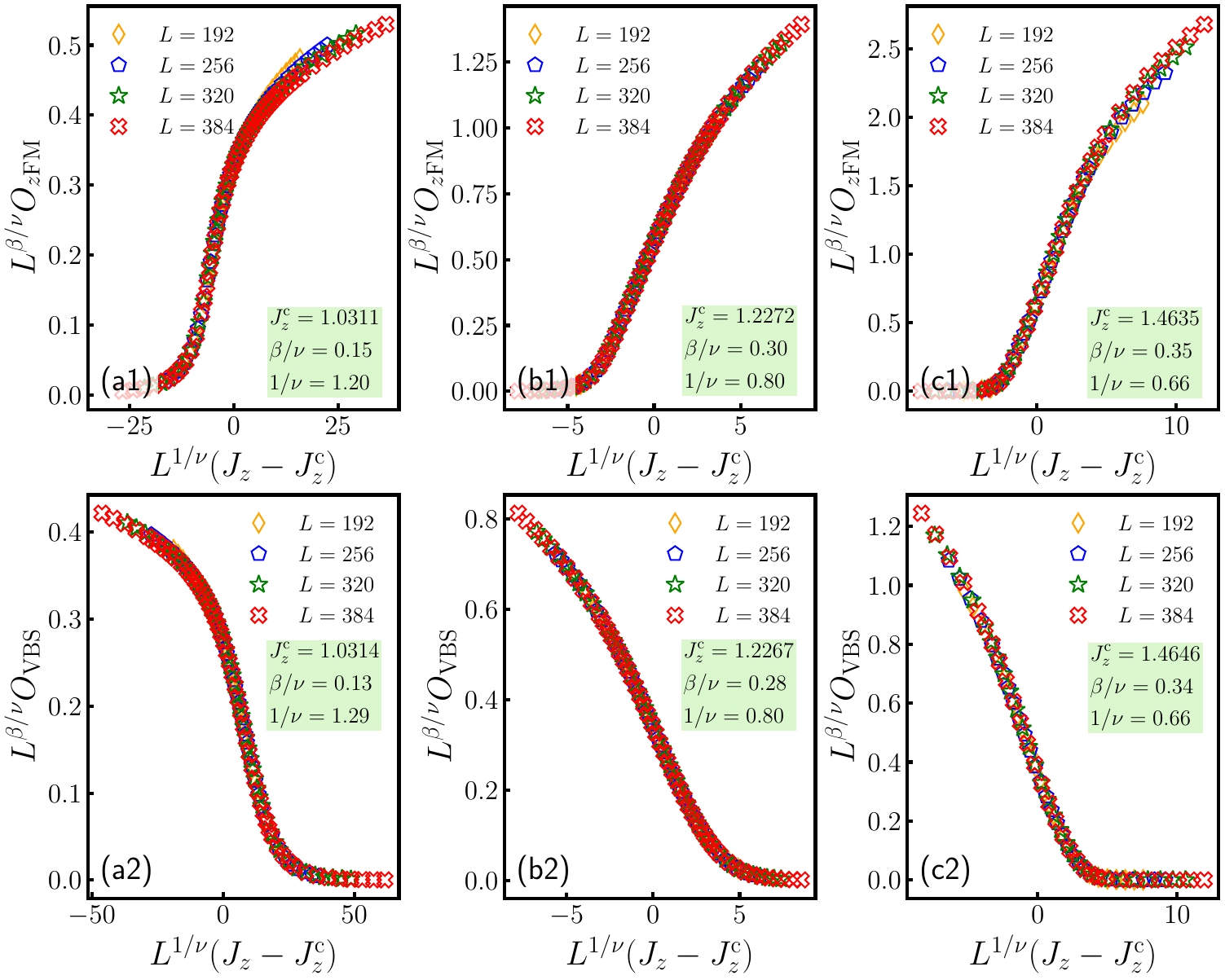}
\caption{The finite-size scaling analysis of order parameters $O_{z\rm FM}$ and $O_{\rm VBS}$ for the long-range JM model with $\alpha=2.50$ [(a1)-(a2)], $\alpha=3.33$ [(b1)-(b2)], and $\alpha=+\infty$ [(c1)-(c2)].
}
\label{fig:ffs_alpha}
\end{figure}

As elaborated in Fig.~\ref{fig:ffs_alpha}, we have performed conventional finite-size scaling analysis for several representative $\alpha$ values. It is noted that we have added a pinning field of strength $1$ at both boundaries when we calculate the $z{\rm FM}$ order parameter, and we also further restrict the summation in Eq.~\eqref{eq:order} within the central two sites to minimize the boundary effect as we can. Similar to Ref.~\cite{luo2019prb,luo2023prb}, we first adjust the exponent $\eta=\beta/\nu$ such that the curves $L^{\eta}\langle{O_{x}}\rangle$ as a function of $J_{z}$ intersect with each other for all system sizes, and $J_{z}^{\rm c}$ can be estimated by the crossing point. Then we adjust the exponent $1/\nu$ until a good collapse of $L^{\eta}\langle{O_{x}}\rangle$ versus $L^{1/\nu}(J_{z}-J_{z}^{\rm c})$ for all $L$ is achieved. 

Following the detailed procedure, we present final data collapses of the order parameters in Fig.~\ref{fig:ffs_alpha}. It is evident that both order parameters obey the standard scaling relation~\eqref{eq:fss} quite well and the extracted critical exponents $\beta/\nu$ are agreed with each other within numerical accuracy, corroborating that the 1+1D DQCP hosted in the original JM model is robust against the long-range $S^{z}-S^{z}$ interaction when $\alpha>\alpha_{\rm c}$. It is also interesting to notice that the exponents $\eta\equiv\beta/\nu$ and $\nu$ both decrease gradually as increasing $1/\alpha$, but the equality $2\nu(1-2\eta)=1$ still holds roughly for all the $\alpha$ values examined here, which is consistent with the prediction from the dual Luttinger-like theory calculations shown below.

\subsection{First-order phase transition at $\alpha<\alpha_{\rm c}$}
Different from the large $\alpha$ regime, the VBS-to-$z{\rm FM}$ phase transition evolves from continuous to first order as the power $\alpha$ is decreased smaller than a certain critical value $\alpha_{\rm c}$, which is beyond the conventional classification of the critical behaviors affected by long-range interactions (see Sec.~\ref{sec:introduction} or Ref.~\cite{defenu2021longrange}). 

A faithful quantity commonly used to distinguish between the continuous and first-order phase transitions is the Binder ratio $U$~\cite{Binder1981}, which is defined by (for the $z$FM order here)
\begin{equation}
\label{eq:Binder_ratio}
U_{z\rm FM} = \frac{1}{2}\left(3- \frac{\langle{O_{z\rm FM}^{4}}\rangle}{\langle{O_{z\rm FM}^{2}}\rangle^{2}}\right)\,.
\end{equation}
This observable has a vanishing scaling dimension and hence can give reliable information on the nature and position of the QCP. For continuous phase transitions, $U_{z\rm FM}$ typically shows a monotonic behavior, while for first-order phase transitions, $U_{z\rm FM}$ instead displays a nonmonotonic behavior and exhibits a diverged negative peak near the QCP with increasing system size~\cite{Binder1993,sandvik2010aip,wang2017prb}.

\begin{figure}[tb]
\includegraphics[width=\linewidth]{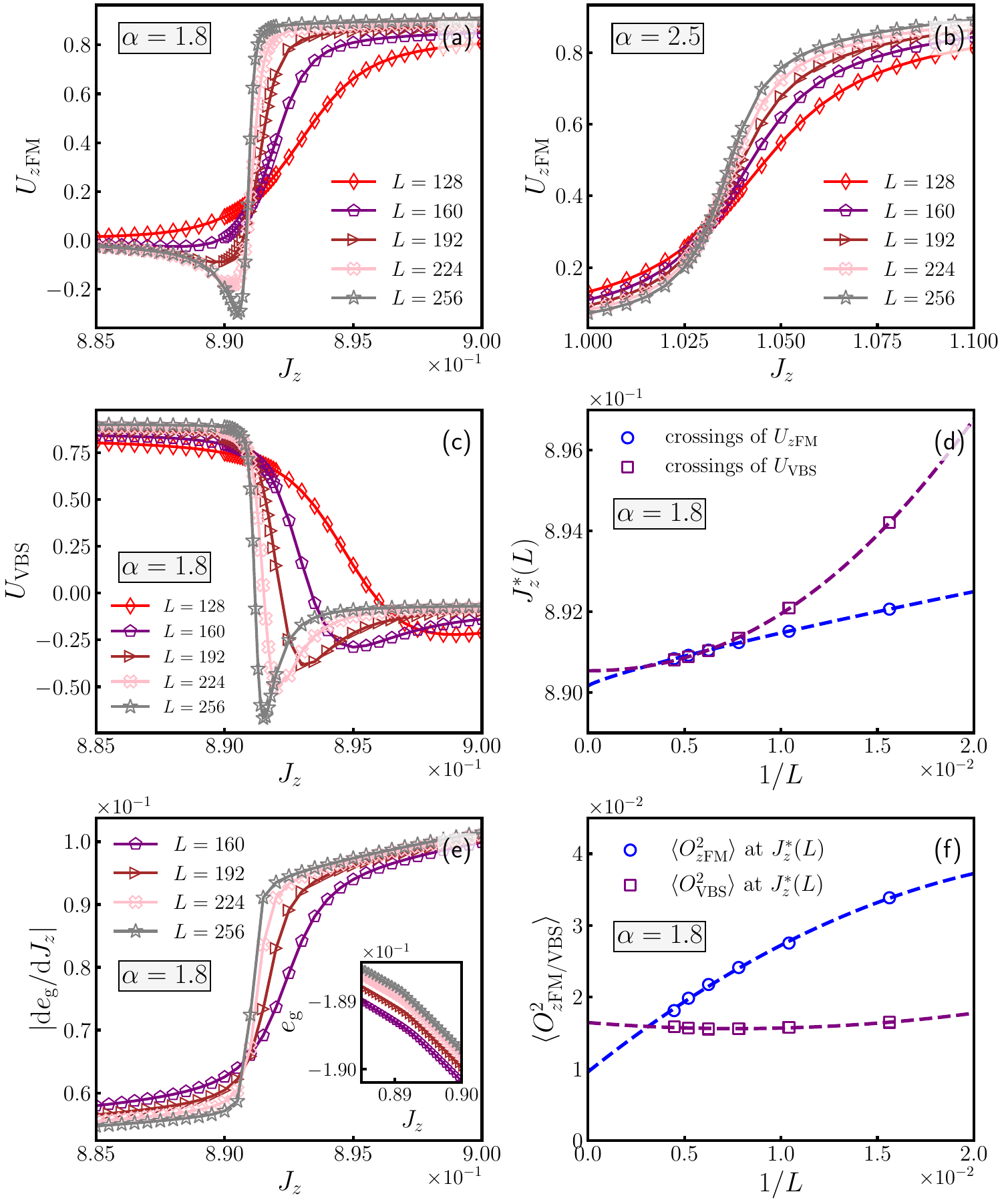}
\caption{(a)-(b) The Binder ratio of the magnetization $U_{z\rm FM}$ versus $J_{z}$. (c) The Binder ratio of the VBS order $U_{\rm VBS}$ versus $J_{z}$. (d) The crossing points $J_{z}^{*}(L)$ of $U_{x}(L)$ and $U_{x}(L+32)$ ($x = {z\rm FM}$ or ${\rm VBS}$) are displayed versus $1/L$. The dashed curves are least-squares fitting according to $J_{z}^{*}(L) = J_{z}^{\rm c} + aL^{-b}$. (e) The ground-state energy per site and its first derivative with respect to $J_{z}$. (f) The squared order parameter $\langle{O_{x}^{2}}\rangle$ ($x = {z\rm FM}$ or ${\rm VBS}$) at the finite-size pseudocritical points $J_{z}^{*}(L)$. The curves are second-order polynomial fits. (a) and (c-f) are plotted for $\alpha=1.8$, but (b)
 is plotted for $\alpha=2.5$\,.}
\label{fig:second_vs_first}
\end{figure}

As shown in Fig.~\ref{fig:second_vs_first}(a)-(b) and Fig.~\ref{fig:Binder_all}, it is found that there exists a critical value $\alpha_{\rm c}\approx1.95$, such that when $\alpha>\alpha_{\rm c}$, the Binder ratio $U_{z\rm FM}$ shows a monotonic growth as $J_{z}$ is increased, but when $\alpha<\alpha_{\rm c}$, $U_{z\rm FM}$ exhibits a nonmonotonic behavior with $J_{z}$ and develops a diverged negative peak near the transition point. The distinct behaviors of $U_{z\rm FM}$ imply that the quantum phase transition changes into a first-order type as $\alpha$ is decreased smaller than $\alpha_{\rm c}$. Furthermore, the precise critical point $J_{z}^{\rm c}$ can be determined by extrapolation based on the relation, $J_{z}^{*}(L) = J_{z}^{\rm c} + aL^{-b}$, where $J_{z}^{*}(L)$ is the crossing point of $U_{z\rm FM}(L)$ and $U_{z\rm FM}(L+32)$~\cite{sandvik2010aip}. In Fig.~\ref{fig:second_vs_first}(c)-(d), one can see a similar diverged negative peak developed in the VBS Binder ratio, $U_{\rm VBS}=(3-\langle{O_{\rm VBS}^{4}}\rangle/\langle{O_{\rm VBS}^{2}}\rangle^{2})/2$, and the critical points obtained independently from $U_{z\rm FM}$ and $U_{\rm VBS}$ are close to each other. Other results of such least-squares fitting are included in Appendix~\ref{sec:appC}, and the estimated critical points are used to demarcate the phase boundaries in Fig.~\ref{fig:phase_diagram}(b). 

To further confirm the first-order phase transition happened at $\alpha<\alpha_{\rm c}$, in Fig.~\ref{fig:second_vs_first}(e), we also calculate the ground-state energy density $e_{\rm g}$ and its corresponding first derivative $\partial{e_{\rm g}}/\partial{J_{z}}$ near the transition point for $\alpha=1.8$\,. It is found that the first-derivative curves are more and more steep and a distinct jump is expected in the thermodynamic limit, which is a definite evidence for first-order phase transitions. Moreover, order parameters at their respective pseudocritical points $J_{z}^{*}(L)$ (i.e., crossing points of $U_{z\rm FM}$ and $U_{\rm VBS}$) are displayed versus $1/L$ in Fig.~\ref{fig:second_vs_first}(f). The coexistence of $z{\rm FM}$ and VBS orders at the critical point can be another decisive indicator of the first-order transition. 

In summary, all the elaborated results consistently corroborate a first-order phase transition in the small $\alpha$ regime. As we have examined the observed phase transition from three different perspectives, each of which has been used as the main evidence for first-order transitions in many works, the first-order transition found here should be reliable with these self-consistent results. It is also worth mentioning that a bimodal histogram of energy or certain quantities may give other evidence for the first-order transition, however, such an illustration seems not practicable within the adopted DMRG framework fundamentally distinct from the sampling-based Monte Carlo simulation. On the other hand, in the present work, we only focus on the properties of the first-order and continuous transitions. The tricritical point $\alpha_{\rm c}$ is very interesting and can be studied by the flowgram method developed in~\cite{kuklov2006deconfined}, but we leave it for future investigations.

\subsection{Enlarged symmetry on the critical line}

\begin{figure}[tb]
\includegraphics[width=\linewidth]{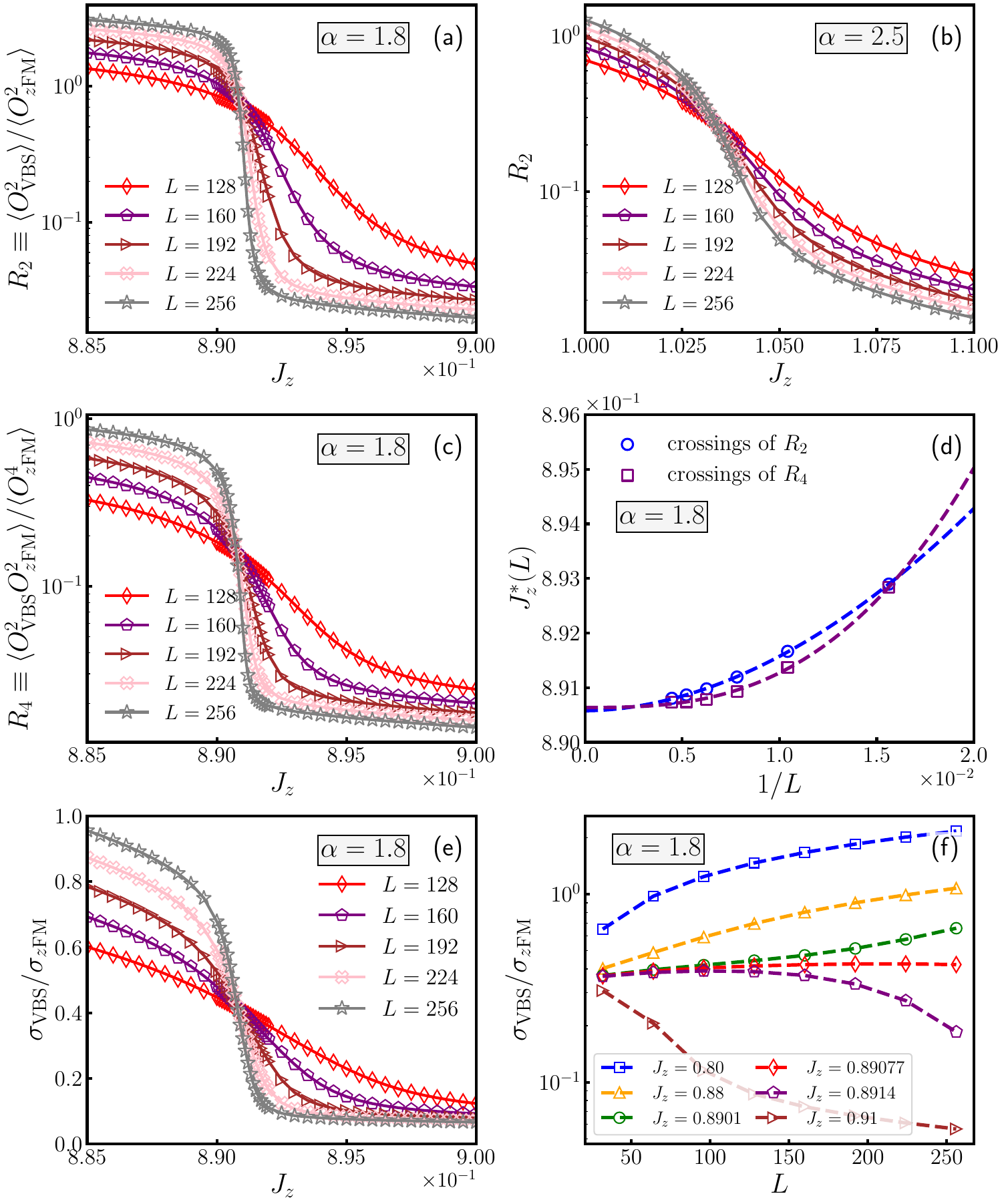}
\caption{(a)-(b) The ratio of the squared order parameters $R_{2}$ versus $J_{z}$. (c) The cross ratio of the squared order parameters $R_{4}$ versus $J_{z}$. (d) The crossing locations $J_{z}^{*}(L)$ of $R_{2/4}(L)$ and $R_{2/4}(L+32)$ are shown versus $1/L$. The dashed curves are least-squares fitting according to the relation, $J_{z}^{*}(L) = J_{z}^{\rm c} + aL^{-b}$. (e) The variance ratio $\sigma_{\rm VBS}/\sigma_{z\rm FM}$, where $\sigma_{x}\equiv(\langle{O_{x}^{4}}\rangle-\langle{O_{x}^{2}}\rangle^{2})^{1/2}$ ($x = {z\rm FM}$ or ${\rm VBS}$), as a function of $J_{z}$ for several system sizes. (f) The ratio $\sigma_{\rm VBS}/\sigma_{z\rm FM}$ as a function of $L$ for several $J_{z}$ near the critical point $J_{z}^{\rm c}$. (a) and (c-f) are plotted for $\alpha=1.8$, but (b)
 is plotted for $\alpha=2.5$\,.}
\label{fig:symmetry}
\end{figure}

One of the most significant features of the 1+1D DQCP in the original short-range JM model is the O(2)$\times$O(2) symmetry emerged exactly at the deconfined critical point~\cite{jiang2019_1ddqcp,roberts2019dqcp1d,huang2019_1ddqcp}. Therefore, it is natural to ask whether this enlarged symmetry still exists at the QCPs of the long-range JM model. 

For this purpose, we calculate the ratio of the squared order parameters defined by $R_{2}=\langle{O_{\rm VBS}^{2}}\rangle/\langle{O_{z\rm FM}^{2}}\rangle$. According to Refs.~\cite{nahum2015prl,nahum2019prl,nahum2019prb,liu2022emergent}, if the VBS and $z$FM order parameters have the same scaling dimension, the ratio $R_{2}$ should be size-independent at the transition point, and the QCP would have an enlarged symmetry that rotates these two orders.
The results of $R_{2}$ are detailed and summarized in Fig.~\ref{fig:symmetry}(a)-(b) and Fig.~\ref{fig:r2_all}; it is obvious that $R_{2}$ becomes size-independent near the QCP for all $\alpha$, indicating that the VBS-to-$z{\rm FM}$ transition still hosts the O(2)$\times$O(2) symmetry even when its nature has been driven into first-order. Similarly, as shown in Fig.~\ref{fig:symmetry}(c)-(d), the cross ratio of order parameters, $R_{4}=\langle{O_{\rm VBS}^{2}O_{z\rm FM}^{2}}\rangle/\langle{O_{z\rm FM}^{4}}\rangle$, of different system sizes also intersect with each other roughly at a single point, and the extrapolation of the pseudocritical point is also close to the one obtained from the ratio $R_{2}$.

In Fig.~\ref{fig:symmetry}(e), we also examine the variance ratio $\sigma_{\rm VBS}/\sigma_{z\rm FM}$ where $\sigma_{x}=(\langle{O_{x}^{4}}\rangle-\langle{O_{x}^{2}}\rangle^{2})^{1/2}$~\cite{nahum2015prl,sato2017prl}, which is another useful detector for symmetries at QCPs. Similar to $R_{2}$, we can indeed see an intersection of $\sigma_{\rm VBS}/\sigma_{z\rm FM}$ curves of all $L$ at the transition point. Fig.~\ref{fig:symmetry}(f) explicitly shows the dependence of $\sigma_{\rm VBS}/\sigma_{z\rm FM}$ on $L$ near the critical point. The universal behavior of $\sigma_{\rm VBS}/\sigma_{z\rm FM}$ around $J_{z}\approx0.8908$ gives another evidence for the enlarged O(2)$\times$O(2) symmetry at the first-order phase transition. 

Till now, the presented numerical simulations are consistent with each other and points to a first-order phase transition with enlarged O(2)$\times$O(2) symmetry beyond conventional understandings. Therefore, it is necessary to explain our results, and the key point is the emergent self-duality survived from the long-range interaction which preserves the O(2)$\times$O(2) symmetry.

\subsection{Transition nature of the original JM model with $K_{z}\neq1/2$}

\begin{figure}[tb]
\includegraphics[width=\linewidth]{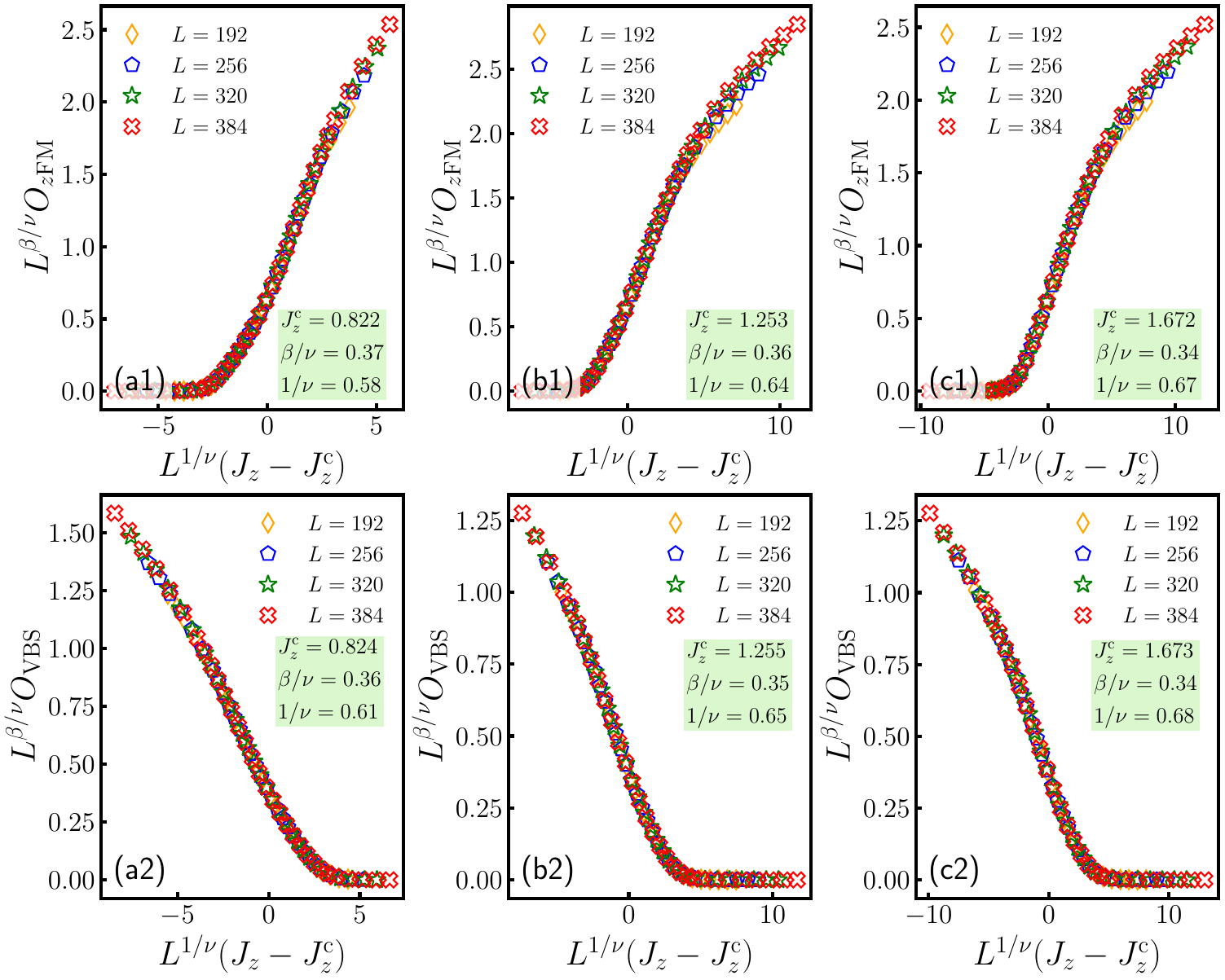}
\caption{The finite-size scaling analysis of order parameters $O_{z\rm FM}$ and $O_{\rm VBS}$ for the original JM model with $K_{z}=0.2$ [(a1)-(a2)], $K_{z}=0.4$ [(b1)-(b2)], and $K_{z}=0.6$ [(c1)-(c2)].}
\label{fig:ffs_kz}
\end{figure}

Before the presentation of the low-energy field theory analysis, it is also necessary to verify that the observed first-order phase transition is not induced by the naive modification of $K_{z}$, since the inclusion of the long-range $S^{z}-S^{z}$ interaction in the JM model can effectively change the value of $K_{z}$. For this purpose, we investigate the quantum phase transition of the \textit{original} JM model with the parameter setting, $J_{x}=1$, $K_{x}=1/2$, and $K_{z}\ne1/2$. 

Following the same procedure explained in Sec.~\ref{subsec:continuous}, we utilize the standard finite-size scaling analysis according to Eq.~\eqref{eq:fss} to extract the quantum critical point $J_{z}^{\rm c}$ and related critical exponents, $\beta$ and $\nu$. As summarized in Fig.~\ref{fig:ffs_kz}, it is clear that the obtained critical exponents $\beta/\nu$ are almost identical for $z{\rm FM}$ and VBS orders, which is a key property of the DQCP theory~\cite{jiang2019_1ddqcp}, indicating that the quantum phase transition is still continuous. As the effective value of $K_{z}$ modified by the long-range $S^{z}-S^{z}$ interaction, $K_{z}^{\rm eff}=1/2-J_{z}/[2^{\alpha}N(\alpha)]$, is larger than $0.2$ at the critical point for $\alpha=1.8$, the results shown here can support that the first-order phase transition found in the long-range JM model at $\alpha=1.8$ is indeed induced by the long-range $S^{z}-S^{z}$ interaction, thus ruling out the possibility that the first-order transition is caused by a naive modification of the coupling $K_{z}$ in the original JM model.

\section{Low-energy effective theory}
\label{sec:theory}

The phase transition between the $z$FM and VBS orders is second order when $\alpha$ is large and first order when $\alpha$ is small. 
The continuous to first order transition happens at the critical power $\alpha_{\rm c}$\,. 
This continuous to first-order transition is driven by the long-range $S^z$-$S^z$ interaction. 
According to the bosonization method\cite{giamarchi2003_book,jiang2019_1ddqcp,mudry2019_1dqdcp} (see Section~\ref{sec:SM5} of Appendix for the details), the spin operators can be represented by a bosonic field $\phi$ in the continuous limit, 
\begin{equation}
S_j^z \sim \cos\phi(x)/2,\quad
S_j^x \sim -\sin\phi(x)/2,
\end{equation}
where the discrete coordinate is replaced by its continuous version, $x_j\rightarrow x$.
Thus, the 1D long-range $S^z$-$S^z$ interaction takes the following form in the effective continuum theory \cite{longrangeboson2017},
\begin{equation}
\sum_{i,j}\frac{S^{z}_{i}S^{z}_{j}}{|i-j|^{\alpha}}\sim \int dx dy \frac{\cos[\phi (x)] \cos[\phi(y)]}{4|x-y|^{\alpha}},
\label{eq:LRInt}
\end{equation}
where $x,y$ are the 1D continuous coordinates.
The effective action in the Euclidean path integral formulation under bosonization is given by \cite{giamarchi2003_book,jiang2019_1ddqcp},
\begin{equation}
\begin{aligned}
S =&\int d\tau\,dx\ \Big[ \frac{i}{\pi} \partial_{\tau}\phi \partial_x\theta
+\frac{v}{2\pi} \Big( \frac{1}{g} (\partial_x\theta)^2 +g(\partial_x\phi)^2  \Big) \Big]    \\
+& \int d\tau\,dx\ \Big[ \lambda_u \cos(4\theta) 
+\lambda_a \cos(2\phi) \Big]
+S_{LR}.
\end{aligned}
\end{equation}
with imaginary time $\tau$, spatial coordinates $x$, velocity $v$ and Luttinger parameter $g$. 
$\lambda_u$ and $\lambda_a$ are the most relevant short-range interaction which persevering the symmetry of the JM model.
The long-range part in the Lagrangian is deduced from Eq.~\eqref{eq:LRInt},
\begin{equation}
\begin{aligned}
&S_{LR}=\frac{\lambda_+}{2}\int d\tau dxdr \frac{1}{|r|^{\alpha}} \cos\big[\phi(x+r,\tau)+\phi(x,\tau)\big]   \\
+&\frac{\lambda_-}{2}\int d\tau dxdr \frac{1}{|r|^{\alpha}} \cos\big[\phi(x+r,\tau)-\phi(x,\tau)\big]\Big).
\end{aligned}
\label{eq:LRIntBare}
\end{equation}
where $r$ denotes the relative distance of the fields.
Here, the $\cos$-$\cos$ correlation in Eq.~\eqref{eq:LRInt} has been separated into two parts, whose effects would be different.
For smaller interaction range $r$, $\lambda_-$ will contribute to the renormalization of the Luttinger parameters.
We would focus on the renormalization of the long-range contribution and can omit this shorter range $r$ contribution at this stage.
Moreover, it should also be noticed that the Luttinger parameter $g$ could be a non-universal quantity at the critical point.
While for the pure XXZ model, the explicit value of $g$ could be deduced from the microscopic parameters based on the Bethe ansatz \cite{yang1966aXXZ,yang1966bXXZ,yang1966cXXZ},
for the general spin model, e.g., for the JM model with long-range interaction, it would be hard to handle out the explicit value of $g$ from the microscopic parameters.

Based on the RG analysis (in Section~\ref{sec:SM6} of Appendix), 
the long-range $S^z$-$S^z$ interaction Eq.~\eqref{eq:LRIntBare} is irrelevant or less relevant than the short-range one when the exponent is greater than some critical value, $\alpha>\alpha_c$. 
In the large $\alpha$ regime, the continuous transition between the VBS and $z$FM phases driven by short-range interaction \cite{jiang2019_1ddqcp} is stable against the long-range perturbation. 
On the other hand, when $\alpha$ is smaller than the critical value, the long-range $S^z$-$S^z$ interaction becomes the most relevant operator of the system and drives the continuous transition to a first order transition. 
To decode the non-trivial transition between the VBS and $z$FM phases, it will be more sufficient to work in the effective dual theory formulation, as done in the original JM model \cite{jiang2019_1ddqcp,roberts2019dqcp1d,mudry2019_1dqdcp,huang2019_1ddqcp}.

\subsection{Dual field theory}
The dual field theory description plays important role for understanding the 1D DQCP nature of the JM model\cite{jiang2019_1ddqcp}.
Especially, the continuous transition between VBS and $z$FM phases in the original JM model has emergent self-duality as shown in \cite{jiang2019_1ddqcp,roberts2019dqcp1d,mudry2019_1dqdcp,huang2019_1ddqcp}, since the scaling dimensions of the two order parameters are the same in both numerical and theoretical calculations. 
While the construction of the dual theory field requires much effort,
the final formulation is direct and simple, that is, the $z$-FM and VBS order parameters could be represented by the continuous dual field $\tilde{\theta}$ as $\Psi_{z\text{FM}} \sim  \sin(\tilde{\theta})$ and
$\Psi_{\text{VBS}} \sim  \cos(\tilde{\theta})$ \cite{jiang2019_1ddqcp}.
The dual field theory unifies $z$FM and VBS order parameters together by the single field $\tilde{\theta}$.
The self-duality manifests in the dual Luttinger liquid theory in the imaginary-time path integral action \cite{giamarchi2003_book,jiang2019_1ddqcp},
\begin{equation}
\begin{aligned}
&\tilde{S}=\int d\tau\,dx\ \Big[ \frac{i}{\pi} \partial_{\tau}\tilde{\phi} \partial_x\tilde{\theta}
+\frac{\tilde{v}}{2\pi} \Big( \frac{1}{\tilde{g}} (\partial_x\tilde{\theta})^2 +\tilde{g}(\partial_x\tilde{\phi})^2  \Big) \Big]    \\
&+ \int d\tau\,dx\ \Big[ 
\tilde{\lambda} \cos(2\tilde{\theta})  \Big]
+\tilde{S}_{LR},
\label{eq:DirectLuttinger}
\end{aligned}
\end{equation}
with the $(1+1)$D spatial-time coordinates $(x,\tau)$, effective velocity $\tilde{v}$ and Luttinger parameter $\tilde{g}$. 
The tilde symbol is used to emphasize the difference from the original field theory in Eq.~(\ref{eq:DirectLuttinger}). 
$(\tilde{\theta},\tilde{\phi})$ is a pair of conjugate fields in the dual field theory \cite{jiang2019_1ddqcp}.
We first summarized some basic results in the short range system without $\tilde{S}_{LR}$ \cite{jiang2019_1ddqcp,roberts2019dqcp1d,mudry2019_1dqdcp,huang2019_1ddqcp}.
Since the relevant problem lying in the parameter regime $\tilde{g}\in(1/2,2)$,
$\tilde{\lambda}$ is the only relevant short-ranged operator, which could drive the phase transition between the $z$-FM and VBS orders. 
A relevant positive (negative) $\tilde{\lambda}$ will pin down the dual field $\tilde{\theta}=\frac{\pi}{2}$ ($\tilde{\theta}=0$), which corresponds to the $z$FM (VBS) phase, $\Psi_{z\text{FM}}\neq 0$ ($\Psi_{\text{VBS}}\neq 0$).
On the contrary, $\tilde{\phi}$ field could be fully integrated out in the path integral since its interaction term is irrelevant in the critical theory, leading to a pure sine-Gordon theory for the field $\tilde{\theta}$.
Instantly, $z$FM and VBS order parameters have the same scaling dimensions $\mathrm{dim}[\Psi_{z\text{FM}}]=\mathrm{dim}[\Psi_{\text{VBS}}]=\tilde{g}/4$ \cite{jiang2019_1ddqcp,roberts2019dqcp1d}. 
The emergent self-duality permutes VBS and $z$FM order parameters. 
Combined with the global symmetry O(2), the emergent self-duality promotes the global symmetry to O(2)$\times$ O(2).

From the above numerical simulations, the substantial evidence shows that the emergent self-duality still persists in the first order transition when the long-range $S^z$-$S^z$ interaction is relevant. 
Indeed, the long-range $S^z$-$S^z$ interaction is effectively self-dual preserving at the critical point. 
Based on the dual Bosonization approach, the long-range $S^z$-$S^z$ interaction in the dual theory is represented as,
\begin{equation}
\begin{aligned}
\tilde{S}_{LR}
=&\frac{\tilde{\lambda}_-}{2} \int d\tau dxdr \frac{1}{|r|^{\alpha}} \cos\big[\tilde{\theta}(x+r,\tau)-\tilde{\theta}(x,\tau)\big]  \\
-&\frac{\tilde{\lambda}_+}{2}\int d\tau dxdr \frac{1}{|r|^{\alpha}}\cos\big[\tilde{\theta}(x+r,\tau)+\tilde{\theta}(x,\tau)\big].
\end{aligned}
\end{equation}
Here, the effective coupling $\tilde{\lambda}_-$ drives the system to a spatial uniform pattern, 
while the sign of $\tilde{\lambda}_+$ leads to the VBS or $z$FM order.
In the infrared limit, long-range $\tilde{\lambda}_+$ has a similar effect as the short-range $\tilde{\lambda}$ and a combination of them leads to an renormalized driving coupling, which will tune the phase transition between VBS and $z$FM order.
This effective tuning coupling also accounts for the shift of the phase boundary to the left in Fig.~\ref{fig:phase_diagram} (b). 
Under the RG flow (in Section ~\ref{sec:SM6} of SM), $\tilde{\lambda}_-$ becomes more relevant than $\tilde{\lambda}_+$, while the VBS-$z$FM phase transition is still tuned by $\tilde{\lambda}_+$.
When the power of long-range interaction becomes smaller than the critical value $\alpha<\alpha_c$, $\tilde{\lambda}_-$ becomes the most relevant operator of the system, driving the second order transition into a first order one. 
Therefore, only $\tilde{\lambda}_-$ term affects the infrared fate along the critical line.

Under the self-duality $\sin(\tilde{\theta}) \leftrightarrow \cos(\Tilde{\theta})$, the long-range interaction transforms as $\tilde{\lambda}_- \rightarrow \tilde{\lambda}_-, \tilde{\lambda}_+ \rightarrow -\tilde{\lambda}_+$. 
The $\tilde{\lambda}_-$ term is manifestly self-dual invariant. 
On the contrary, the $\tilde{\lambda}_+$ term breaks the self-duality and transforms the same as $\tilde{\lambda}$.
Along the critical line, the VBS to $z$FM tuning coupling tends to zero and the system is self-dual invariant in the low-energy field theory.
The emergent self-duality persists along the transition line from continuous to first order transition. 
This can be understood as the self-duality protected criticality \cite{lu2021prb}. 
Since the self-duality permutes the two phases VBS and $z$FM, the self-duality invariant region should be the interface between them, and that is the phase boundary in Fig.~\ref{fig:phase_diagram} (b).

\section{Discussions and Conclusions}
\label{sec:con}
%\dcl{may not discuss the brown part then:}{\color{brown} We noticed recent works~\cite{zhao2020prl,Zhao_2020} introduced an additional relevant perturbation, which drives the second order into a first-order transition. Consequently, we can identify the DQCP as a multicritical Lifshitz point and provide a possible explanation for the previously inconsistent critical-exponent bounds obtained from the conformal-bootstrap~\cite{yu2016prl}. Moreover, Lu et al~\cite{lu2021prb} used the field theory method to demonstrate that the introduction of long-range current-current interactions drives the continuous DQCP to the first-order transition.} On the other hand, 
We noticed that some researchers have discovered emergent symmetry at special~\cite{zhao2019symmetry,Sun_2021} or weakly first-order transitions~\cite{sreejith2019prl,serna2019prb,wildeboer2020prb,wildeboer2020prb2}. They explained that the absence of a free energy barrier 
%for coexisting orders 
allows different orders to transform into each other. However, we emphasize that our model exhibits an unambiguous enlarged O(2)$\times$O(2) symmetry at the strong first-order phase transition, which is different from the previous cases. Our findings reveal that, in the low-energy effective field theory, the self-dual invariant long-range operator changes from being irrelevant to relevant as $\alpha$ decreases, resulting in a first-order critical point with emergent self-duality, which leads to enlarged O(2)$\times$O(2) global symmetry. Additionally, emergent supersymmetry~\cite{yu2019prb} has also been discovered at first-order critical points.

Regarding experimental realization, Lee et al~\cite{lee2022landauforbidden} recently proposed a Landau-forbidden quantum phase transition with an emergent symmetry in a one-dimensional strongly interacting array of trapped neutral Rydberg atoms. This can be experimentally observed with measurement snapshots on a standard computational basis.

%\section{Conclusion}
In conclusion, we perform large-scale DMRG simulations to decipher the critical properties of the JM model with long-range interactions. Our numerical simulation unambiguously demonstrates that the emergent self-duality appears along the critical line, from the continuous transition to the first-order transition. And the self-duality enlarges the global symmetry to O(2)$\times$O(2). This finding aligns with the Luttinger-liquid theory calculations, where part of the long-range spin-spin interaction becomes the self-dual invariant relevant operator and drives the continuous transition to the first order transition. This is reminiscent of the tricritical Ising model, where the self-dual invariant operator can drive the tricritical point to either the Ising transition or the first order transition. In particular, the first order transition is a gapped phase with 3 groundstate degeneracies due to the anomalous self-duality \cite{chang2019topological,thorngren2021fusion}. 

%For the first time, we unambiguously demonstrate that, in addition to the DQCP with emergent symmetry, the first-order critical point also exhibits emergent symmetry, protected by self-duality in the infrared (IR) limit. This deviates from the conventional understanding of emergent symmetry typically occurring at continuous phase transitions. 

We leave for future work the determination of a precise value of $\alpha_{c}$, comparisons of universal quantities with long-range interactions through renormalization group analysis, and a comparative study of the quantum critical behavior at the multicritical point. Our work paves a new way for understanding the interplay between unconventional quantum critical points and long-range physics through experimental and theoretically controlled manner.

\begin{acknowledgements}
We thanks Yi-Zhuang You, Zi Yang Meng, Ruben Verresen, and Hai-Qing Lin for very helpful discussions. Numerical simulations were carried out with the ITENSOR package~\cite{itensor} on the Kirin No.2 High Performance Cluster supported by the Institute for Fusion Theory and Simulation (IFTS) at Zhejiang University. Z.P. is supported by National Natural Science Foundation of China (No. 12147104). X.-J.Yu acknowledges support from the start-up grant of Fuzhou University.

\end{acknowledgements}

\bibliography{main}

%merlin.mbs apsrev4-1.bst 2010-07-25 4.21a (PWD, AO, DPC) hacked
%Control: key (0)
%Control: author (0) dotless jnrlst
%Control: editor formatted (1) identically to author
%Control: production of article title (0) allowed
%Control: page (1) range
%Control: year (0) verbatim
%Control: production of eprint (0) enabled
\begin{thebibliography}{150}%
\makeatletter
\providecommand \@ifxundefined [1]{%
 \@ifx{#1\undefined}
}%
\providecommand \@ifnum [1]{%
 \ifnum #1\expandafter \@firstoftwo
 \else \expandafter \@secondoftwo
 \fi
}%
\providecommand \@ifx [1]{%
 \ifx #1\expandafter \@firstoftwo
 \else \expandafter \@secondoftwo
 \fi
}%
\providecommand \natexlab [1]{#1}%
\providecommand \enquote  [1]{``#1''}%
\providecommand \bibnamefont  [1]{#1}%
\providecommand \bibfnamefont [1]{#1}%
\providecommand \citenamefont [1]{#1}%
\providecommand \href@noop [0]{\@secondoftwo}%
\providecommand \href [0]{\begingroup \@sanitize@url \@href}%
\providecommand \@href[1]{\@@startlink{#1}\@@href}%
\providecommand \@@href[1]{\endgroup#1\@@endlink}%
\providecommand \@sanitize@url [0]{\catcode `\\12\catcode `\$12\catcode
  `\&12\catcode `\#12\catcode `\^12\catcode `\_12\catcode `\%12\relax}%
\providecommand \@@startlink[1]{}%
\providecommand \@@endlink[0]{}%
\providecommand \url  [0]{\begingroup\@sanitize@url \@url }%
\providecommand \@url [1]{\endgroup\@href {#1}{\urlprefix }}%
\providecommand \urlprefix  [0]{URL }%
\providecommand \Eprint [0]{\href }%
\providecommand \doibase [0]{http://dx.doi.org/}%
\providecommand \selectlanguage [0]{\@gobble}%
\providecommand \bibinfo  [0]{\@secondoftwo}%
\providecommand \bibfield  [0]{\@secondoftwo}%
\providecommand \translation [1]{[#1]}%
\providecommand \BibitemOpen [0]{}%
\providecommand \bibitemStop [0]{}%
\providecommand \bibitemNoStop [0]{.\EOS\space}%
\providecommand \EOS [0]{\spacefactor3000\relax}%
\providecommand \BibitemShut  [1]{\csname bibitem#1\endcsname}%
\let\auto@bib@innerbib\@empty
%</preamble>
\bibitem [{\citenamefont {Sachdev}(2023)}]{sachdev2023quantum}%
  \BibitemOpen
  \bibfield  {author} {\bibinfo {author} {\bibfnamefont {Subir}\ \bibnamefont
  {Sachdev}},\ }\href@noop {} {\emph {\bibinfo {title} {Quantum Phases of
  Matter}}}\ (\bibinfo  {publisher} {Cambridge University Press},\ \bibinfo
  {year} {2023})\BibitemShut {NoStop}%
\bibitem [{\citenamefont {Sachdev}(2011)}]{sachdev_2011}%
  \BibitemOpen
  \bibfield  {author} {\bibinfo {author} {\bibfnamefont {Subir}\ \bibnamefont
  {Sachdev}},\ }\href {\doibase 10.1017/CBO9780511973765} {\emph {\bibinfo
  {title} {Quantum Phase Transitions}}},\ \bibinfo {edition} {2nd}\ ed.\
  (\bibinfo  {publisher} {Cambridge University Press},\ \bibinfo {year}
  {2011})\BibitemShut {NoStop}%
\bibitem [{\citenamefont {Wilson}(1983)}]{wilson1983rmp}%
  \BibitemOpen
  \bibfield  {author} {\bibinfo {author} {\bibfnamefont {Kenneth~G.}\
  \bibnamefont {Wilson}},\ }\bibfield  {title} {\enquote {\bibinfo {title} {The
  renormalization group and critical phenomena},}\ }\href {\doibase
  10.1103/RevModPhys.55.583} {\bibfield  {journal} {\bibinfo  {journal} {Rev.
  Mod. Phys.}\ }\textbf {\bibinfo {volume} {55}},\ \bibinfo {pages} {583--600}
  (\bibinfo {year} {1983})}\BibitemShut {NoStop}%
\bibitem [{\citenamefont {Kramers}\ and\ \citenamefont
  {Wannier}(1941)}]{Kramers1941pr}%
  \BibitemOpen
  \bibfield  {author} {\bibinfo {author} {\bibfnamefont {H.~A.}\ \bibnamefont
  {Kramers}}\ and\ \bibinfo {author} {\bibfnamefont {G.~H.}\ \bibnamefont
  {Wannier}},\ }\bibfield  {title} {\enquote {\bibinfo {title} {Statistics of
  the two-dimensional ferromagnet. part i},}\ }\href {\doibase
  10.1103/PhysRev.60.252} {\bibfield  {journal} {\bibinfo  {journal} {Phys.
  Rev.}\ }\textbf {\bibinfo {volume} {60}},\ \bibinfo {pages} {252--262}
  (\bibinfo {year} {1941})}\BibitemShut {NoStop}%
\bibitem [{\citenamefont {Kadanoff}\ and\ \citenamefont
  {Ceva}(1971)}]{kadanoff1971prb}%
  \BibitemOpen
  \bibfield  {author} {\bibinfo {author} {\bibfnamefont {Leo~P.}\ \bibnamefont
  {Kadanoff}}\ and\ \bibinfo {author} {\bibfnamefont {Horacio}\ \bibnamefont
  {Ceva}},\ }\bibfield  {title} {\enquote {\bibinfo {title} {Determination of
  an operator algebra for the two-dimensional ising model},}\ }\href {\doibase
  10.1103/PhysRevB.3.3918} {\bibfield  {journal} {\bibinfo  {journal} {Phys.
  Rev. B}\ }\textbf {\bibinfo {volume} {3}},\ \bibinfo {pages} {3918--3939}
  (\bibinfo {year} {1971})}\BibitemShut {NoStop}%
\bibitem [{\citenamefont {Xu}(2012)}]{xu2012unconventional}%
  \BibitemOpen
  \bibfield  {author} {\bibinfo {author} {\bibfnamefont {Cenke}\ \bibnamefont
  {Xu}},\ }\bibfield  {title} {\enquote {\bibinfo {title} {Unconventional
  quantum critical points},}\ }\href@noop {} {\bibfield  {journal} {\bibinfo
  {journal} {International Journal of Modern Physics B}\ }\textbf {\bibinfo
  {volume} {26}},\ \bibinfo {pages} {1230007} (\bibinfo {year}
  {2012})}\BibitemShut {NoStop}%
\bibitem [{\citenamefont {Senthil}\ \emph {et~al.}(2004)\citenamefont
  {Senthil}, \citenamefont {Balents}, \citenamefont {Sachdev}, \citenamefont
  {Vishwanath},\ and\ \citenamefont {Fisher}}]{senthil2004prb}%
  \BibitemOpen
  \bibfield  {author} {\bibinfo {author} {\bibfnamefont {T.}~\bibnamefont
  {Senthil}}, \bibinfo {author} {\bibfnamefont {Leon}\ \bibnamefont {Balents}},
  \bibinfo {author} {\bibfnamefont {Subir}\ \bibnamefont {Sachdev}}, \bibinfo
  {author} {\bibfnamefont {Ashvin}\ \bibnamefont {Vishwanath}}, \ and\ \bibinfo
  {author} {\bibfnamefont {Matthew P.~A.}\ \bibnamefont {Fisher}},\ }\bibfield
  {title} {\enquote {\bibinfo {title} {Quantum criticality beyond the
  landau-ginzburg-wilson paradigm},}\ }\href {\doibase
  10.1103/PhysRevB.70.144407} {\bibfield  {journal} {\bibinfo  {journal} {Phys.
  Rev. B}\ }\textbf {\bibinfo {volume} {70}},\ \bibinfo {pages} {144407}
  (\bibinfo {year} {2004})}\BibitemShut {NoStop}%
\bibitem [{\citenamefont {Senthil}\ \emph {et~al.}(2005)\citenamefont
  {Senthil}, \citenamefont {Balents}, \citenamefont {Sachdev}, \citenamefont
  {Vishwanath},\ and\ \citenamefont {P.~A.~Fisher}}]{Senthil_2005}%
  \BibitemOpen
  \bibfield  {author} {\bibinfo {author} {\bibfnamefont {Todadri}\ \bibnamefont
  {Senthil}}, \bibinfo {author} {\bibfnamefont {Leon}\ \bibnamefont {Balents}},
  \bibinfo {author} {\bibfnamefont {Subir}\ \bibnamefont {Sachdev}}, \bibinfo
  {author} {\bibfnamefont {Ashvin}\ \bibnamefont {Vishwanath}}, \ and\ \bibinfo
  {author} {\bibfnamefont {Matthew}\ \bibnamefont {P.~A.~Fisher}},\ }\bibfield
  {title} {\enquote {\bibinfo {title} {Deconfined criticality critically
  defined},}\ }\href {\doibase 10.1143/jpsjs.74s.1} {\bibfield  {journal}
  {\bibinfo  {journal} {Journal of the Physical Society of Japan}\ }\textbf
  {\bibinfo {volume} {74}},\ \bibinfo {pages} {1–9} (\bibinfo {year}
  {2005})}\BibitemShut {NoStop}%
\bibitem [{\citenamefont {Levin}\ and\ \citenamefont
  {Senthil}(2004)}]{levin2004prb}%
  \BibitemOpen
  \bibfield  {author} {\bibinfo {author} {\bibfnamefont {Michael}\ \bibnamefont
  {Levin}}\ and\ \bibinfo {author} {\bibfnamefont {T.}~\bibnamefont
  {Senthil}},\ }\bibfield  {title} {\enquote {\bibinfo {title} {Deconfined
  quantum criticality and n\'eel order via dimer disorder},}\ }\href {\doibase
  10.1103/PhysRevB.70.220403} {\bibfield  {journal} {\bibinfo  {journal} {Phys.
  Rev. B}\ }\textbf {\bibinfo {volume} {70}},\ \bibinfo {pages} {220403}
  (\bibinfo {year} {2004})}\BibitemShut {NoStop}%
\bibitem [{\citenamefont {Senthil}\ and\ \citenamefont
  {Fisher}(2006)}]{senthil2005}%
  \BibitemOpen
  \bibfield  {author} {\bibinfo {author} {\bibfnamefont {T.}~\bibnamefont
  {Senthil}}\ and\ \bibinfo {author} {\bibfnamefont {Matthew P.~A.}\
  \bibnamefont {Fisher}},\ }\bibfield  {title} {\enquote {\bibinfo {title}
  {Competing orders, nonlinear sigma models, and topological terms in quantum
  magnets},}\ }\href {\doibase 10.1103/PhysRevB.74.064405} {\bibfield
  {journal} {\bibinfo  {journal} {Phys. Rev. B}\ }\textbf {\bibinfo {volume}
  {74}},\ \bibinfo {pages} {064405} (\bibinfo {year} {2006})}\BibitemShut
  {NoStop}%
\bibitem [{\citenamefont {Swingle}\ and\ \citenamefont
  {Senthil}(2012)}]{swingle2012prb}%
  \BibitemOpen
  \bibfield  {author} {\bibinfo {author} {\bibfnamefont {Brian}\ \bibnamefont
  {Swingle}}\ and\ \bibinfo {author} {\bibfnamefont {T.}~\bibnamefont
  {Senthil}},\ }\bibfield  {title} {\enquote {\bibinfo {title} {Structure of
  entanglement at deconfined quantum critical points},}\ }\href {\doibase
  10.1103/PhysRevB.86.155131} {\bibfield  {journal} {\bibinfo  {journal} {Phys.
  Rev. B}\ }\textbf {\bibinfo {volume} {86}},\ \bibinfo {pages} {155131}
  (\bibinfo {year} {2012})}\BibitemShut {NoStop}%
\bibitem [{\citenamefont {Wang}\ \emph
  {et~al.}(2017{\natexlab{a}})\citenamefont {Wang}, \citenamefont {Nahum},
  \citenamefont {Metlitski}, \citenamefont {Xu},\ and\ \citenamefont
  {Senthil}}]{wang2017prx}%
  \BibitemOpen
  \bibfield  {author} {\bibinfo {author} {\bibfnamefont {Chong}\ \bibnamefont
  {Wang}}, \bibinfo {author} {\bibfnamefont {Adam}\ \bibnamefont {Nahum}},
  \bibinfo {author} {\bibfnamefont {Max~A.}\ \bibnamefont {Metlitski}},
  \bibinfo {author} {\bibfnamefont {Cenke}\ \bibnamefont {Xu}}, \ and\ \bibinfo
  {author} {\bibfnamefont {T.}~\bibnamefont {Senthil}},\ }\bibfield  {title}
  {\enquote {\bibinfo {title} {Deconfined quantum critical points: Symmetries
  and dualities},}\ }\href {\doibase 10.1103/PhysRevX.7.031051} {\bibfield
  {journal} {\bibinfo  {journal} {Phys. Rev. X}\ }\textbf {\bibinfo {volume}
  {7}},\ \bibinfo {pages} {031051} (\bibinfo {year}
  {2017}{\natexlab{a}})}\BibitemShut {NoStop}%
\bibitem [{\citenamefont {Bi}\ and\ \citenamefont {Senthil}(2019)}]{bi2019prx}%
  \BibitemOpen
  \bibfield  {author} {\bibinfo {author} {\bibfnamefont {Zhen}\ \bibnamefont
  {Bi}}\ and\ \bibinfo {author} {\bibfnamefont {T.}~\bibnamefont {Senthil}},\
  }\bibfield  {title} {\enquote {\bibinfo {title} {Adventure in topological
  phase transitions in $3+1$-d: Non-abelian deconfined quantum criticalities
  and a possible duality},}\ }\href {\doibase 10.1103/PhysRevX.9.021034}
  {\bibfield  {journal} {\bibinfo  {journal} {Phys. Rev. X}\ }\textbf {\bibinfo
  {volume} {9}},\ \bibinfo {pages} {021034} (\bibinfo {year}
  {2019})}\BibitemShut {NoStop}%
\bibitem [{\citenamefont {Bi}\ \emph {et~al.}(2020)\citenamefont {Bi},
  \citenamefont {Lake},\ and\ \citenamefont {Senthil}}]{bi2020prr}%
  \BibitemOpen
  \bibfield  {author} {\bibinfo {author} {\bibfnamefont {Zhen}\ \bibnamefont
  {Bi}}, \bibinfo {author} {\bibfnamefont {Ethan}\ \bibnamefont {Lake}}, \ and\
  \bibinfo {author} {\bibfnamefont {T.}~\bibnamefont {Senthil}},\ }\bibfield
  {title} {\enquote {\bibinfo {title} {Landau ordering phase transitions beyond
  the landau paradigm},}\ }\href {\doibase 10.1103/PhysRevResearch.2.023031}
  {\bibfield  {journal} {\bibinfo  {journal} {Phys. Rev. Res.}\ }\textbf
  {\bibinfo {volume} {2}},\ \bibinfo {pages} {023031} (\bibinfo {year}
  {2020})}\BibitemShut {NoStop}%
\bibitem [{\citenamefont {Jiang}\ and\ \citenamefont
  {Motrunich}(2019)}]{jiang2019_1ddqcp}%
  \BibitemOpen
  \bibfield  {author} {\bibinfo {author} {\bibfnamefont {Shenghan}\
  \bibnamefont {Jiang}}\ and\ \bibinfo {author} {\bibfnamefont {Olexei}\
  \bibnamefont {Motrunich}},\ }\bibfield  {title} {\enquote {\bibinfo {title}
  {Ising ferromagnet to valence bond solid transition in a one-dimensional spin
  chain: Analogies to deconfined quantum critical points},}\ }\href@noop {}
  {\bibfield  {journal} {\bibinfo  {journal} {Physical Review B}\ }\textbf
  {\bibinfo {volume} {99}},\ \bibinfo {pages} {075103} (\bibinfo {year}
  {2019})}\BibitemShut {NoStop}%
\bibitem [{\citenamefont {Senthil}(2023)}]{senthil2023deconfined}%
  \BibitemOpen
  \bibfield  {author} {\bibinfo {author} {\bibfnamefont {T.}~\bibnamefont
  {Senthil}},\ }\href@noop {} {\enquote {\bibinfo {title} {Deconfined quantum
  critical points: a review},}\ } (\bibinfo {year} {2023}),\ \Eprint
  {http://arxiv.org/abs/2306.12638} {arXiv:2306.12638 [cond-mat.str-el]}
  \BibitemShut {NoStop}%
\bibitem [{\citenamefont {Ömer M.~Aksoy}\ \emph {et~al.}(2023)\citenamefont
  {Ömer M.~Aksoy}, \citenamefont {Mudry}, \citenamefont {Furusaki},\ and\
  \citenamefont {Tiwari}}]{aksoy2023liebschultzmattis}%
  \BibitemOpen
  \bibfield  {author} {\bibinfo {author} {\bibnamefont {Ömer M.~Aksoy}},
  \bibinfo {author} {\bibfnamefont {Christopher}\ \bibnamefont {Mudry}},
  \bibinfo {author} {\bibfnamefont {Akira}\ \bibnamefont {Furusaki}}, \ and\
  \bibinfo {author} {\bibfnamefont {Apoorv}\ \bibnamefont {Tiwari}},\
  }\href@noop {} {\enquote {\bibinfo {title} {Lieb-schultz-mattis anomalies and
  web of dualities induced by gauging in quantum spin chains},}\ } (\bibinfo
  {year} {2023}),\ \Eprint {http://arxiv.org/abs/2308.00743} {arXiv:2308.00743
  [cond-mat.str-el]} \BibitemShut {NoStop}%
\bibitem [{\citenamefont {Metlitski}\ and\ \citenamefont
  {Thorngren}(2018)}]{max2018prb}%
  \BibitemOpen
  \bibfield  {author} {\bibinfo {author} {\bibfnamefont {Max~A.}\ \bibnamefont
  {Metlitski}}\ and\ \bibinfo {author} {\bibfnamefont {Ryan}\ \bibnamefont
  {Thorngren}},\ }\bibfield  {title} {\enquote {\bibinfo {title} {Intrinsic and
  emergent anomalies at deconfined critical points},}\ }\href {\doibase
  10.1103/PhysRevB.98.085140} {\bibfield  {journal} {\bibinfo  {journal} {Phys.
  Rev. B}\ }\textbf {\bibinfo {volume} {98}},\ \bibinfo {pages} {085140}
  (\bibinfo {year} {2018})}\BibitemShut {NoStop}%
\bibitem [{\citenamefont {Ma}\ \emph {et~al.}(2018)\citenamefont {Ma},
  \citenamefont {Sun}, \citenamefont {You}, \citenamefont {Xu}, \citenamefont
  {Vishwanath}, \citenamefont {Sandvik},\ and\ \citenamefont
  {Meng}}]{ma2018prb}%
  \BibitemOpen
  \bibfield  {author} {\bibinfo {author} {\bibfnamefont {Nvsen}\ \bibnamefont
  {Ma}}, \bibinfo {author} {\bibfnamefont {Guang-Yu}\ \bibnamefont {Sun}},
  \bibinfo {author} {\bibfnamefont {Yi-Zhuang}\ \bibnamefont {You}}, \bibinfo
  {author} {\bibfnamefont {Cenke}\ \bibnamefont {Xu}}, \bibinfo {author}
  {\bibfnamefont {Ashvin}\ \bibnamefont {Vishwanath}}, \bibinfo {author}
  {\bibfnamefont {Anders~W.}\ \bibnamefont {Sandvik}}, \ and\ \bibinfo {author}
  {\bibfnamefont {Zi~Yang}\ \bibnamefont {Meng}},\ }\bibfield  {title}
  {\enquote {\bibinfo {title} {Dynamical signature of fractionalization at a
  deconfined quantum critical point},}\ }\href {\doibase
  10.1103/PhysRevB.98.174421} {\bibfield  {journal} {\bibinfo  {journal} {Phys.
  Rev. B}\ }\textbf {\bibinfo {volume} {98}},\ \bibinfo {pages} {174421}
  (\bibinfo {year} {2018})}\BibitemShut {NoStop}%
\bibitem [{\citenamefont {Qin}\ \emph {et~al.}(2017)\citenamefont {Qin},
  \citenamefont {He}, \citenamefont {You}, \citenamefont {Lu}, \citenamefont
  {Sen}, \citenamefont {Sandvik}, \citenamefont {Xu},\ and\ \citenamefont
  {Meng}}]{qin2017prx}%
  \BibitemOpen
  \bibfield  {author} {\bibinfo {author} {\bibfnamefont {Yan~Qi}\ \bibnamefont
  {Qin}}, \bibinfo {author} {\bibfnamefont {Yuan-Yao}\ \bibnamefont {He}},
  \bibinfo {author} {\bibfnamefont {Yi-Zhuang}\ \bibnamefont {You}}, \bibinfo
  {author} {\bibfnamefont {Zhong-Yi}\ \bibnamefont {Lu}}, \bibinfo {author}
  {\bibfnamefont {Arnab}\ \bibnamefont {Sen}}, \bibinfo {author} {\bibfnamefont
  {Anders~W.}\ \bibnamefont {Sandvik}}, \bibinfo {author} {\bibfnamefont
  {Cenke}\ \bibnamefont {Xu}}, \ and\ \bibinfo {author} {\bibfnamefont
  {Zi~Yang}\ \bibnamefont {Meng}},\ }\bibfield  {title} {\enquote {\bibinfo
  {title} {Duality between the deconfined quantum-critical point and the
  bosonic topological transition},}\ }\href {\doibase
  10.1103/PhysRevX.7.031052} {\bibfield  {journal} {\bibinfo  {journal} {Phys.
  Rev. X}\ }\textbf {\bibinfo {volume} {7}},\ \bibinfo {pages} {031052}
  (\bibinfo {year} {2017})}\BibitemShut {NoStop}%
\bibitem [{\citenamefont {Sandvik}(2007)}]{sandvik2007prl}%
  \BibitemOpen
  \bibfield  {author} {\bibinfo {author} {\bibfnamefont {Anders~W.}\
  \bibnamefont {Sandvik}},\ }\bibfield  {title} {\enquote {\bibinfo {title}
  {Evidence for deconfined quantum criticality in a two-dimensional heisenberg
  model with four-spin interactions},}\ }\href {\doibase
  10.1103/PhysRevLett.98.227202} {\bibfield  {journal} {\bibinfo  {journal}
  {Phys. Rev. Lett.}\ }\textbf {\bibinfo {volume} {98}},\ \bibinfo {pages}
  {227202} (\bibinfo {year} {2007})}\BibitemShut {NoStop}%
\bibitem [{\citenamefont {Nahum}\ \emph
  {et~al.}(2015{\natexlab{a}})\citenamefont {Nahum}, \citenamefont {Serna},
  \citenamefont {Chalker}, \citenamefont {Ortu\~no},\ and\ \citenamefont
  {Somoza}}]{nahum2015prl}%
  \BibitemOpen
  \bibfield  {author} {\bibinfo {author} {\bibfnamefont {Adam}\ \bibnamefont
  {Nahum}}, \bibinfo {author} {\bibfnamefont {P.}~\bibnamefont {Serna}},
  \bibinfo {author} {\bibfnamefont {J.~T.}\ \bibnamefont {Chalker}}, \bibinfo
  {author} {\bibfnamefont {M.}~\bibnamefont {Ortu\~no}}, \ and\ \bibinfo
  {author} {\bibfnamefont {A.~M.}\ \bibnamefont {Somoza}},\ }\bibfield  {title}
  {\enquote {\bibinfo {title} {Emergent so(5) symmetry at the n\'eel to
  valence-bond-solid transition},}\ }\href {\doibase
  10.1103/PhysRevLett.115.267203} {\bibfield  {journal} {\bibinfo  {journal}
  {Phys. Rev. Lett.}\ }\textbf {\bibinfo {volume} {115}},\ \bibinfo {pages}
  {267203} (\bibinfo {year} {2015}{\natexlab{a}})}\BibitemShut {NoStop}%
\bibitem [{\citenamefont {Ma}\ \emph {et~al.}(2019)\citenamefont {Ma},
  \citenamefont {You},\ and\ \citenamefont {Meng}}]{ma2019prl}%
  \BibitemOpen
  \bibfield  {author} {\bibinfo {author} {\bibfnamefont {Nvsen}\ \bibnamefont
  {Ma}}, \bibinfo {author} {\bibfnamefont {Yi-Zhuang}\ \bibnamefont {You}}, \
  and\ \bibinfo {author} {\bibfnamefont {Zi~Yang}\ \bibnamefont {Meng}},\
  }\bibfield  {title} {\enquote {\bibinfo {title} {Role of noether's theorem at
  the deconfined quantum critical point},}\ }\href {\doibase
  10.1103/PhysRevLett.122.175701} {\bibfield  {journal} {\bibinfo  {journal}
  {Phys. Rev. Lett.}\ }\textbf {\bibinfo {volume} {122}},\ \bibinfo {pages}
  {175701} (\bibinfo {year} {2019})}\BibitemShut {NoStop}%
\bibitem [{\citenamefont {Ma}\ and\ \citenamefont {Wang}(2020)}]{ma2020prb}%
  \BibitemOpen
  \bibfield  {author} {\bibinfo {author} {\bibfnamefont {Ruochen}\ \bibnamefont
  {Ma}}\ and\ \bibinfo {author} {\bibfnamefont {Chong}\ \bibnamefont {Wang}},\
  }\bibfield  {title} {\enquote {\bibinfo {title} {Theory of deconfined
  pseudocriticality},}\ }\href {\doibase 10.1103/PhysRevB.102.020407}
  {\bibfield  {journal} {\bibinfo  {journal} {Phys. Rev. B}\ }\textbf {\bibinfo
  {volume} {102}},\ \bibinfo {pages} {020407} (\bibinfo {year}
  {2020})}\BibitemShut {NoStop}%
\bibitem [{\citenamefont {Nahum}(2020)}]{nahum2020prb}%
  \BibitemOpen
  \bibfield  {author} {\bibinfo {author} {\bibfnamefont {Adam}\ \bibnamefont
  {Nahum}},\ }\bibfield  {title} {\enquote {\bibinfo {title} {Note on
  wess-zumino-witten models and quasiuniversality in $2+1$ dimensions},}\
  }\href {\doibase 10.1103/PhysRevB.102.201116} {\bibfield  {journal} {\bibinfo
   {journal} {Phys. Rev. B}\ }\textbf {\bibinfo {volume} {102}},\ \bibinfo
  {pages} {201116} (\bibinfo {year} {2020})}\BibitemShut {NoStop}%
\bibitem [{\citenamefont {Lu}(2022)}]{lu2022nonlinear}%
  \BibitemOpen
  \bibfield  {author} {\bibinfo {author} {\bibfnamefont {Da-Chuan}\
  \bibnamefont {Lu}},\ }\href@noop {} {\enquote {\bibinfo {title} {Nonlinear
  sigma model description of deconfined quantum criticality in arbitrary
  dimensions},}\ } (\bibinfo {year} {2022}),\ \Eprint
  {http://arxiv.org/abs/2209.00670} {arXiv:2209.00670 [cond-mat.str-el]}
  \BibitemShut {NoStop}%
\bibitem [{\citenamefont {Ji}\ \emph {et~al.}(2022)\citenamefont {Ji},
  \citenamefont {Tantivasadakarn},\ and\ \citenamefont {Xu}}]{ji2022boundary}%
  \BibitemOpen
  \bibfield  {author} {\bibinfo {author} {\bibfnamefont {Wenjie}\ \bibnamefont
  {Ji}}, \bibinfo {author} {\bibfnamefont {Nathanan}\ \bibnamefont
  {Tantivasadakarn}}, \ and\ \bibinfo {author} {\bibfnamefont {Cenke}\
  \bibnamefont {Xu}},\ }\href@noop {} {\enquote {\bibinfo {title} {Boundary
  states of three dimensional topological order and the deconfined quantum
  critical point},}\ } (\bibinfo {year} {2022}),\ \Eprint
  {http://arxiv.org/abs/2212.09754} {arXiv:2212.09754 [cond-mat.str-el]}
  \BibitemShut {NoStop}%
\bibitem [{\citenamefont {Prembabu}\ \emph {et~al.}(2022)\citenamefont
  {Prembabu}, \citenamefont {Thorngren},\ and\ \citenamefont
  {Verresen}}]{prembabu2022boundary}%
  \BibitemOpen
  \bibfield  {author} {\bibinfo {author} {\bibfnamefont {Saranesh}\
  \bibnamefont {Prembabu}}, \bibinfo {author} {\bibfnamefont {Ryan}\
  \bibnamefont {Thorngren}}, \ and\ \bibinfo {author} {\bibfnamefont {Ruben}\
  \bibnamefont {Verresen}},\ }\href@noop {} {\enquote {\bibinfo {title}
  {Boundary deconfined quantum criticality at transitions between
  symmetry-protected topological chains},}\ } (\bibinfo {year} {2022}),\
  \Eprint {http://arxiv.org/abs/2208.12258} {arXiv:2208.12258
  [cond-mat.str-el]} \BibitemShut {NoStop}%
\bibitem [{\citenamefont {Zhang}\ and\ \citenamefont
  {Levin}(2023)}]{zhang2023prl}%
  \BibitemOpen
  \bibfield  {author} {\bibinfo {author} {\bibfnamefont {Carolyn}\ \bibnamefont
  {Zhang}}\ and\ \bibinfo {author} {\bibfnamefont {Michael}\ \bibnamefont
  {Levin}},\ }\bibfield  {title} {\enquote {\bibinfo {title} {Exactly solvable
  model for a deconfined quantum critical point in 1d},}\ }\href {\doibase
  10.1103/PhysRevLett.130.026801} {\bibfield  {journal} {\bibinfo  {journal}
  {Phys. Rev. Lett.}\ }\textbf {\bibinfo {volume} {130}},\ \bibinfo {pages}
  {026801} (\bibinfo {year} {2023})}\BibitemShut {NoStop}%
\bibitem [{\citenamefont {Shyta}\ \emph {et~al.}(2022)\citenamefont {Shyta},
  \citenamefont {van~den Brink},\ and\ \citenamefont
  {Nogueira}}]{shyta2022prl}%
  \BibitemOpen
  \bibfield  {author} {\bibinfo {author} {\bibfnamefont {Vira}\ \bibnamefont
  {Shyta}}, \bibinfo {author} {\bibfnamefont {Jeroen}\ \bibnamefont {van~den
  Brink}}, \ and\ \bibinfo {author} {\bibfnamefont {Flavio~S.}\ \bibnamefont
  {Nogueira}},\ }\bibfield  {title} {\enquote {\bibinfo {title} {Frozen
  deconfined quantum criticality},}\ }\href {\doibase
  10.1103/PhysRevLett.129.227203} {\bibfield  {journal} {\bibinfo  {journal}
  {Phys. Rev. Lett.}\ }\textbf {\bibinfo {volume} {129}},\ \bibinfo {pages}
  {227203} (\bibinfo {year} {2022})}\BibitemShut {NoStop}%
\bibitem [{\citenamefont {Joshi}\ \emph {et~al.}(2020)\citenamefont {Joshi},
  \citenamefont {Li}, \citenamefont {Tarnopolsky}, \citenamefont {Georges},\
  and\ \citenamefont {Sachdev}}]{Joshi2020prx}%
  \BibitemOpen
  \bibfield  {author} {\bibinfo {author} {\bibfnamefont {Darshan~G.}\
  \bibnamefont {Joshi}}, \bibinfo {author} {\bibfnamefont {Chenyuan}\
  \bibnamefont {Li}}, \bibinfo {author} {\bibfnamefont {Grigory}\ \bibnamefont
  {Tarnopolsky}}, \bibinfo {author} {\bibfnamefont {Antoine}\ \bibnamefont
  {Georges}}, \ and\ \bibinfo {author} {\bibfnamefont {Subir}\ \bibnamefont
  {Sachdev}},\ }\bibfield  {title} {\enquote {\bibinfo {title} {Deconfined
  critical point in a doped random quantum heisenberg magnet},}\ }\href
  {\doibase 10.1103/PhysRevX.10.021033} {\bibfield  {journal} {\bibinfo
  {journal} {Phys. Rev. X}\ }\textbf {\bibinfo {volume} {10}},\ \bibinfo
  {pages} {021033} (\bibinfo {year} {2020})}\BibitemShut {NoStop}%
\bibitem [{\citenamefont {Lee}\ \emph {et~al.}(2019)\citenamefont {Lee},
  \citenamefont {You}, \citenamefont {Sachdev},\ and\ \citenamefont
  {Vishwanath}}]{lee2019prx}%
  \BibitemOpen
  \bibfield  {author} {\bibinfo {author} {\bibfnamefont {Jong~Yeon}\
  \bibnamefont {Lee}}, \bibinfo {author} {\bibfnamefont {Yi-Zhuang}\
  \bibnamefont {You}}, \bibinfo {author} {\bibfnamefont {Subir}\ \bibnamefont
  {Sachdev}}, \ and\ \bibinfo {author} {\bibfnamefont {Ashvin}\ \bibnamefont
  {Vishwanath}},\ }\bibfield  {title} {\enquote {\bibinfo {title} {Signatures
  of a deconfined phase transition on the shastry-sutherland lattice:
  Applications to quantum critical
  ${\mathrm{srcu}}_{2}({\mathrm{bo}}_{3}{)}_{2}$},}\ }\href {\doibase
  10.1103/PhysRevX.9.041037} {\bibfield  {journal} {\bibinfo  {journal} {Phys.
  Rev. X}\ }\textbf {\bibinfo {volume} {9}},\ \bibinfo {pages} {041037}
  (\bibinfo {year} {2019})}\BibitemShut {NoStop}%
\bibitem [{\citenamefont {You}\ \emph {et~al.}(2018)\citenamefont {You},
  \citenamefont {He}, \citenamefont {Xu},\ and\ \citenamefont
  {Vishwanath}}]{you2018prx}%
  \BibitemOpen
  \bibfield  {author} {\bibinfo {author} {\bibfnamefont {Yi-Zhuang}\
  \bibnamefont {You}}, \bibinfo {author} {\bibfnamefont {Yin-Chen}\
  \bibnamefont {He}}, \bibinfo {author} {\bibfnamefont {Cenke}\ \bibnamefont
  {Xu}}, \ and\ \bibinfo {author} {\bibfnamefont {Ashvin}\ \bibnamefont
  {Vishwanath}},\ }\bibfield  {title} {\enquote {\bibinfo {title} {Symmetric
  fermion mass generation as deconfined quantum criticality},}\ }\href
  {\doibase 10.1103/PhysRevX.8.011026} {\bibfield  {journal} {\bibinfo
  {journal} {Phys. Rev. X}\ }\textbf {\bibinfo {volume} {8}},\ \bibinfo {pages}
  {011026} (\bibinfo {year} {2018})}\BibitemShut {NoStop}%
\bibitem [{\citenamefont {Jian}\ \emph {et~al.}(2018)\citenamefont {Jian},
  \citenamefont {Thomson}, \citenamefont {Rasmussen}, \citenamefont {Bi},\ and\
  \citenamefont {Xu}}]{jian2018prb}%
  \BibitemOpen
  \bibfield  {author} {\bibinfo {author} {\bibfnamefont {Chao-Ming}\
  \bibnamefont {Jian}}, \bibinfo {author} {\bibfnamefont {Alex}\ \bibnamefont
  {Thomson}}, \bibinfo {author} {\bibfnamefont {Alex}\ \bibnamefont
  {Rasmussen}}, \bibinfo {author} {\bibfnamefont {Zhen}\ \bibnamefont {Bi}}, \
  and\ \bibinfo {author} {\bibfnamefont {Cenke}\ \bibnamefont {Xu}},\
  }\bibfield  {title} {\enquote {\bibinfo {title} {Deconfined quantum critical
  point on the triangular lattice},}\ }\href {\doibase
  10.1103/PhysRevB.97.195115} {\bibfield  {journal} {\bibinfo  {journal} {Phys.
  Rev. B}\ }\textbf {\bibinfo {volume} {97}},\ \bibinfo {pages} {195115}
  (\bibinfo {year} {2018})}\BibitemShut {NoStop}%
\bibitem [{\citenamefont {Jian}\ \emph {et~al.}(2017)\citenamefont {Jian},
  \citenamefont {Rasmussen}, \citenamefont {You},\ and\ \citenamefont
  {Xu}}]{jian2017emergent}%
  \BibitemOpen
  \bibfield  {author} {\bibinfo {author} {\bibfnamefont {Chao-Ming}\
  \bibnamefont {Jian}}, \bibinfo {author} {\bibfnamefont {Alex}\ \bibnamefont
  {Rasmussen}}, \bibinfo {author} {\bibfnamefont {Yi-Zhuang}\ \bibnamefont
  {You}}, \ and\ \bibinfo {author} {\bibfnamefont {Cenke}\ \bibnamefont {Xu}},\
  }\href@noop {} {\enquote {\bibinfo {title} {Emergent symmetry and tricritical
  points near the deconfined quantum critical point},}\ } (\bibinfo {year}
  {2017}),\ \Eprint {http://arxiv.org/abs/1708.03050} {arXiv:1708.03050
  [cond-mat.str-el]} \BibitemShut {NoStop}%
\bibitem [{\citenamefont {Zou}\ \emph {et~al.}(2021)\citenamefont {Zou},
  \citenamefont {He},\ and\ \citenamefont {Wang}}]{zou2021prx}%
  \BibitemOpen
  \bibfield  {author} {\bibinfo {author} {\bibfnamefont {Liujun}\ \bibnamefont
  {Zou}}, \bibinfo {author} {\bibfnamefont {Yin-Chen}\ \bibnamefont {He}}, \
  and\ \bibinfo {author} {\bibfnamefont {Chong}\ \bibnamefont {Wang}},\
  }\bibfield  {title} {\enquote {\bibinfo {title} {Stiefel liquids: Possible
  non-lagrangian quantum criticality from intertwined orders},}\ }\href
  {\doibase 10.1103/PhysRevX.11.031043} {\bibfield  {journal} {\bibinfo
  {journal} {Phys. Rev. X}\ }\textbf {\bibinfo {volume} {11}},\ \bibinfo
  {pages} {031043} (\bibinfo {year} {2021})}\BibitemShut {NoStop}%
\bibitem [{\citenamefont {Song}\ \emph {et~al.}(2019)\citenamefont {Song},
  \citenamefont {Wang}, \citenamefont {Vishwanath},\ and\ \citenamefont
  {He}}]{song2019unifying}%
  \BibitemOpen
  \bibfield  {author} {\bibinfo {author} {\bibfnamefont {Xue-Yang}\
  \bibnamefont {Song}}, \bibinfo {author} {\bibfnamefont {Chong}\ \bibnamefont
  {Wang}}, \bibinfo {author} {\bibfnamefont {Ashvin}\ \bibnamefont
  {Vishwanath}}, \ and\ \bibinfo {author} {\bibfnamefont {Yin-Chen}\
  \bibnamefont {He}},\ }\bibfield  {title} {\enquote {\bibinfo {title}
  {Unifying description of competing orders in two-dimensional quantum
  magnets},}\ }\href@noop {} {\bibfield  {journal} {\bibinfo  {journal} {Nature
  communications}\ }\textbf {\bibinfo {volume} {10}},\ \bibinfo {pages} {4254}
  (\bibinfo {year} {2019})}\BibitemShut {NoStop}%
\bibitem [{\citenamefont {Song}\ \emph {et~al.}(2020)\citenamefont {Song},
  \citenamefont {He}, \citenamefont {Vishwanath},\ and\ \citenamefont
  {Wang}}]{song2020prx}%
  \BibitemOpen
  \bibfield  {author} {\bibinfo {author} {\bibfnamefont {Xue-Yang}\
  \bibnamefont {Song}}, \bibinfo {author} {\bibfnamefont {Yin-Chen}\
  \bibnamefont {He}}, \bibinfo {author} {\bibfnamefont {Ashvin}\ \bibnamefont
  {Vishwanath}}, \ and\ \bibinfo {author} {\bibfnamefont {Chong}\ \bibnamefont
  {Wang}},\ }\bibfield  {title} {\enquote {\bibinfo {title} {From spinon band
  topology to the symmetry quantum numbers of monopoles in dirac spin
  liquids},}\ }\href {\doibase 10.1103/PhysRevX.10.011033} {\bibfield
  {journal} {\bibinfo  {journal} {Phys. Rev. X}\ }\textbf {\bibinfo {volume}
  {10}},\ \bibinfo {pages} {011033} (\bibinfo {year} {2020})}\BibitemShut
  {NoStop}%
\bibitem [{\citenamefont {Janssen}\ and\ \citenamefont
  {He}(2017)}]{lukas2017prb}%
  \BibitemOpen
  \bibfield  {author} {\bibinfo {author} {\bibfnamefont {Lukas}\ \bibnamefont
  {Janssen}}\ and\ \bibinfo {author} {\bibfnamefont {Yin-Chen}\ \bibnamefont
  {He}},\ }\bibfield  {title} {\enquote {\bibinfo {title} {Critical behavior of
  the ${\mathrm{qed}}_{3}$-gross-neveu model: Duality and deconfined
  criticality},}\ }\href {\doibase 10.1103/PhysRevB.96.205113} {\bibfield
  {journal} {\bibinfo  {journal} {Phys. Rev. B}\ }\textbf {\bibinfo {volume}
  {96}},\ \bibinfo {pages} {205113} (\bibinfo {year} {2017})}\BibitemShut
  {NoStop}%
\bibitem [{\citenamefont {Sandvik}(2010{\natexlab{a}})}]{sandvik2010prl}%
  \BibitemOpen
  \bibfield  {author} {\bibinfo {author} {\bibfnamefont {Anders~W.}\
  \bibnamefont {Sandvik}},\ }\bibfield  {title} {\enquote {\bibinfo {title}
  {Continuous quantum phase transition between an antiferromagnet and a
  valence-bond solid in two dimensions: Evidence for logarithmic corrections to
  scaling},}\ }\href {\doibase 10.1103/PhysRevLett.104.177201} {\bibfield
  {journal} {\bibinfo  {journal} {Phys. Rev. Lett.}\ }\textbf {\bibinfo
  {volume} {104}},\ \bibinfo {pages} {177201} (\bibinfo {year}
  {2010}{\natexlab{a}})}\BibitemShut {NoStop}%
\bibitem [{\citenamefont {Kaul}\ and\ \citenamefont
  {Sandvik}(2012)}]{kaul2012prl}%
  \BibitemOpen
  \bibfield  {author} {\bibinfo {author} {\bibfnamefont {Ribhu~K.}\
  \bibnamefont {Kaul}}\ and\ \bibinfo {author} {\bibfnamefont {Anders~W.}\
  \bibnamefont {Sandvik}},\ }\bibfield  {title} {\enquote {\bibinfo {title}
  {Lattice model for the $\mathrm{SU}(n)$ n\'eel to valence-bond solid quantum
  phase transition at large $n$},}\ }\href {\doibase
  10.1103/PhysRevLett.108.137201} {\bibfield  {journal} {\bibinfo  {journal}
  {Phys. Rev. Lett.}\ }\textbf {\bibinfo {volume} {108}},\ \bibinfo {pages}
  {137201} (\bibinfo {year} {2012})}\BibitemShut {NoStop}%
\bibitem [{\citenamefont {Sandvik}\ and\ \citenamefont
  {Zhao}(2020)}]{Sandvik_2020}%
  \BibitemOpen
  \bibfield  {author} {\bibinfo {author} {\bibfnamefont {Anders~W.}\
  \bibnamefont {Sandvik}}\ and\ \bibinfo {author} {\bibfnamefont {Bowen}\
  \bibnamefont {Zhao}},\ }\bibfield  {title} {\enquote {\bibinfo {title}
  {Consistent scaling exponents at the deconfined quantum-critical point*},}\
  }\href {\doibase 10.1088/0256-307X/37/5/057502} {\bibfield  {journal}
  {\bibinfo  {journal} {Chinese Physics Letters}\ }\textbf {\bibinfo {volume}
  {37}},\ \bibinfo {pages} {057502} (\bibinfo {year} {2020})}\BibitemShut
  {NoStop}%
\bibitem [{\citenamefont {Shao}\ \emph {et~al.}(2016)\citenamefont {Shao},
  \citenamefont {Guo},\ and\ \citenamefont {Sandvik}}]{shao2016quantum}%
  \BibitemOpen
  \bibfield  {author} {\bibinfo {author} {\bibfnamefont {Hui}\ \bibnamefont
  {Shao}}, \bibinfo {author} {\bibfnamefont {Wenan}\ \bibnamefont {Guo}}, \
  and\ \bibinfo {author} {\bibfnamefont {Anders~W}\ \bibnamefont {Sandvik}},\
  }\bibfield  {title} {\enquote {\bibinfo {title} {Quantum criticality with two
  length scales},}\ }\href@noop {} {\bibfield  {journal} {\bibinfo  {journal}
  {Science}\ }\textbf {\bibinfo {volume} {352}},\ \bibinfo {pages} {213--216}
  (\bibinfo {year} {2016})}\BibitemShut {NoStop}%
\bibitem [{\citenamefont {Block}\ \emph {et~al.}(2013)\citenamefont {Block},
  \citenamefont {Melko},\ and\ \citenamefont {Kaul}}]{block2013prl}%
  \BibitemOpen
  \bibfield  {author} {\bibinfo {author} {\bibfnamefont {Matthew~S.}\
  \bibnamefont {Block}}, \bibinfo {author} {\bibfnamefont {Roger~G.}\
  \bibnamefont {Melko}}, \ and\ \bibinfo {author} {\bibfnamefont {Ribhu~K.}\
  \bibnamefont {Kaul}},\ }\bibfield  {title} {\enquote {\bibinfo {title} {Fate
  of $\mathbb{C}{\mathbb{p}}^{N\ensuremath{-}1}$ fixed points with $q$
  monopoles},}\ }\href {\doibase 10.1103/PhysRevLett.111.137202} {\bibfield
  {journal} {\bibinfo  {journal} {Phys. Rev. Lett.}\ }\textbf {\bibinfo
  {volume} {111}},\ \bibinfo {pages} {137202} (\bibinfo {year}
  {2013})}\BibitemShut {NoStop}%
\bibitem [{\citenamefont {Kaul}\ and\ \citenamefont
  {Melko}(2008)}]{kaul2008prb}%
  \BibitemOpen
  \bibfield  {author} {\bibinfo {author} {\bibfnamefont {Ribhu~K.}\
  \bibnamefont {Kaul}}\ and\ \bibinfo {author} {\bibfnamefont {Roger~G.}\
  \bibnamefont {Melko}},\ }\bibfield  {title} {\enquote {\bibinfo {title}
  {Large-$n$ estimates of universal amplitudes of the
  $\mathbb{C}{\mathbb{p}}^{N\ensuremath{-}1}$ theory and comparison with a
  $s=\frac{1}{2}$ square-lattice model with competing four-spin
  interactions},}\ }\href {\doibase 10.1103/PhysRevB.78.014417} {\bibfield
  {journal} {\bibinfo  {journal} {Phys. Rev. B}\ }\textbf {\bibinfo {volume}
  {78}},\ \bibinfo {pages} {014417} (\bibinfo {year} {2008})}\BibitemShut
  {NoStop}%
\bibitem [{\citenamefont {Nakayama}\ and\ \citenamefont
  {Ohtsuki}(2016)}]{yu2016prl}%
  \BibitemOpen
  \bibfield  {author} {\bibinfo {author} {\bibfnamefont {Yu}~\bibnamefont
  {Nakayama}}\ and\ \bibinfo {author} {\bibfnamefont {Tomoki}\ \bibnamefont
  {Ohtsuki}},\ }\bibfield  {title} {\enquote {\bibinfo {title} {Necessary
  condition for emergent symmetry from the conformal bootstrap},}\ }\href
  {\doibase 10.1103/PhysRevLett.117.131601} {\bibfield  {journal} {\bibinfo
  {journal} {Phys. Rev. Lett.}\ }\textbf {\bibinfo {volume} {117}},\ \bibinfo
  {pages} {131601} (\bibinfo {year} {2016})}\BibitemShut {NoStop}%
\bibitem [{\citenamefont {He}\ \emph {et~al.}(2021)\citenamefont {He},
  \citenamefont {Rong},\ and\ \citenamefont {Su}}]{he2021scipost}%
  \BibitemOpen
  \bibfield  {author} {\bibinfo {author} {\bibfnamefont {Yin-Chen}\
  \bibnamefont {He}}, \bibinfo {author} {\bibfnamefont {Junchen}\ \bibnamefont
  {Rong}}, \ and\ \bibinfo {author} {\bibfnamefont {Ning}\ \bibnamefont {Su}},\
  }\bibfield  {title} {\enquote {\bibinfo {title} {{Non-Wilson-Fisher kinks of
  $O(N)$ numerical bootstrap: from the deconfined phase transition to a
  putative new family of CFTs}},}\ }\href {\doibase
  10.21468/SciPostPhys.10.5.115} {\bibfield  {journal} {\bibinfo  {journal}
  {SciPost Phys.}\ }\textbf {\bibinfo {volume} {10}},\ \bibinfo {pages} {115}
  (\bibinfo {year} {2021})}\BibitemShut {NoStop}%
\bibitem [{\citenamefont {Shu}\ \emph {et~al.}(2023)\citenamefont {Shu},
  \citenamefont {Jian}, \citenamefont {Sandvik},\ and\ \citenamefont
  {Yin}}]{shu2023equilibration}%
  \BibitemOpen
  \bibfield  {author} {\bibinfo {author} {\bibfnamefont {Yu-Rong}\ \bibnamefont
  {Shu}}, \bibinfo {author} {\bibfnamefont {Shao-Kai}\ \bibnamefont {Jian}},
  \bibinfo {author} {\bibfnamefont {Anders~W.}\ \bibnamefont {Sandvik}}, \ and\
  \bibinfo {author} {\bibfnamefont {Shuai}\ \bibnamefont {Yin}},\ }\href@noop
  {} {\enquote {\bibinfo {title} {Equilibration of topological defects at the
  deconfined quantum critical point},}\ } (\bibinfo {year} {2023}),\ \Eprint
  {http://arxiv.org/abs/2305.04771} {arXiv:2305.04771 [cond-mat.str-el]}
  \BibitemShut {NoStop}%
\bibitem [{\citenamefont {Li}\ \emph {et~al.}(2019)\citenamefont {Li},
  \citenamefont {Jian},\ and\ \citenamefont {Yao}}]{li2019deconfined}%
  \BibitemOpen
  \bibfield  {author} {\bibinfo {author} {\bibfnamefont {Zi-Xiang}\
  \bibnamefont {Li}}, \bibinfo {author} {\bibfnamefont {Shao-Kai}\ \bibnamefont
  {Jian}}, \ and\ \bibinfo {author} {\bibfnamefont {Hong}\ \bibnamefont
  {Yao}},\ }\href@noop {} {\enquote {\bibinfo {title} {Deconfined quantum
  criticality and emergent so(5) symmetry in fermionic systems},}\ } (\bibinfo
  {year} {2019}),\ \Eprint {http://arxiv.org/abs/1904.10975} {arXiv:1904.10975
  [cond-mat.str-el]} \BibitemShut {NoStop}%
\bibitem [{\citenamefont {Liu}\ \emph {et~al.}(2023{\natexlab{a}})\citenamefont
  {Liu}, \citenamefont {Xiong}, \citenamefont {Xu},\ and\ \citenamefont
  {Zhang}}]{liu2023does}%
  \BibitemOpen
  \bibfield  {author} {\bibinfo {author} {\bibfnamefont {Dong-Xu}\ \bibnamefont
  {Liu}}, \bibinfo {author} {\bibfnamefont {Zijian}\ \bibnamefont {Xiong}},
  \bibinfo {author} {\bibfnamefont {Yining}\ \bibnamefont {Xu}}, \ and\
  \bibinfo {author} {\bibfnamefont {Xue-Feng}\ \bibnamefont {Zhang}},\
  }\href@noop {} {\enquote {\bibinfo {title} {Does deconfined quantum phase
  transition have to keep lorentz symmetry? two velocities of spinon and
  string},}\ } (\bibinfo {year} {2023}{\natexlab{a}}),\ \Eprint
  {http://arxiv.org/abs/2301.12864} {arXiv:2301.12864 [cond-mat.str-el]}
  \BibitemShut {NoStop}%
\bibitem [{\citenamefont {Zhang}\ \emph {et~al.}(2018)\citenamefont {Zhang},
  \citenamefont {He}, \citenamefont {Eggert}, \citenamefont {Moessner},\ and\
  \citenamefont {Pollmann}}]{zhang2018prl}%
  \BibitemOpen
  \bibfield  {author} {\bibinfo {author} {\bibfnamefont {Xue-Feng}\
  \bibnamefont {Zhang}}, \bibinfo {author} {\bibfnamefont {Yin-Chen}\
  \bibnamefont {He}}, \bibinfo {author} {\bibfnamefont {Sebastian}\
  \bibnamefont {Eggert}}, \bibinfo {author} {\bibfnamefont {Roderich}\
  \bibnamefont {Moessner}}, \ and\ \bibinfo {author} {\bibfnamefont {Frank}\
  \bibnamefont {Pollmann}},\ }\bibfield  {title} {\enquote {\bibinfo {title}
  {Continuous easy-plane deconfined phase transition on the kagome lattice},}\
  }\href {\doibase 10.1103/PhysRevLett.120.115702} {\bibfield  {journal}
  {\bibinfo  {journal} {Phys. Rev. Lett.}\ }\textbf {\bibinfo {volume} {120}},\
  \bibinfo {pages} {115702} (\bibinfo {year} {2018})}\BibitemShut {NoStop}%
\bibitem [{\citenamefont {Chen}\ \emph {et~al.}(2023)\citenamefont {Chen},
  \citenamefont {Zhang}, \citenamefont {Wang}, \citenamefont {Sun},\ and\
  \citenamefont {Meng}}]{chen2023phases}%
  \BibitemOpen
  \bibfield  {author} {\bibinfo {author} {\bibfnamefont {Bin-Bin}\ \bibnamefont
  {Chen}}, \bibinfo {author} {\bibfnamefont {Xu}~\bibnamefont {Zhang}},
  \bibinfo {author} {\bibfnamefont {Yuxuan}\ \bibnamefont {Wang}}, \bibinfo
  {author} {\bibfnamefont {Kai}\ \bibnamefont {Sun}}, \ and\ \bibinfo {author}
  {\bibfnamefont {Zi~Yang}\ \bibnamefont {Meng}},\ }\href@noop {} {\enquote
  {\bibinfo {title} {Phases of (2+1)d so(5) non-linear sigma model with a
  topological term on a sphere: multicritical point and disorder phase},}\ }
  (\bibinfo {year} {2023}),\ \Eprint {http://arxiv.org/abs/2307.05307}
  {arXiv:2307.05307 [cond-mat.str-el]} \BibitemShut {NoStop}%
\bibitem [{\citenamefont {Song}\ \emph
  {et~al.}(2023{\natexlab{a}})\citenamefont {Song}, \citenamefont {Zhao},
  \citenamefont {Janssen}, \citenamefont {Scherer},\ and\ \citenamefont
  {Meng}}]{song2023deconfined}%
  \BibitemOpen
  \bibfield  {author} {\bibinfo {author} {\bibfnamefont {Menghan}\ \bibnamefont
  {Song}}, \bibinfo {author} {\bibfnamefont {Jiarui}\ \bibnamefont {Zhao}},
  \bibinfo {author} {\bibfnamefont {Lukas}\ \bibnamefont {Janssen}}, \bibinfo
  {author} {\bibfnamefont {Michael~M.}\ \bibnamefont {Scherer}}, \ and\
  \bibinfo {author} {\bibfnamefont {Zi~Yang}\ \bibnamefont {Meng}},\
  }\href@noop {} {\enquote {\bibinfo {title} {Deconfined quantum criticality
  lost},}\ } (\bibinfo {year} {2023}{\natexlab{a}}),\ \Eprint
  {http://arxiv.org/abs/2307.02547} {arXiv:2307.02547 [cond-mat.str-el]}
  \BibitemShut {NoStop}%
\bibitem [{\citenamefont {Liu}\ \emph {et~al.}(2023{\natexlab{b}})\citenamefont
  {Liu}, \citenamefont {Jiang}, \citenamefont {Chen}, \citenamefont {Rong},
  \citenamefont {Cheng}, \citenamefont {Sun}, \citenamefont {Meng},\ and\
  \citenamefont {Assaad}}]{liu2023prl}%
  \BibitemOpen
  \bibfield  {author} {\bibinfo {author} {\bibfnamefont {Zi~Hong}\ \bibnamefont
  {Liu}}, \bibinfo {author} {\bibfnamefont {Weilun}\ \bibnamefont {Jiang}},
  \bibinfo {author} {\bibfnamefont {Bin-Bin}\ \bibnamefont {Chen}}, \bibinfo
  {author} {\bibfnamefont {Junchen}\ \bibnamefont {Rong}}, \bibinfo {author}
  {\bibfnamefont {Meng}\ \bibnamefont {Cheng}}, \bibinfo {author}
  {\bibfnamefont {Kai}\ \bibnamefont {Sun}}, \bibinfo {author} {\bibfnamefont
  {Zi~Yang}\ \bibnamefont {Meng}}, \ and\ \bibinfo {author} {\bibfnamefont
  {Fakher~F.}\ \bibnamefont {Assaad}},\ }\bibfield  {title} {\enquote {\bibinfo
  {title} {Fermion disorder operator at gross-neveu and deconfined quantum
  criticalities},}\ }\href {\doibase 10.1103/PhysRevLett.130.266501} {\bibfield
   {journal} {\bibinfo  {journal} {Phys. Rev. Lett.}\ }\textbf {\bibinfo
  {volume} {130}},\ \bibinfo {pages} {266501} (\bibinfo {year}
  {2023}{\natexlab{b}})}\BibitemShut {NoStop}%
\bibitem [{\citenamefont {Da~Liao}\ \emph {et~al.}(2022)\citenamefont
  {Da~Liao}, \citenamefont {Xu}, \citenamefont {Meng},\ and\ \citenamefont
  {Qi}}]{liao2022prb}%
  \BibitemOpen
  \bibfield  {author} {\bibinfo {author} {\bibfnamefont {Yuan}\ \bibnamefont
  {Da~Liao}}, \bibinfo {author} {\bibfnamefont {Xiao~Yan}\ \bibnamefont {Xu}},
  \bibinfo {author} {\bibfnamefont {Zi~Yang}\ \bibnamefont {Meng}}, \ and\
  \bibinfo {author} {\bibfnamefont {Yang}\ \bibnamefont {Qi}},\ }\bibfield
  {title} {\enquote {\bibinfo {title} {Dirac fermions with plaquette
  interactions. i. su(2) phase diagram with gross-neveu and deconfined quantum
  criticalities},}\ }\href {\doibase 10.1103/PhysRevB.106.075111} {\bibfield
  {journal} {\bibinfo  {journal} {Phys. Rev. B}\ }\textbf {\bibinfo {volume}
  {106}},\ \bibinfo {pages} {075111} (\bibinfo {year} {2022})}\BibitemShut
  {NoStop}%
\bibitem [{\citenamefont {Zhao}\ \emph {et~al.}(2022)\citenamefont {Zhao},
  \citenamefont {Wang}, \citenamefont {Yan}, \citenamefont {Cheng},\ and\
  \citenamefont {Meng}}]{zhao2022prl}%
  \BibitemOpen
  \bibfield  {author} {\bibinfo {author} {\bibfnamefont {Jiarui}\ \bibnamefont
  {Zhao}}, \bibinfo {author} {\bibfnamefont {Yan-Cheng}\ \bibnamefont {Wang}},
  \bibinfo {author} {\bibfnamefont {Zheng}\ \bibnamefont {Yan}}, \bibinfo
  {author} {\bibfnamefont {Meng}\ \bibnamefont {Cheng}}, \ and\ \bibinfo
  {author} {\bibfnamefont {Zi~Yang}\ \bibnamefont {Meng}},\ }\bibfield  {title}
  {\enquote {\bibinfo {title} {Scaling of entanglement entropy at deconfined
  quantum criticality},}\ }\href {\doibase 10.1103/PhysRevLett.128.010601}
  {\bibfield  {journal} {\bibinfo  {journal} {Phys. Rev. Lett.}\ }\textbf
  {\bibinfo {volume} {128}},\ \bibinfo {pages} {010601} (\bibinfo {year}
  {2022})}\BibitemShut {NoStop}%
\bibitem [{\citenamefont {Wang}\ \emph {et~al.}(2022)\citenamefont {Wang},
  \citenamefont {Ma}, \citenamefont {Cheng},\ and\ \citenamefont
  {Meng}}]{wang2022scipost}%
  \BibitemOpen
  \bibfield  {author} {\bibinfo {author} {\bibfnamefont {Yan-Cheng}\
  \bibnamefont {Wang}}, \bibinfo {author} {\bibfnamefont {Nvsen}\ \bibnamefont
  {Ma}}, \bibinfo {author} {\bibfnamefont {Meng}\ \bibnamefont {Cheng}}, \ and\
  \bibinfo {author} {\bibfnamefont {Zi~Yang}\ \bibnamefont {Meng}},\ }\bibfield
   {title} {\enquote {\bibinfo {title} {{Scaling of the disorder operator at
  deconfined quantum criticality}},}\ }\href {\doibase
  10.21468/SciPostPhys.13.6.123} {\bibfield  {journal} {\bibinfo  {journal}
  {SciPost Phys.}\ }\textbf {\bibinfo {volume} {13}},\ \bibinfo {pages} {123}
  (\bibinfo {year} {2022})}\BibitemShut {NoStop}%
\bibitem [{\citenamefont {Liao}\ \emph {et~al.}(2023)\citenamefont {Liao},
  \citenamefont {Pan}, \citenamefont {Jiang}, \citenamefont {Qi},\ and\
  \citenamefont {Meng}}]{daliao2023teaching}%
  \BibitemOpen
  \bibfield  {author} {\bibinfo {author} {\bibfnamefont {Yuan~Da}\ \bibnamefont
  {Liao}}, \bibinfo {author} {\bibfnamefont {Gaopei}\ \bibnamefont {Pan}},
  \bibinfo {author} {\bibfnamefont {Weilun}\ \bibnamefont {Jiang}}, \bibinfo
  {author} {\bibfnamefont {Yang}\ \bibnamefont {Qi}}, \ and\ \bibinfo {author}
  {\bibfnamefont {Zi~Yang}\ \bibnamefont {Meng}},\ }\href@noop {} {\enquote
  {\bibinfo {title} {The teaching from entanglement: 2d su(2) antiferromagnet
  to valence bond solid deconfined quantum critical points are not
  conformal},}\ } (\bibinfo {year} {2023}),\ \Eprint
  {http://arxiv.org/abs/2302.11742} {arXiv:2302.11742 [cond-mat.str-el]}
  \BibitemShut {NoStop}%
\bibitem [{\citenamefont {Liu}\ \emph {et~al.}(2022{\natexlab{a}})\citenamefont
  {Liu}, \citenamefont {Gong}, \citenamefont {Chen},\ and\ \citenamefont
  {Gu}}]{liu2022emergent}%
  \BibitemOpen
  \bibfield  {author} {\bibinfo {author} {\bibfnamefont {Wen-Yuan}\
  \bibnamefont {Liu}}, \bibinfo {author} {\bibfnamefont {Shou-Shu}\
  \bibnamefont {Gong}}, \bibinfo {author} {\bibfnamefont {Wei-Qiang}\
  \bibnamefont {Chen}}, \ and\ \bibinfo {author} {\bibfnamefont {Zheng-Cheng}\
  \bibnamefont {Gu}},\ }\href@noop {} {\enquote {\bibinfo {title} {Emergent
  symmetry in frustrated magnets: From deconfined quantum critical point to
  gapless quantum spin liquid},}\ } (\bibinfo {year} {2022}{\natexlab{a}}),\
  \Eprint {http://arxiv.org/abs/2212.00707} {arXiv:2212.00707
  [cond-mat.str-el]} \BibitemShut {NoStop}%
\bibitem [{\citenamefont {You}\ and\ \citenamefont
  {Wang}(2022)}]{you2022deconfined}%
  \BibitemOpen
  \bibfield  {author} {\bibinfo {author} {\bibfnamefont {Yi-Zhuang}\
  \bibnamefont {You}}\ and\ \bibinfo {author} {\bibfnamefont {Juven}\
  \bibnamefont {Wang}},\ }\bibfield  {title} {\enquote {\bibinfo {title}
  {Deconfined quantum criticality among grant unified theories},}\ }\href@noop
  {} {\bibfield  {journal} {\bibinfo  {journal} {A Festschrift in Honor of the
  CN Yang Centenary: Scientific Papers}\ ,\ \bibinfo {pages} {367--383}}
  (\bibinfo {year} {2022})}\BibitemShut {NoStop}%
\bibitem [{\citenamefont {Xi}\ and\ \citenamefont {Yu}(2022)}]{Xi_2022}%
  \BibitemOpen
  \bibfield  {author} {\bibinfo {author} {\bibfnamefont {Ning}\ \bibnamefont
  {Xi}}\ and\ \bibinfo {author} {\bibfnamefont {Rong}\ \bibnamefont {Yu}},\
  }\bibfield  {title} {\enquote {\bibinfo {title} {Dynamical signatures of the
  one-dimensional deconfined quantum critical point},}\ }\href {\doibase
  10.1088/1674-1056/ac5987} {\bibfield  {journal} {\bibinfo  {journal} {Chinese
  Physics B}\ }\textbf {\bibinfo {volume} {31}},\ \bibinfo {pages} {057501}
  (\bibinfo {year} {2022})}\BibitemShut {NoStop}%
\bibitem [{\citenamefont {Shu}\ and\ \citenamefont {Yin}(2022)}]{shu2022prb}%
  \BibitemOpen
  \bibfield  {author} {\bibinfo {author} {\bibfnamefont {Yu-Rong}\ \bibnamefont
  {Shu}}\ and\ \bibinfo {author} {\bibfnamefont {Shuai}\ \bibnamefont {Yin}},\
  }\bibfield  {title} {\enquote {\bibinfo {title} {Dual dynamic scaling in
  deconfined quantum criticality},}\ }\href {\doibase
  10.1103/PhysRevB.105.104420} {\bibfield  {journal} {\bibinfo  {journal}
  {Phys. Rev. B}\ }\textbf {\bibinfo {volume} {105}},\ \bibinfo {pages}
  {104420} (\bibinfo {year} {2022})}\BibitemShut {NoStop}%
\bibitem [{\citenamefont {Liu}\ \emph {et~al.}(2022{\natexlab{b}})\citenamefont
  {Liu}, \citenamefont {Hasik}, \citenamefont {Gong}, \citenamefont
  {Poilblanc}, \citenamefont {Chen},\ and\ \citenamefont {Gu}}]{liu2022prx}%
  \BibitemOpen
  \bibfield  {author} {\bibinfo {author} {\bibfnamefont {Wen-Yuan}\
  \bibnamefont {Liu}}, \bibinfo {author} {\bibfnamefont {Juraj}\ \bibnamefont
  {Hasik}}, \bibinfo {author} {\bibfnamefont {Shou-Shu}\ \bibnamefont {Gong}},
  \bibinfo {author} {\bibfnamefont {Didier}\ \bibnamefont {Poilblanc}},
  \bibinfo {author} {\bibfnamefont {Wei-Qiang}\ \bibnamefont {Chen}}, \ and\
  \bibinfo {author} {\bibfnamefont {Zheng-Cheng}\ \bibnamefont {Gu}},\
  }\bibfield  {title} {\enquote {\bibinfo {title} {Emergence of gapless quantum
  spin liquid from deconfined quantum critical point},}\ }\href {\doibase
  10.1103/PhysRevX.12.031039} {\bibfield  {journal} {\bibinfo  {journal} {Phys.
  Rev. X}\ }\textbf {\bibinfo {volume} {12}},\ \bibinfo {pages} {031039}
  (\bibinfo {year} {2022}{\natexlab{b}})}\BibitemShut {NoStop}%
\bibitem [{\citenamefont {Liu}\ \emph {et~al.}(2022{\natexlab{c}})\citenamefont
  {Liu}, \citenamefont {Vojta}, \citenamefont {Assaad},\ and\ \citenamefont
  {Janssen}}]{liu2022prl}%
  \BibitemOpen
  \bibfield  {author} {\bibinfo {author} {\bibfnamefont {Zi~Hong}\ \bibnamefont
  {Liu}}, \bibinfo {author} {\bibfnamefont {Matthias}\ \bibnamefont {Vojta}},
  \bibinfo {author} {\bibfnamefont {Fakher~F.}\ \bibnamefont {Assaad}}, \ and\
  \bibinfo {author} {\bibfnamefont {Lukas}\ \bibnamefont {Janssen}},\
  }\bibfield  {title} {\enquote {\bibinfo {title} {Metallic and deconfined
  quantum criticality in dirac systems},}\ }\href {\doibase
  10.1103/PhysRevLett.128.087201} {\bibfield  {journal} {\bibinfo  {journal}
  {Phys. Rev. Lett.}\ }\textbf {\bibinfo {volume} {128}},\ \bibinfo {pages}
  {087201} (\bibinfo {year} {2022}{\natexlab{c}})}\BibitemShut {NoStop}%
\bibitem [{\citenamefont {Shu}\ \emph {et~al.}(2022)\citenamefont {Shu},
  \citenamefont {Jian},\ and\ \citenamefont {Yin}}]{shu2022prl}%
  \BibitemOpen
  \bibfield  {author} {\bibinfo {author} {\bibfnamefont {Yu-Rong}\ \bibnamefont
  {Shu}}, \bibinfo {author} {\bibfnamefont {Shao-Kai}\ \bibnamefont {Jian}}, \
  and\ \bibinfo {author} {\bibfnamefont {Shuai}\ \bibnamefont {Yin}},\
  }\bibfield  {title} {\enquote {\bibinfo {title} {Nonequilibrium dynamics of
  deconfined quantum critical point in imaginary time},}\ }\href {\doibase
  10.1103/PhysRevLett.128.020601} {\bibfield  {journal} {\bibinfo  {journal}
  {Phys. Rev. Lett.}\ }\textbf {\bibinfo {volume} {128}},\ \bibinfo {pages}
  {020601} (\bibinfo {year} {2022})}\BibitemShut {NoStop}%
\bibitem [{\citenamefont {Huang}\ and\ \citenamefont
  {Yin}(2020)}]{huang2020prr}%
  \BibitemOpen
  \bibfield  {author} {\bibinfo {author} {\bibfnamefont {Rui-Zhen}\
  \bibnamefont {Huang}}\ and\ \bibinfo {author} {\bibfnamefont {Shuai}\
  \bibnamefont {Yin}},\ }\bibfield  {title} {\enquote {\bibinfo {title}
  {Kibble-zurek mechanism for a one-dimensional incarnation of a deconfined
  quantum critical point},}\ }\href {\doibase 10.1103/PhysRevResearch.2.023175}
  {\bibfield  {journal} {\bibinfo  {journal} {Phys. Rev. Res.}\ }\textbf
  {\bibinfo {volume} {2}},\ \bibinfo {pages} {023175} (\bibinfo {year}
  {2020})}\BibitemShut {NoStop}%
\bibitem [{\citenamefont {Zeng}\ \emph {et~al.}(2020)\citenamefont {Zeng},
  \citenamefont {Sheng},\ and\ \citenamefont {Zhu}}]{zeng2020prb}%
  \BibitemOpen
  \bibfield  {author} {\bibinfo {author} {\bibfnamefont {Tian-Sheng}\
  \bibnamefont {Zeng}}, \bibinfo {author} {\bibfnamefont {D.~N.}\ \bibnamefont
  {Sheng}}, \ and\ \bibinfo {author} {\bibfnamefont {W.}~\bibnamefont {Zhu}},\
  }\bibfield  {title} {\enquote {\bibinfo {title} {Continuous phase transition
  between bosonic integer quantum hall liquid and a trivial insulator: Evidence
  for deconfined quantum criticality},}\ }\href {\doibase
  10.1103/PhysRevB.101.035138} {\bibfield  {journal} {\bibinfo  {journal}
  {Phys. Rev. B}\ }\textbf {\bibinfo {volume} {101}},\ \bibinfo {pages}
  {035138} (\bibinfo {year} {2020})}\BibitemShut {NoStop}%
\bibitem [{\citenamefont {Sun}\ \emph {et~al.}(2019)\citenamefont {Sun},
  \citenamefont {Wei},\ and\ \citenamefont {Kou}}]{sun2019prb}%
  \BibitemOpen
  \bibfield  {author} {\bibinfo {author} {\bibfnamefont {Gaoyong}\ \bibnamefont
  {Sun}}, \bibinfo {author} {\bibfnamefont {Bo-Bo}\ \bibnamefont {Wei}}, \ and\
  \bibinfo {author} {\bibfnamefont {Su-Peng}\ \bibnamefont {Kou}},\ }\bibfield
  {title} {\enquote {\bibinfo {title} {Fidelity as a probe for a deconfined
  quantum critical point},}\ }\href {\doibase 10.1103/PhysRevB.100.064427}
  {\bibfield  {journal} {\bibinfo  {journal} {Phys. Rev. B}\ }\textbf {\bibinfo
  {volume} {100}},\ \bibinfo {pages} {064427} (\bibinfo {year}
  {2019})}\BibitemShut {NoStop}%
\bibitem [{\citenamefont {Roberts}\ \emph
  {et~al.}(2019{\natexlab{a}})\citenamefont {Roberts}, \citenamefont {Jiang},\
  and\ \citenamefont {Motrunich}}]{robert2019prb}%
  \BibitemOpen
  \bibfield  {author} {\bibinfo {author} {\bibfnamefont {Brenden}\ \bibnamefont
  {Roberts}}, \bibinfo {author} {\bibfnamefont {Shenghan}\ \bibnamefont
  {Jiang}}, \ and\ \bibinfo {author} {\bibfnamefont {Olexei~I.}\ \bibnamefont
  {Motrunich}},\ }\bibfield  {title} {\enquote {\bibinfo {title} {Deconfined
  quantum critical point in one dimension},}\ }\href {\doibase
  10.1103/PhysRevB.99.165143} {\bibfield  {journal} {\bibinfo  {journal} {Phys.
  Rev. B}\ }\textbf {\bibinfo {volume} {99}},\ \bibinfo {pages} {165143}
  (\bibinfo {year} {2019}{\natexlab{a}})}\BibitemShut {NoStop}%
\bibitem [{\citenamefont {Nahum}\ \emph
  {et~al.}(2015{\natexlab{b}})\citenamefont {Nahum}, \citenamefont {Chalker},
  \citenamefont {Serna}, \citenamefont {Ortu\~no},\ and\ \citenamefont
  {Somoza}}]{nahum2015px}%
  \BibitemOpen
  \bibfield  {author} {\bibinfo {author} {\bibfnamefont {Adam}\ \bibnamefont
  {Nahum}}, \bibinfo {author} {\bibfnamefont {J.~T.}\ \bibnamefont {Chalker}},
  \bibinfo {author} {\bibfnamefont {P.}~\bibnamefont {Serna}}, \bibinfo
  {author} {\bibfnamefont {M.}~\bibnamefont {Ortu\~no}}, \ and\ \bibinfo
  {author} {\bibfnamefont {A.~M.}\ \bibnamefont {Somoza}},\ }\bibfield  {title}
  {\enquote {\bibinfo {title} {Deconfined quantum criticality, scaling
  violations, and classical loop models},}\ }\href {\doibase
  10.1103/PhysRevX.5.041048} {\bibfield  {journal} {\bibinfo  {journal} {Phys.
  Rev. X}\ }\textbf {\bibinfo {volume} {5}},\ \bibinfo {pages} {041048}
  (\bibinfo {year} {2015}{\natexlab{b}})}\BibitemShut {NoStop}%
\bibitem [{\citenamefont {Zhou}\ \emph {et~al.}(2023)\citenamefont {Zhou},
  \citenamefont {Hu}, \citenamefont {Zhu},\ and\ \citenamefont
  {He}}]{zhou2023mathrmso5}%
  \BibitemOpen
  \bibfield  {author} {\bibinfo {author} {\bibfnamefont {Zheng}\ \bibnamefont
  {Zhou}}, \bibinfo {author} {\bibfnamefont {Liangdong}\ \bibnamefont {Hu}},
  \bibinfo {author} {\bibfnamefont {W.}~\bibnamefont {Zhu}}, \ and\ \bibinfo
  {author} {\bibfnamefont {Yin-Chen}\ \bibnamefont {He}},\ }\href@noop {}
  {\enquote {\bibinfo {title} {The $\mathrm{SO}(5)$ deconfined phase transition
  under the fuzzy sphere microscope: Approximate conformal symmetry,
  pseudo-criticality, and operator spectrum},}\ } (\bibinfo {year} {2023}),\
  \Eprint {http://arxiv.org/abs/2306.16435} {arXiv:2306.16435
  [cond-mat.str-el]} \BibitemShut {NoStop}%
\bibitem [{\citenamefont {Liu}\ \emph {et~al.}(2019)\citenamefont {Liu},
  \citenamefont {Wang}, \citenamefont {Sato}, \citenamefont {Hohenadler},
  \citenamefont {Wang}, \citenamefont {Guo},\ and\ \citenamefont
  {Assaad}}]{liu2019superconductivity}%
  \BibitemOpen
  \bibfield  {author} {\bibinfo {author} {\bibfnamefont {Yuhai}\ \bibnamefont
  {Liu}}, \bibinfo {author} {\bibfnamefont {Zhenjiu}\ \bibnamefont {Wang}},
  \bibinfo {author} {\bibfnamefont {Toshihiro}\ \bibnamefont {Sato}}, \bibinfo
  {author} {\bibfnamefont {Martin}\ \bibnamefont {Hohenadler}}, \bibinfo
  {author} {\bibfnamefont {Chong}\ \bibnamefont {Wang}}, \bibinfo {author}
  {\bibfnamefont {Wenan}\ \bibnamefont {Guo}}, \ and\ \bibinfo {author}
  {\bibfnamefont {Fakher~F}\ \bibnamefont {Assaad}},\ }\bibfield  {title}
  {\enquote {\bibinfo {title} {Superconductivity from the condensation of
  topological defects in a quantum spin-hall insulator},}\ }\href@noop {}
  {\bibfield  {journal} {\bibinfo  {journal} {Nature Communications}\ }\textbf
  {\bibinfo {volume} {10}},\ \bibinfo {pages} {2658} (\bibinfo {year}
  {2019})}\BibitemShut {NoStop}%
\bibitem [{\citenamefont {D'Emidio}(2023)}]{demidio2023leeyang}%
  \BibitemOpen
  \bibfield  {author} {\bibinfo {author} {\bibfnamefont {Jonathan}\
  \bibnamefont {D'Emidio}},\ }\href@noop {} {\enquote {\bibinfo {title}
  {Lee-yang zeros at $o(3)$ and deconfined quantum critical points},}\ }
  (\bibinfo {year} {2023}),\ \Eprint {http://arxiv.org/abs/2308.00575}
  {arXiv:2308.00575 [cond-mat.str-el]} \BibitemShut {NoStop}%
\bibitem [{\citenamefont {D'Emidio}\ and\ \citenamefont
  {Kaul}(2017)}]{jonathan2017prl}%
  \BibitemOpen
  \bibfield  {author} {\bibinfo {author} {\bibfnamefont {Jonathan}\
  \bibnamefont {D'Emidio}}\ and\ \bibinfo {author} {\bibfnamefont {Ribhu~K.}\
  \bibnamefont {Kaul}},\ }\bibfield  {title} {\enquote {\bibinfo {title} {New
  easy-plane $\mathbb{C}{\mathbb{p}}^{N\ensuremath{-}1}$ fixed points},}\
  }\href {\doibase 10.1103/PhysRevLett.118.187202} {\bibfield  {journal}
  {\bibinfo  {journal} {Phys. Rev. Lett.}\ }\textbf {\bibinfo {volume} {118}},\
  \bibinfo {pages} {187202} (\bibinfo {year} {2017})}\BibitemShut {NoStop}%
\bibitem [{\citenamefont {Chen}\ \emph {et~al.}(2013)\citenamefont {Chen},
  \citenamefont {Huang}, \citenamefont {Deng}, \citenamefont {Kuklov},
  \citenamefont {Prokof'ev},\ and\ \citenamefont {Svistunov}}]{chen2013prl}%
  \BibitemOpen
  \bibfield  {author} {\bibinfo {author} {\bibfnamefont {Kun}\ \bibnamefont
  {Chen}}, \bibinfo {author} {\bibfnamefont {Yuan}\ \bibnamefont {Huang}},
  \bibinfo {author} {\bibfnamefont {Youjin}\ \bibnamefont {Deng}}, \bibinfo
  {author} {\bibfnamefont {A.~B.}\ \bibnamefont {Kuklov}}, \bibinfo {author}
  {\bibfnamefont {N.~V.}\ \bibnamefont {Prokof'ev}}, \ and\ \bibinfo {author}
  {\bibfnamefont {B.~V.}\ \bibnamefont {Svistunov}},\ }\bibfield  {title}
  {\enquote {\bibinfo {title} {Deconfined criticality flow in the heisenberg
  model with ring-exchange interactions},}\ }\href {\doibase
  10.1103/PhysRevLett.110.185701} {\bibfield  {journal} {\bibinfo  {journal}
  {Phys. Rev. Lett.}\ }\textbf {\bibinfo {volume} {110}},\ \bibinfo {pages}
  {185701} (\bibinfo {year} {2013})}\BibitemShut {NoStop}%
\bibitem [{\citenamefont {Guo}\ \emph {et~al.}(2020)\citenamefont {Guo},
  \citenamefont {Sun}, \citenamefont {Zhao}, \citenamefont {Wang},
  \citenamefont {Hong}, \citenamefont {Sidorov}, \citenamefont {Ma},
  \citenamefont {Wu}, \citenamefont {Li}, \citenamefont {Meng}, \citenamefont
  {Sandvik},\ and\ \citenamefont {Sun}}]{guo2020prl}%
  \BibitemOpen
  \bibfield  {author} {\bibinfo {author} {\bibfnamefont {Jing}\ \bibnamefont
  {Guo}}, \bibinfo {author} {\bibfnamefont {Guangyu}\ \bibnamefont {Sun}},
  \bibinfo {author} {\bibfnamefont {Bowen}\ \bibnamefont {Zhao}}, \bibinfo
  {author} {\bibfnamefont {Ling}\ \bibnamefont {Wang}}, \bibinfo {author}
  {\bibfnamefont {Wenshan}\ \bibnamefont {Hong}}, \bibinfo {author}
  {\bibfnamefont {Vladimir~A.}\ \bibnamefont {Sidorov}}, \bibinfo {author}
  {\bibfnamefont {Nvsen}\ \bibnamefont {Ma}}, \bibinfo {author} {\bibfnamefont
  {Qi}~\bibnamefont {Wu}}, \bibinfo {author} {\bibfnamefont {Shiliang}\
  \bibnamefont {Li}}, \bibinfo {author} {\bibfnamefont {Zi~Yang}\ \bibnamefont
  {Meng}}, \bibinfo {author} {\bibfnamefont {Anders~W.}\ \bibnamefont
  {Sandvik}}, \ and\ \bibinfo {author} {\bibfnamefont {Liling}\ \bibnamefont
  {Sun}},\ }\bibfield  {title} {\enquote {\bibinfo {title} {Quantum phases of
  ${\mathrm{srcu}}_{2}({\mathrm{bo}}_{3}{)}_{2}$ from high-pressure
  thermodynamics},}\ }\href {\doibase 10.1103/PhysRevLett.124.206602}
  {\bibfield  {journal} {\bibinfo  {journal} {Phys. Rev. Lett.}\ }\textbf
  {\bibinfo {volume} {124}},\ \bibinfo {pages} {206602} (\bibinfo {year}
  {2020})}\BibitemShut {NoStop}%
\bibitem [{\citenamefont {Cui}\ \emph {et~al.}(2023)\citenamefont {Cui},
  \citenamefont {Liu}, \citenamefont {Lin}, \citenamefont {Wu}, \citenamefont
  {Hong}, \citenamefont {Liu}, \citenamefont {Li}, \citenamefont {Hu},
  \citenamefont {Xi}, \citenamefont {Li} \emph {et~al.}}]{cui2023proximate}%
  \BibitemOpen
  \bibfield  {author} {\bibinfo {author} {\bibfnamefont {Yi}~\bibnamefont
  {Cui}}, \bibinfo {author} {\bibfnamefont {Lu}~\bibnamefont {Liu}}, \bibinfo
  {author} {\bibfnamefont {Huihang}\ \bibnamefont {Lin}}, \bibinfo {author}
  {\bibfnamefont {Kai-Hsin}\ \bibnamefont {Wu}}, \bibinfo {author}
  {\bibfnamefont {Wenshan}\ \bibnamefont {Hong}}, \bibinfo {author}
  {\bibfnamefont {Xuefei}\ \bibnamefont {Liu}}, \bibinfo {author}
  {\bibfnamefont {Cong}\ \bibnamefont {Li}}, \bibinfo {author} {\bibfnamefont
  {Ze}~\bibnamefont {Hu}}, \bibinfo {author} {\bibfnamefont {Ning}\
  \bibnamefont {Xi}}, \bibinfo {author} {\bibfnamefont {Shiliang}\ \bibnamefont
  {Li}},  \emph {et~al.},\ }\bibfield  {title} {\enquote {\bibinfo {title}
  {Proximate deconfined quantum critical point in srcu2 (bo3) 2},}\ }\href@noop
  {} {\bibfield  {journal} {\bibinfo  {journal} {Science}\ }\textbf {\bibinfo
  {volume} {380}},\ \bibinfo {pages} {1179--1184} (\bibinfo {year}
  {2023})}\BibitemShut {NoStop}%
\bibitem [{\citenamefont {Song}\ \emph
  {et~al.}(2023{\natexlab{b}})\citenamefont {Song}, \citenamefont {Jia},
  \citenamefont {Yu}, \citenamefont {Tang}, \citenamefont {Wang}, \citenamefont
  {Singha}, \citenamefont {Gui}, \citenamefont {Uzan}, \citenamefont
  {Onyszczak}, \citenamefont {Watanabe}, \citenamefont {Taniguchi},
  \citenamefont {Cava}, \citenamefont {Schoop}, \citenamefont {Ong},\ and\
  \citenamefont {Wu}}]{song2023unconventional}%
  \BibitemOpen
  \bibfield  {author} {\bibinfo {author} {\bibfnamefont {Tiancheng}\
  \bibnamefont {Song}}, \bibinfo {author} {\bibfnamefont {Yanyu}\ \bibnamefont
  {Jia}}, \bibinfo {author} {\bibfnamefont {Guo}\ \bibnamefont {Yu}}, \bibinfo
  {author} {\bibfnamefont {Yue}\ \bibnamefont {Tang}}, \bibinfo {author}
  {\bibfnamefont {Pengjie}\ \bibnamefont {Wang}}, \bibinfo {author}
  {\bibfnamefont {Ratnadwip}\ \bibnamefont {Singha}}, \bibinfo {author}
  {\bibfnamefont {Xin}\ \bibnamefont {Gui}}, \bibinfo {author} {\bibfnamefont
  {Ayelet~J.}\ \bibnamefont {Uzan}}, \bibinfo {author} {\bibfnamefont
  {Michael}\ \bibnamefont {Onyszczak}}, \bibinfo {author} {\bibfnamefont
  {Kenji}\ \bibnamefont {Watanabe}}, \bibinfo {author} {\bibfnamefont
  {Takashi}\ \bibnamefont {Taniguchi}}, \bibinfo {author} {\bibfnamefont
  {Robert~J.}\ \bibnamefont {Cava}}, \bibinfo {author} {\bibfnamefont
  {Leslie~M.}\ \bibnamefont {Schoop}}, \bibinfo {author} {\bibfnamefont
  {N.~P.}\ \bibnamefont {Ong}}, \ and\ \bibinfo {author} {\bibfnamefont
  {Sanfeng}\ \bibnamefont {Wu}},\ }\href@noop {} {\enquote {\bibinfo {title}
  {Unconventional superconducting quantum criticality in monolayer wte2},}\ }
  (\bibinfo {year} {2023}{\natexlab{b}}),\ \Eprint
  {http://arxiv.org/abs/2303.06540} {arXiv:2303.06540 [cond-mat.mes-hall]}
  \BibitemShut {NoStop}%
\bibitem [{\citenamefont {Saffman}\ \emph {et~al.}(2010)\citenamefont
  {Saffman}, \citenamefont {Walker},\ and\ \citenamefont
  {M\o{}lmer}}]{Saffman2010rmp}%
  \BibitemOpen
  \bibfield  {author} {\bibinfo {author} {\bibfnamefont {M.}~\bibnamefont
  {Saffman}}, \bibinfo {author} {\bibfnamefont {T.~G.}\ \bibnamefont {Walker}},
  \ and\ \bibinfo {author} {\bibfnamefont {K.}~\bibnamefont {M\o{}lmer}},\
  }\bibfield  {title} {\enquote {\bibinfo {title} {Quantum information with
  rydberg atoms},}\ }\href {\doibase 10.1103/RevModPhys.82.2313} {\bibfield
  {journal} {\bibinfo  {journal} {Rev. Mod. Phys.}\ }\textbf {\bibinfo {volume}
  {82}},\ \bibinfo {pages} {2313--2363} (\bibinfo {year} {2010})}\BibitemShut
  {NoStop}%
\bibitem [{\citenamefont {Deng}\ \emph {et~al.}(2005)\citenamefont {Deng},
  \citenamefont {Porras},\ and\ \citenamefont {Cirac}}]{deng2005pra}%
  \BibitemOpen
  \bibfield  {author} {\bibinfo {author} {\bibfnamefont {X.-L.}\ \bibnamefont
  {Deng}}, \bibinfo {author} {\bibfnamefont {D.}~\bibnamefont {Porras}}, \ and\
  \bibinfo {author} {\bibfnamefont {J.~I.}\ \bibnamefont {Cirac}},\ }\bibfield
  {title} {\enquote {\bibinfo {title} {Effective spin quantum phases in systems
  of trapped ions},}\ }\href {\doibase 10.1103/PhysRevA.72.063407} {\bibfield
  {journal} {\bibinfo  {journal} {Phys. Rev. A}\ }\textbf {\bibinfo {volume}
  {72}},\ \bibinfo {pages} {063407} (\bibinfo {year} {2005})}\BibitemShut
  {NoStop}%
\bibitem [{\citenamefont {Lahaye}\ \emph {et~al.}(2009)\citenamefont {Lahaye},
  \citenamefont {Menotti}, \citenamefont {Santos}, \citenamefont {Lewenstein},\
  and\ \citenamefont {Pfau}}]{Lahaye_2009}%
  \BibitemOpen
  \bibfield  {author} {\bibinfo {author} {\bibfnamefont {T}~\bibnamefont
  {Lahaye}}, \bibinfo {author} {\bibfnamefont {C}~\bibnamefont {Menotti}},
  \bibinfo {author} {\bibfnamefont {L}~\bibnamefont {Santos}}, \bibinfo
  {author} {\bibfnamefont {M}~\bibnamefont {Lewenstein}}, \ and\ \bibinfo
  {author} {\bibfnamefont {T}~\bibnamefont {Pfau}},\ }\bibfield  {title}
  {\enquote {\bibinfo {title} {The physics of dipolar bosonic quantum gases},}\
  }\href {\doibase 10.1088/0034-4885/72/12/126401} {\bibfield  {journal}
  {\bibinfo  {journal} {Reports on Progress in Physics}\ }\textbf {\bibinfo
  {volume} {72}},\ \bibinfo {pages} {126401} (\bibinfo {year}
  {2009})}\BibitemShut {NoStop}%
\bibitem [{\citenamefont {Ritsch}\ \emph {et~al.}(2013)\citenamefont {Ritsch},
  \citenamefont {Domokos}, \citenamefont {Brennecke},\ and\ \citenamefont
  {Esslinger}}]{ritsch2013rmp}%
  \BibitemOpen
  \bibfield  {author} {\bibinfo {author} {\bibfnamefont {Helmut}\ \bibnamefont
  {Ritsch}}, \bibinfo {author} {\bibfnamefont {Peter}\ \bibnamefont {Domokos}},
  \bibinfo {author} {\bibfnamefont {Ferdinand}\ \bibnamefont {Brennecke}}, \
  and\ \bibinfo {author} {\bibfnamefont {Tilman}\ \bibnamefont {Esslinger}},\
  }\bibfield  {title} {\enquote {\bibinfo {title} {Cold atoms in
  cavity-generated dynamical optical potentials},}\ }\href {\doibase
  10.1103/RevModPhys.85.553} {\bibfield  {journal} {\bibinfo  {journal} {Rev.
  Mod. Phys.}\ }\textbf {\bibinfo {volume} {85}},\ \bibinfo {pages} {553--601}
  (\bibinfo {year} {2013})}\BibitemShut {NoStop}%
\bibitem [{\citenamefont {Carr}\ \emph {et~al.}(2009)\citenamefont {Carr},
  \citenamefont {DeMille}, \citenamefont {Krems},\ and\ \citenamefont
  {Ye}}]{Carr_2009}%
  \BibitemOpen
  \bibfield  {author} {\bibinfo {author} {\bibfnamefont {Lincoln~D}\
  \bibnamefont {Carr}}, \bibinfo {author} {\bibfnamefont {David}\ \bibnamefont
  {DeMille}}, \bibinfo {author} {\bibfnamefont {Roman~V}\ \bibnamefont
  {Krems}}, \ and\ \bibinfo {author} {\bibfnamefont {Jun}\ \bibnamefont {Ye}},\
  }\bibfield  {title} {\enquote {\bibinfo {title} {Cold and ultracold
  molecules: science, technology and applications},}\ }\href {\doibase
  10.1088/1367-2630/11/5/055049} {\bibfield  {journal} {\bibinfo  {journal}
  {New Journal of Physics}\ }\textbf {\bibinfo {volume} {11}},\ \bibinfo
  {pages} {055049} (\bibinfo {year} {2009})}\BibitemShut {NoStop}%
\bibitem [{\citenamefont {Blatt}\ and\ \citenamefont
  {Roos}(2012)}]{blatt2012quantum}%
  \BibitemOpen
  \bibfield  {author} {\bibinfo {author} {\bibfnamefont {Rainer}\ \bibnamefont
  {Blatt}}\ and\ \bibinfo {author} {\bibfnamefont {Christian~F}\ \bibnamefont
  {Roos}},\ }\bibfield  {title} {\enquote {\bibinfo {title} {Quantum
  simulations with trapped ions},}\ }\href@noop {} {\bibfield  {journal}
  {\bibinfo  {journal} {Nature Physics}\ }\textbf {\bibinfo {volume} {8}},\
  \bibinfo {pages} {277--284} (\bibinfo {year} {2012})}\BibitemShut {NoStop}%
\bibitem [{\citenamefont {Britton}\ \emph {et~al.}(2012)\citenamefont
  {Britton}, \citenamefont {Sawyer}, \citenamefont {Keith}, \citenamefont
  {Wang}, \citenamefont {Freericks}, \citenamefont {Uys}, \citenamefont
  {Biercuk},\ and\ \citenamefont {Bollinger}}]{britton2012engineered}%
  \BibitemOpen
  \bibfield  {author} {\bibinfo {author} {\bibfnamefont {Joseph~W}\
  \bibnamefont {Britton}}, \bibinfo {author} {\bibfnamefont {Brian~C}\
  \bibnamefont {Sawyer}}, \bibinfo {author} {\bibfnamefont {Adam~C}\
  \bibnamefont {Keith}}, \bibinfo {author} {\bibfnamefont {C-C~Joseph}\
  \bibnamefont {Wang}}, \bibinfo {author} {\bibfnamefont {James~K}\
  \bibnamefont {Freericks}}, \bibinfo {author} {\bibfnamefont {Hermann}\
  \bibnamefont {Uys}}, \bibinfo {author} {\bibfnamefont {Michael~J}\
  \bibnamefont {Biercuk}}, \ and\ \bibinfo {author} {\bibfnamefont {John~J}\
  \bibnamefont {Bollinger}},\ }\bibfield  {title} {\enquote {\bibinfo {title}
  {Engineered two-dimensional ising interactions in a trapped-ion quantum
  simulator with hundreds of spins},}\ }\href@noop {} {\bibfield  {journal}
  {\bibinfo  {journal} {Nature}\ }\textbf {\bibinfo {volume} {484}},\ \bibinfo
  {pages} {489--492} (\bibinfo {year} {2012})}\BibitemShut {NoStop}%
\bibitem [{\citenamefont {Islam}\ \emph {et~al.}(2013)\citenamefont {Islam},
  \citenamefont {Senko}, \citenamefont {Campbell}, \citenamefont {Korenblit},
  \citenamefont {Smith}, \citenamefont {Lee}, \citenamefont {Edwards},
  \citenamefont {Wang}, \citenamefont {Freericks},\ and\ \citenamefont
  {Monroe}}]{islam2013emergence}%
  \BibitemOpen
  \bibfield  {author} {\bibinfo {author} {\bibfnamefont {R}~\bibnamefont
  {Islam}}, \bibinfo {author} {\bibfnamefont {Crystal}\ \bibnamefont {Senko}},
  \bibinfo {author} {\bibfnamefont {Wes~C}\ \bibnamefont {Campbell}}, \bibinfo
  {author} {\bibfnamefont {S}~\bibnamefont {Korenblit}}, \bibinfo {author}
  {\bibfnamefont {J}~\bibnamefont {Smith}}, \bibinfo {author} {\bibfnamefont
  {A}~\bibnamefont {Lee}}, \bibinfo {author} {\bibfnamefont {EE}~\bibnamefont
  {Edwards}}, \bibinfo {author} {\bibfnamefont {C-CJ}\ \bibnamefont {Wang}},
  \bibinfo {author} {\bibfnamefont {JK}~\bibnamefont {Freericks}}, \ and\
  \bibinfo {author} {\bibfnamefont {C}~\bibnamefont {Monroe}},\ }\bibfield
  {title} {\enquote {\bibinfo {title} {Emergence and frustration of magnetism
  with variable-range interactions in a quantum simulator},}\ }\href@noop {}
  {\bibfield  {journal} {\bibinfo  {journal} {science}\ }\textbf {\bibinfo
  {volume} {340}},\ \bibinfo {pages} {583--587} (\bibinfo {year}
  {2013})}\BibitemShut {NoStop}%
\bibitem [{\citenamefont {Richerme}\ \emph {et~al.}(2014)\citenamefont
  {Richerme}, \citenamefont {Gong}, \citenamefont {Lee}, \citenamefont {Senko},
  \citenamefont {Smith}, \citenamefont {Foss-Feig}, \citenamefont {Michalakis},
  \citenamefont {Gorshkov},\ and\ \citenamefont {Monroe}}]{richerme2014non}%
  \BibitemOpen
  \bibfield  {author} {\bibinfo {author} {\bibfnamefont {Philip}\ \bibnamefont
  {Richerme}}, \bibinfo {author} {\bibfnamefont {Zhe-Xuan}\ \bibnamefont
  {Gong}}, \bibinfo {author} {\bibfnamefont {Aaron}\ \bibnamefont {Lee}},
  \bibinfo {author} {\bibfnamefont {Crystal}\ \bibnamefont {Senko}}, \bibinfo
  {author} {\bibfnamefont {Jacob}\ \bibnamefont {Smith}}, \bibinfo {author}
  {\bibfnamefont {Michael}\ \bibnamefont {Foss-Feig}}, \bibinfo {author}
  {\bibfnamefont {Spyridon}\ \bibnamefont {Michalakis}}, \bibinfo {author}
  {\bibfnamefont {Alexey~V}\ \bibnamefont {Gorshkov}}, \ and\ \bibinfo {author}
  {\bibfnamefont {Christopher}\ \bibnamefont {Monroe}},\ }\bibfield  {title}
  {\enquote {\bibinfo {title} {Non-local propagation of correlations in quantum
  systems with long-range interactions},}\ }\href@noop {} {\bibfield  {journal}
  {\bibinfo  {journal} {Nature}\ }\textbf {\bibinfo {volume} {511}},\ \bibinfo
  {pages} {198--201} (\bibinfo {year} {2014})}\BibitemShut {NoStop}%
\bibitem [{\citenamefont {Jurcevic}\ \emph {et~al.}(2014)\citenamefont
  {Jurcevic}, \citenamefont {Lanyon}, \citenamefont {Hauke}, \citenamefont
  {Hempel}, \citenamefont {Zoller}, \citenamefont {Blatt},\ and\ \citenamefont
  {Roos}}]{jurcevic2014quasiparticle}%
  \BibitemOpen
  \bibfield  {author} {\bibinfo {author} {\bibfnamefont {Petar}\ \bibnamefont
  {Jurcevic}}, \bibinfo {author} {\bibfnamefont {Ben~P}\ \bibnamefont
  {Lanyon}}, \bibinfo {author} {\bibfnamefont {Philipp}\ \bibnamefont {Hauke}},
  \bibinfo {author} {\bibfnamefont {Cornelius}\ \bibnamefont {Hempel}},
  \bibinfo {author} {\bibfnamefont {Peter}\ \bibnamefont {Zoller}}, \bibinfo
  {author} {\bibfnamefont {Rainer}\ \bibnamefont {Blatt}}, \ and\ \bibinfo
  {author} {\bibfnamefont {Christian~F}\ \bibnamefont {Roos}},\ }\bibfield
  {title} {\enquote {\bibinfo {title} {Quasiparticle engineering and
  entanglement propagation in a quantum many-body system},}\ }\href@noop {}
  {\bibfield  {journal} {\bibinfo  {journal} {Nature}\ }\textbf {\bibinfo
  {volume} {511}},\ \bibinfo {pages} {202--205} (\bibinfo {year}
  {2014})}\BibitemShut {NoStop}%
\bibitem [{\citenamefont {Yu}\ \emph {et~al.}(2022)\citenamefont {Yu},
  \citenamefont {Yang}, \citenamefont {Xu},\ and\ \citenamefont
  {Xu}}]{yu2022prb}%
  \BibitemOpen
  \bibfield  {author} {\bibinfo {author} {\bibfnamefont {Xue-Jia}\ \bibnamefont
  {Yu}}, \bibinfo {author} {\bibfnamefont {Sheng}\ \bibnamefont {Yang}},
  \bibinfo {author} {\bibfnamefont {Jing-Bo}\ \bibnamefont {Xu}}, \ and\
  \bibinfo {author} {\bibfnamefont {Limei}\ \bibnamefont {Xu}},\ }\bibfield
  {title} {\enquote {\bibinfo {title} {Fidelity susceptibility as a diagnostic
  of the commensurate-incommensurate transition: A revisit of the programmable
  rydberg chain},}\ }\href {\doibase 10.1103/PhysRevB.106.165124} {\bibfield
  {journal} {\bibinfo  {journal} {Phys. Rev. B}\ }\textbf {\bibinfo {volume}
  {106}},\ \bibinfo {pages} {165124} (\bibinfo {year} {2022})}\BibitemShut
  {NoStop}%
\bibitem [{\citenamefont {Defenu}\ \emph {et~al.}(2018)\citenamefont {Defenu},
  \citenamefont {Enss}, \citenamefont {Kastner},\ and\ \citenamefont
  {Morigi}}]{defenu2018prl}%
  \BibitemOpen
  \bibfield  {author} {\bibinfo {author} {\bibfnamefont {Nicol\`o}\
  \bibnamefont {Defenu}}, \bibinfo {author} {\bibfnamefont {Tilman}\
  \bibnamefont {Enss}}, \bibinfo {author} {\bibfnamefont {Michael}\
  \bibnamefont {Kastner}}, \ and\ \bibinfo {author} {\bibfnamefont {Giovanna}\
  \bibnamefont {Morigi}},\ }\bibfield  {title} {\enquote {\bibinfo {title}
  {Dynamical critical scaling of long-range interacting quantum magnets},}\
  }\href {\doibase 10.1103/PhysRevLett.121.240403} {\bibfield  {journal}
  {\bibinfo  {journal} {Phys. Rev. Lett.}\ }\textbf {\bibinfo {volume} {121}},\
  \bibinfo {pages} {240403} (\bibinfo {year} {2018})}\BibitemShut {NoStop}%
\bibitem [{\citenamefont {Defenu}\ \emph {et~al.}(2021)\citenamefont {Defenu},
  \citenamefont {Donner}, \citenamefont {Macrì}, \citenamefont {Pagano},
  \citenamefont {Ruffo},\ and\ \citenamefont
  {Trombettoni}}]{defenu2021longrange}%
  \BibitemOpen
  \bibfield  {author} {\bibinfo {author} {\bibfnamefont {Nicolò}\ \bibnamefont
  {Defenu}}, \bibinfo {author} {\bibfnamefont {Tobias}\ \bibnamefont {Donner}},
  \bibinfo {author} {\bibfnamefont {Tommaso}\ \bibnamefont {Macrì}}, \bibinfo
  {author} {\bibfnamefont {Guido}\ \bibnamefont {Pagano}}, \bibinfo {author}
  {\bibfnamefont {Stefano}\ \bibnamefont {Ruffo}}, \ and\ \bibinfo {author}
  {\bibfnamefont {Andrea}\ \bibnamefont {Trombettoni}},\ }\href@noop {}
  {\enquote {\bibinfo {title} {Long-range interacting quantum systems},}\ }
  (\bibinfo {year} {2021}),\ \Eprint {http://arxiv.org/abs/2109.01063}
  {arXiv:2109.01063 [cond-mat.quant-gas]} \BibitemShut {NoStop}%
\bibitem [{\citenamefont {Defenu}\ \emph {et~al.}(2023)\citenamefont {Defenu},
  \citenamefont {Lerose},\ and\ \citenamefont
  {Pappalardi}}]{defenu2023outofequilibrium}%
  \BibitemOpen
  \bibfield  {author} {\bibinfo {author} {\bibfnamefont {Nicolò}\ \bibnamefont
  {Defenu}}, \bibinfo {author} {\bibfnamefont {Alessio}\ \bibnamefont
  {Lerose}}, \ and\ \bibinfo {author} {\bibfnamefont {Silvia}\ \bibnamefont
  {Pappalardi}},\ }\href@noop {} {\enquote {\bibinfo {title}
  {Out-of-equilibrium dynamics of quantum many-body systems with long-range
  interactions},}\ } (\bibinfo {year} {2023}),\ \Eprint
  {http://arxiv.org/abs/2307.04802} {arXiv:2307.04802 [cond-mat.quant-gas]}
  \BibitemShut {NoStop}%
\bibitem [{\citenamefont {Syed}\ \emph {et~al.}(2021)\citenamefont {Syed},
  \citenamefont {Enss},\ and\ \citenamefont {Defenu}}]{syed2021prb}%
  \BibitemOpen
  \bibfield  {author} {\bibinfo {author} {\bibfnamefont {Marvin}\ \bibnamefont
  {Syed}}, \bibinfo {author} {\bibfnamefont {Tilman}\ \bibnamefont {Enss}}, \
  and\ \bibinfo {author} {\bibfnamefont {Nicol\`o}\ \bibnamefont {Defenu}},\
  }\bibfield  {title} {\enquote {\bibinfo {title} {Dynamical quantum phase
  transition in a bosonic system with long-range interactions},}\ }\href
  {\doibase 10.1103/PhysRevB.103.064306} {\bibfield  {journal} {\bibinfo
  {journal} {Phys. Rev. B}\ }\textbf {\bibinfo {volume} {103}},\ \bibinfo
  {pages} {064306} (\bibinfo {year} {2021})}\BibitemShut {NoStop}%
\bibitem [{\citenamefont {Defenu}\ \emph {et~al.}(2016)\citenamefont {Defenu},
  \citenamefont {Trombettoni},\ and\ \citenamefont {Ruffo}}]{defenu2016prb}%
  \BibitemOpen
  \bibfield  {author} {\bibinfo {author} {\bibfnamefont {Nicol\`o}\
  \bibnamefont {Defenu}}, \bibinfo {author} {\bibfnamefont {Andrea}\
  \bibnamefont {Trombettoni}}, \ and\ \bibinfo {author} {\bibfnamefont
  {Stefano}\ \bibnamefont {Ruffo}},\ }\bibfield  {title} {\enquote {\bibinfo
  {title} {Anisotropic long-range spin systems},}\ }\href {\doibase
  10.1103/PhysRevB.94.224411} {\bibfield  {journal} {\bibinfo  {journal} {Phys.
  Rev. B}\ }\textbf {\bibinfo {volume} {94}},\ \bibinfo {pages} {224411}
  (\bibinfo {year} {2016})}\BibitemShut {NoStop}%
\bibitem [{\citenamefont {Gong}\ \emph
  {et~al.}(2016{\natexlab{a}})\citenamefont {Gong}, \citenamefont {Maghrebi},
  \citenamefont {Hu}, \citenamefont {Wall}, \citenamefont {Foss-Feig},\ and\
  \citenamefont {Gorshkov}}]{gong2016prb}%
  \BibitemOpen
  \bibfield  {author} {\bibinfo {author} {\bibfnamefont {Z.-X.}\ \bibnamefont
  {Gong}}, \bibinfo {author} {\bibfnamefont {M.~F.}\ \bibnamefont {Maghrebi}},
  \bibinfo {author} {\bibfnamefont {A.}~\bibnamefont {Hu}}, \bibinfo {author}
  {\bibfnamefont {M.~L.}\ \bibnamefont {Wall}}, \bibinfo {author}
  {\bibfnamefont {M.}~\bibnamefont {Foss-Feig}}, \ and\ \bibinfo {author}
  {\bibfnamefont {A.~V.}\ \bibnamefont {Gorshkov}},\ }\bibfield  {title}
  {\enquote {\bibinfo {title} {Topological phases with long-range
  interactions},}\ }\href {\doibase 10.1103/PhysRevB.93.041102} {\bibfield
  {journal} {\bibinfo  {journal} {Phys. Rev. B}\ }\textbf {\bibinfo {volume}
  {93}},\ \bibinfo {pages} {041102} (\bibinfo {year}
  {2016}{\natexlab{a}})}\BibitemShut {NoStop}%
\bibitem [{\citenamefont {Gong}\ \emph
  {et~al.}(2016{\natexlab{b}})\citenamefont {Gong}, \citenamefont {Maghrebi},
  \citenamefont {Hu}, \citenamefont {Foss-Feig}, \citenamefont {Richerme},
  \citenamefont {Monroe},\ and\ \citenamefont {Gorshkov}}]{gong2016prb2}%
  \BibitemOpen
  \bibfield  {author} {\bibinfo {author} {\bibfnamefont {Z.-X.}\ \bibnamefont
  {Gong}}, \bibinfo {author} {\bibfnamefont {M.~F.}\ \bibnamefont {Maghrebi}},
  \bibinfo {author} {\bibfnamefont {A.}~\bibnamefont {Hu}}, \bibinfo {author}
  {\bibfnamefont {M.}~\bibnamefont {Foss-Feig}}, \bibinfo {author}
  {\bibfnamefont {P.}~\bibnamefont {Richerme}}, \bibinfo {author}
  {\bibfnamefont {C.}~\bibnamefont {Monroe}}, \ and\ \bibinfo {author}
  {\bibfnamefont {A.~V.}\ \bibnamefont {Gorshkov}},\ }\bibfield  {title}
  {\enquote {\bibinfo {title} {Kaleidoscope of quantum phases in a long-range
  interacting spin-1 chain},}\ }\href {\doibase 10.1103/PhysRevB.93.205115}
  {\bibfield  {journal} {\bibinfo  {journal} {Phys. Rev. B}\ }\textbf {\bibinfo
  {volume} {93}},\ \bibinfo {pages} {205115} (\bibinfo {year}
  {2016}{\natexlab{b}})}\BibitemShut {NoStop}%
\bibitem [{\citenamefont {Shen}(2023)}]{shen2023long}%
  \BibitemOpen
  \bibfield  {author} {\bibinfo {author} {\bibfnamefont {Xiaoyang}\
  \bibnamefont {Shen}},\ }\href@noop {} {\enquote {\bibinfo {title} {Long range
  syk model and boundary syk model},}\ } (\bibinfo {year} {2023}),\ \Eprint
  {http://arxiv.org/abs/2308.12598} {arXiv:2308.12598 [hep-th]} \BibitemShut
  {NoStop}%
\bibitem [{\citenamefont {Fisher}\ \emph {et~al.}(1972)\citenamefont {Fisher},
  \citenamefont {Ma},\ and\ \citenamefont {Nickel}}]{fisher1972prl}%
  \BibitemOpen
  \bibfield  {author} {\bibinfo {author} {\bibfnamefont {Michael~E.}\
  \bibnamefont {Fisher}}, \bibinfo {author} {\bibfnamefont {Shang-keng}\
  \bibnamefont {Ma}}, \ and\ \bibinfo {author} {\bibfnamefont {B.~G.}\
  \bibnamefont {Nickel}},\ }\bibfield  {title} {\enquote {\bibinfo {title}
  {Critical exponents for long-range interactions},}\ }\href {\doibase
  10.1103/PhysRevLett.29.917} {\bibfield  {journal} {\bibinfo  {journal} {Phys.
  Rev. Lett.}\ }\textbf {\bibinfo {volume} {29}},\ \bibinfo {pages} {917--920}
  (\bibinfo {year} {1972})}\BibitemShut {NoStop}%
\bibitem [{\citenamefont {Defenu}\ \emph {et~al.}(2017)\citenamefont {Defenu},
  \citenamefont {Trombettoni},\ and\ \citenamefont {Ruffo}}]{defenu2017prb}%
  \BibitemOpen
  \bibfield  {author} {\bibinfo {author} {\bibfnamefont {Nicol\`o}\
  \bibnamefont {Defenu}}, \bibinfo {author} {\bibfnamefont {Andrea}\
  \bibnamefont {Trombettoni}}, \ and\ \bibinfo {author} {\bibfnamefont
  {Stefano}\ \bibnamefont {Ruffo}},\ }\bibfield  {title} {\enquote {\bibinfo
  {title} {Criticality and phase diagram of quantum long-range o($n$)
  models},}\ }\href {\doibase 10.1103/PhysRevB.96.104432} {\bibfield  {journal}
  {\bibinfo  {journal} {Phys. Rev. B}\ }\textbf {\bibinfo {volume} {96}},\
  \bibinfo {pages} {104432} (\bibinfo {year} {2017})}\BibitemShut {NoStop}%
\bibitem [{\citenamefont {Defenu}\ \emph {et~al.}(2020)\citenamefont {Defenu},
  \citenamefont {Codello}, \citenamefont {Ruffo},\ and\ \citenamefont
  {Trombettoni}}]{Defenu_2020}%
  \BibitemOpen
  \bibfield  {author} {\bibinfo {author} {\bibfnamefont {Nicolò}\ \bibnamefont
  {Defenu}}, \bibinfo {author} {\bibfnamefont {Alessandro}\ \bibnamefont
  {Codello}}, \bibinfo {author} {\bibfnamefont {Stefano}\ \bibnamefont
  {Ruffo}}, \ and\ \bibinfo {author} {\bibfnamefont {Andrea}\ \bibnamefont
  {Trombettoni}},\ }\bibfield  {title} {\enquote {\bibinfo {title} {Criticality
  of spin systems with weak long-range interactions},}\ }\href {\doibase
  10.1088/1751-8121/ab6a6c} {\bibfield  {journal} {\bibinfo  {journal} {Journal
  of Physics A: Mathematical and Theoretical}\ }\textbf {\bibinfo {volume}
  {53}},\ \bibinfo {pages} {143001} (\bibinfo {year} {2020})}\BibitemShut
  {NoStop}%
\bibitem [{\citenamefont {Giachetti}\ \emph {et~al.}(2021)\citenamefont
  {Giachetti}, \citenamefont {Defenu}, \citenamefont {Ruffo},\ and\
  \citenamefont {Trombettoni}}]{giachetti2021prl}%
  \BibitemOpen
  \bibfield  {author} {\bibinfo {author} {\bibfnamefont {Guido}\ \bibnamefont
  {Giachetti}}, \bibinfo {author} {\bibfnamefont {Nicol\`o}\ \bibnamefont
  {Defenu}}, \bibinfo {author} {\bibfnamefont {Stefano}\ \bibnamefont {Ruffo}},
  \ and\ \bibinfo {author} {\bibfnamefont {Andrea}\ \bibnamefont
  {Trombettoni}},\ }\bibfield  {title} {\enquote {\bibinfo {title}
  {Berezinskii-kosterlitz-thouless phase transitions with long-range
  couplings},}\ }\href {\doibase 10.1103/PhysRevLett.127.156801} {\bibfield
  {journal} {\bibinfo  {journal} {Phys. Rev. Lett.}\ }\textbf {\bibinfo
  {volume} {127}},\ \bibinfo {pages} {156801} (\bibinfo {year}
  {2021})}\BibitemShut {NoStop}%
\bibitem [{\citenamefont {Giachetti}\ \emph {et~al.}(2022)\citenamefont
  {Giachetti}, \citenamefont {Trombettoni}, \citenamefont {Ruffo},\ and\
  \citenamefont {Defenu}}]{giachetti2022prb}%
  \BibitemOpen
  \bibfield  {author} {\bibinfo {author} {\bibfnamefont {Guido}\ \bibnamefont
  {Giachetti}}, \bibinfo {author} {\bibfnamefont {Andrea}\ \bibnamefont
  {Trombettoni}}, \bibinfo {author} {\bibfnamefont {Stefano}\ \bibnamefont
  {Ruffo}}, \ and\ \bibinfo {author} {\bibfnamefont {Nicol\`o}\ \bibnamefont
  {Defenu}},\ }\bibfield  {title} {\enquote {\bibinfo {title}
  {Berezinskii-kosterlitz-thouless transitions in classical and quantum
  long-range systems},}\ }\href {\doibase 10.1103/PhysRevB.106.014106}
  {\bibfield  {journal} {\bibinfo  {journal} {Phys. Rev. B}\ }\textbf {\bibinfo
  {volume} {106}},\ \bibinfo {pages} {014106} (\bibinfo {year}
  {2022})}\BibitemShut {NoStop}%
\bibitem [{\citenamefont {Codello}\ \emph {et~al.}(2015)\citenamefont
  {Codello}, \citenamefont {Defenu},\ and\ \citenamefont
  {D'Odorico}}]{codello2015prd}%
  \BibitemOpen
  \bibfield  {author} {\bibinfo {author} {\bibfnamefont {Alessandro}\
  \bibnamefont {Codello}}, \bibinfo {author} {\bibfnamefont {Nicol\'o}\
  \bibnamefont {Defenu}}, \ and\ \bibinfo {author} {\bibfnamefont {Giulio}\
  \bibnamefont {D'Odorico}},\ }\bibfield  {title} {\enquote {\bibinfo {title}
  {Critical exponents of $o(n)$ models in fractional dimensions},}\ }\href
  {\doibase 10.1103/PhysRevD.91.105003} {\bibfield  {journal} {\bibinfo
  {journal} {Phys. Rev. D}\ }\textbf {\bibinfo {volume} {91}},\ \bibinfo
  {pages} {105003} (\bibinfo {year} {2015})}\BibitemShut {NoStop}%
\bibitem [{\citenamefont {Song}\ \emph
  {et~al.}(2023{\natexlab{c}})\citenamefont {Song}, \citenamefont {Zhao},
  \citenamefont {Zhou},\ and\ \citenamefont {Meng}}]{song2023dynamical}%
  \BibitemOpen
  \bibfield  {author} {\bibinfo {author} {\bibfnamefont {Menghan}\ \bibnamefont
  {Song}}, \bibinfo {author} {\bibfnamefont {Jiarui}\ \bibnamefont {Zhao}},
  \bibinfo {author} {\bibfnamefont {Chengkang}\ \bibnamefont {Zhou}}, \ and\
  \bibinfo {author} {\bibfnamefont {Zi~Yang}\ \bibnamefont {Meng}},\
  }\href@noop {} {\enquote {\bibinfo {title} {Dynamical properties of quantum
  many-body systems with long range interactions},}\ } (\bibinfo {year}
  {2023}{\natexlab{c}}),\ \Eprint {http://arxiv.org/abs/2301.00829}
  {arXiv:2301.00829 [cond-mat.str-el]} \BibitemShut {NoStop}%
\bibitem [{\citenamefont {Song}\ \emph
  {et~al.}(2023{\natexlab{d}})\citenamefont {Song}, \citenamefont {Zhao},
  \citenamefont {Qi}, \citenamefont {Rong},\ and\ \citenamefont
  {Meng}}]{song2023quantum}%
  \BibitemOpen
  \bibfield  {author} {\bibinfo {author} {\bibfnamefont {Menghan}\ \bibnamefont
  {Song}}, \bibinfo {author} {\bibfnamefont {Jiarui}\ \bibnamefont {Zhao}},
  \bibinfo {author} {\bibfnamefont {Yang}\ \bibnamefont {Qi}}, \bibinfo
  {author} {\bibfnamefont {Junchen}\ \bibnamefont {Rong}}, \ and\ \bibinfo
  {author} {\bibfnamefont {Zi~Yang}\ \bibnamefont {Meng}},\ }\href@noop {}
  {\enquote {\bibinfo {title} {Quantum criticality and entanglement for 2d
  long-range heisenberg bilayer},}\ } (\bibinfo {year} {2023}{\natexlab{d}}),\
  \Eprint {http://arxiv.org/abs/2306.05465} {arXiv:2306.05465
  [cond-mat.str-el]} \BibitemShut {NoStop}%
\bibitem [{\citenamefont {Zhao}\ \emph {et~al.}(2023)\citenamefont {Zhao},
  \citenamefont {Song}, \citenamefont {Qi}, \citenamefont {Rong},\ and\
  \citenamefont {Meng}}]{zhao2023finitetemperature}%
  \BibitemOpen
  \bibfield  {author} {\bibinfo {author} {\bibfnamefont {Jiarui}\ \bibnamefont
  {Zhao}}, \bibinfo {author} {\bibfnamefont {Menghan}\ \bibnamefont {Song}},
  \bibinfo {author} {\bibfnamefont {Yang}\ \bibnamefont {Qi}}, \bibinfo
  {author} {\bibfnamefont {Junchen}\ \bibnamefont {Rong}}, \ and\ \bibinfo
  {author} {\bibfnamefont {Zi~Yang}\ \bibnamefont {Meng}},\ }\href@noop {}
  {\enquote {\bibinfo {title} {Finite-temperature critical behaviors in 2d
  long-range quantum heisenberg model},}\ } (\bibinfo {year} {2023}),\ \Eprint
  {http://arxiv.org/abs/2306.01044} {arXiv:2306.01044 [cond-mat.str-el]}
  \BibitemShut {NoStop}%
\bibitem [{\citenamefont {Defenu}\ \emph {et~al.}(2015)\citenamefont {Defenu},
  \citenamefont {Trombettoni},\ and\ \citenamefont {Codello}}]{defenu2015pre}%
  \BibitemOpen
  \bibfield  {author} {\bibinfo {author} {\bibfnamefont {Nicol\'o}\
  \bibnamefont {Defenu}}, \bibinfo {author} {\bibfnamefont {Andrea}\
  \bibnamefont {Trombettoni}}, \ and\ \bibinfo {author} {\bibfnamefont
  {Alessandro}\ \bibnamefont {Codello}},\ }\bibfield  {title} {\enquote
  {\bibinfo {title} {Fixed-point structure and effective fractional
  dimensionality for o($n$) models with long-range interactions},}\ }\href
  {\doibase 10.1103/PhysRevE.92.052113} {\bibfield  {journal} {\bibinfo
  {journal} {Phys. Rev. E}\ }\textbf {\bibinfo {volume} {92}},\ \bibinfo
  {pages} {052113} (\bibinfo {year} {2015})}\BibitemShut {NoStop}%
\bibitem [{\citenamefont {Maghrebi}\ \emph {et~al.}(2016)\citenamefont
  {Maghrebi}, \citenamefont {Gong}, \citenamefont {Foss-Feig},\ and\
  \citenamefont {Gorshkov}}]{maghrebi2016prb}%
  \BibitemOpen
  \bibfield  {author} {\bibinfo {author} {\bibfnamefont {Mohammad~F.}\
  \bibnamefont {Maghrebi}}, \bibinfo {author} {\bibfnamefont {Zhe-Xuan}\
  \bibnamefont {Gong}}, \bibinfo {author} {\bibfnamefont {Michael}\
  \bibnamefont {Foss-Feig}}, \ and\ \bibinfo {author} {\bibfnamefont
  {Alexey~V.}\ \bibnamefont {Gorshkov}},\ }\bibfield  {title} {\enquote
  {\bibinfo {title} {Causality and quantum criticality in long-range lattice
  models},}\ }\href {\doibase 10.1103/PhysRevB.93.125128} {\bibfield  {journal}
  {\bibinfo  {journal} {Phys. Rev. B}\ }\textbf {\bibinfo {volume} {93}},\
  \bibinfo {pages} {125128} (\bibinfo {year} {2016})}\BibitemShut {NoStop}%
\bibitem [{\citenamefont {{Maghrebi}}\ \emph {et~al.}(2017)\citenamefont
  {{Maghrebi}}, \citenamefont {{Gong}},\ and\ \citenamefont
  {{Gorshkov}}}]{longrangeboson2017}%
  \BibitemOpen
  \bibfield  {author} {\bibinfo {author} {\bibfnamefont {Mohammad~F.}\
  \bibnamefont {{Maghrebi}}}, \bibinfo {author} {\bibfnamefont {Zhe-Xuan}\
  \bibnamefont {{Gong}}}, \ and\ \bibinfo {author} {\bibfnamefont {Alexey~V.}\
  \bibnamefont {{Gorshkov}}},\ }\bibfield  {title} {\enquote {\bibinfo {title}
  {{Continuous Symmetry Breaking in 1D Long-Range Interacting Quantum
  Systems}},}\ }\href {\doibase 10.1103/PhysRevLett.119.023001} {\bibfield
  {journal} {\bibinfo  {journal} {\prl}\ }\textbf {\bibinfo {volume} {119}},\
  \bibinfo {eid} {023001} (\bibinfo {year} {2017})}\BibitemShut {NoStop}%
\bibitem [{\citenamefont {Yang}\ \emph {et~al.}(2020)\citenamefont {Yang},
  \citenamefont {Yao},\ and\ \citenamefont {Sandvik}}]{yang2020deconfined}%
  \BibitemOpen
  \bibfield  {author} {\bibinfo {author} {\bibfnamefont {Sibin}\ \bibnamefont
  {Yang}}, \bibinfo {author} {\bibfnamefont {Dao-Xin}\ \bibnamefont {Yao}}, \
  and\ \bibinfo {author} {\bibfnamefont {Anders~W.}\ \bibnamefont {Sandvik}},\
  }\href@noop {} {\enquote {\bibinfo {title} {Deconfined quantum criticality in
  spin-1/2 chains with long-range interactions},}\ } (\bibinfo {year} {2020}),\
  \Eprint {http://arxiv.org/abs/2001.02821} {arXiv:2001.02821
  [physics.comp-ph]} \BibitemShut {NoStop}%
\bibitem [{\citenamefont {Sandvik}(2010{\natexlab{b}})}]{sandvik2010prl2}%
  \BibitemOpen
  \bibfield  {author} {\bibinfo {author} {\bibfnamefont {Anders~W.}\
  \bibnamefont {Sandvik}},\ }\bibfield  {title} {\enquote {\bibinfo {title}
  {Ground states of a frustrated quantum spin chain with long-range
  interactions},}\ }\href {\doibase 10.1103/PhysRevLett.104.137204} {\bibfield
  {journal} {\bibinfo  {journal} {Phys. Rev. Lett.}\ }\textbf {\bibinfo
  {volume} {104}},\ \bibinfo {pages} {137204} (\bibinfo {year}
  {2010}{\natexlab{b}})}\BibitemShut {NoStop}%
\bibitem [{\citenamefont {Yu}\ \emph {et~al.}(2023)\citenamefont {Yu},
  \citenamefont {Ding},\ and\ \citenamefont {Xu}}]{yu2023pre}%
  \BibitemOpen
  \bibfield  {author} {\bibinfo {author} {\bibfnamefont {Xue-Jia}\ \bibnamefont
  {Yu}}, \bibinfo {author} {\bibfnamefont {Chengxiang}\ \bibnamefont {Ding}}, \
  and\ \bibinfo {author} {\bibfnamefont {Limei}\ \bibnamefont {Xu}},\
  }\bibfield  {title} {\enquote {\bibinfo {title} {Quantum criticality of a
  ${\mathbb{z}}_{3}$-symmetric spin chain with long-range interactions},}\
  }\href {\doibase 10.1103/PhysRevE.107.054122} {\bibfield  {journal} {\bibinfo
   {journal} {Phys. Rev. E}\ }\textbf {\bibinfo {volume} {107}},\ \bibinfo
  {pages} {054122} (\bibinfo {year} {2023})}\BibitemShut {NoStop}%
\bibitem [{\citenamefont {Vu}\ \emph {et~al.}(2022)\citenamefont {Vu},
  \citenamefont {Huang}, \citenamefont {Li},\ and\ \citenamefont
  {Das~Sarma}}]{vu2022prl}%
  \BibitemOpen
  \bibfield  {author} {\bibinfo {author} {\bibfnamefont {DinhDuy}\ \bibnamefont
  {Vu}}, \bibinfo {author} {\bibfnamefont {Ke}~\bibnamefont {Huang}}, \bibinfo
  {author} {\bibfnamefont {Xiao}\ \bibnamefont {Li}}, \ and\ \bibinfo {author}
  {\bibfnamefont {S.}~\bibnamefont {Das~Sarma}},\ }\bibfield  {title} {\enquote
  {\bibinfo {title} {Fermionic many-body localization for random and
  quasiperiodic systems in the presence of short- and long-range
  interactions},}\ }\href {\doibase 10.1103/PhysRevLett.128.146601} {\bibfield
  {journal} {\bibinfo  {journal} {Phys. Rev. Lett.}\ }\textbf {\bibinfo
  {volume} {128}},\ \bibinfo {pages} {146601} (\bibinfo {year}
  {2022})}\BibitemShut {NoStop}%
\bibitem [{\citenamefont {Leaw}\ \emph {et~al.}(2019)\citenamefont {Leaw},
  \citenamefont {Tang}, \citenamefont {Trushin}, \citenamefont {Assaad},\ and\
  \citenamefont {Adam}}]{leaw2019universal}%
  \BibitemOpen
  \bibfield  {author} {\bibinfo {author} {\bibfnamefont {Jia~Ning}\
  \bibnamefont {Leaw}}, \bibinfo {author} {\bibfnamefont {Ho-Kin}\ \bibnamefont
  {Tang}}, \bibinfo {author} {\bibfnamefont {Maxim}\ \bibnamefont {Trushin}},
  \bibinfo {author} {\bibfnamefont {Fakher~F}\ \bibnamefont {Assaad}}, \ and\
  \bibinfo {author} {\bibfnamefont {Shaffique}\ \bibnamefont {Adam}},\
  }\bibfield  {title} {\enquote {\bibinfo {title} {Universal fermi-surface
  anisotropy renormalization for interacting dirac fermions with long-range
  interactions},}\ }\href@noop {} {\bibfield  {journal} {\bibinfo  {journal}
  {Proceedings of the National Academy of Sciences}\ }\textbf {\bibinfo
  {volume} {116}},\ \bibinfo {pages} {26431--26434} (\bibinfo {year}
  {2019})}\BibitemShut {NoStop}%
\bibitem [{\citenamefont {Ghazaryan}\ and\ \citenamefont
  {Chakraborty}(2015)}]{ghazaryan2015prb}%
  \BibitemOpen
  \bibfield  {author} {\bibinfo {author} {\bibfnamefont {Areg}\ \bibnamefont
  {Ghazaryan}}\ and\ \bibinfo {author} {\bibfnamefont {Tapash}\ \bibnamefont
  {Chakraborty}},\ }\bibfield  {title} {\enquote {\bibinfo {title} {Long-range
  coulomb interaction and majorana fermions},}\ }\href {\doibase
  10.1103/PhysRevB.92.115138} {\bibfield  {journal} {\bibinfo  {journal} {Phys.
  Rev. B}\ }\textbf {\bibinfo {volume} {92}},\ \bibinfo {pages} {115138}
  (\bibinfo {year} {2015})}\BibitemShut {NoStop}%
\bibitem [{\citenamefont {Vodola}\ \emph {et~al.}(2014)\citenamefont {Vodola},
  \citenamefont {Lepori}, \citenamefont {Ercolessi}, \citenamefont {Gorshkov},\
  and\ \citenamefont {Pupillo}}]{vodola2014prl}%
  \BibitemOpen
  \bibfield  {author} {\bibinfo {author} {\bibfnamefont {Davide}\ \bibnamefont
  {Vodola}}, \bibinfo {author} {\bibfnamefont {Luca}\ \bibnamefont {Lepori}},
  \bibinfo {author} {\bibfnamefont {Elisa}\ \bibnamefont {Ercolessi}}, \bibinfo
  {author} {\bibfnamefont {Alexey~V.}\ \bibnamefont {Gorshkov}}, \ and\
  \bibinfo {author} {\bibfnamefont {Guido}\ \bibnamefont {Pupillo}},\
  }\bibfield  {title} {\enquote {\bibinfo {title} {Kitaev chains with
  long-range pairing},}\ }\href {\doibase 10.1103/PhysRevLett.113.156402}
  {\bibfield  {journal} {\bibinfo  {journal} {Phys. Rev. Lett.}\ }\textbf
  {\bibinfo {volume} {113}},\ \bibinfo {pages} {156402} (\bibinfo {year}
  {2014})}\BibitemShut {NoStop}%
\bibitem [{\citenamefont {Roberts}\ \emph
  {et~al.}(2019{\natexlab{b}})\citenamefont {Roberts}, \citenamefont {Jiang},\
  and\ \citenamefont {Motrunich}}]{roberts2019dqcp1d}%
  \BibitemOpen
  \bibfield  {author} {\bibinfo {author} {\bibfnamefont {Brenden}\ \bibnamefont
  {Roberts}}, \bibinfo {author} {\bibfnamefont {Shenghan}\ \bibnamefont
  {Jiang}}, \ and\ \bibinfo {author} {\bibfnamefont {Olexei~I}\ \bibnamefont
  {Motrunich}},\ }\bibfield  {title} {\enquote {\bibinfo {title} {Deconfined
  quantum critical point in one dimension},}\ }\href@noop {} {\bibfield
  {journal} {\bibinfo  {journal} {Physical Review B}\ }\textbf {\bibinfo
  {volume} {99}},\ \bibinfo {pages} {165143} (\bibinfo {year}
  {2019}{\natexlab{b}})}\BibitemShut {NoStop}%
\bibitem [{\citenamefont {Mudry}\ \emph {et~al.}(2019)\citenamefont {Mudry},
  \citenamefont {Furusaki}, \citenamefont {Morimoto},\ and\ \citenamefont
  {Hikihara}}]{mudry2019_1dqdcp}%
  \BibitemOpen
  \bibfield  {author} {\bibinfo {author} {\bibfnamefont {Christopher}\
  \bibnamefont {Mudry}}, \bibinfo {author} {\bibfnamefont {Akira}\ \bibnamefont
  {Furusaki}}, \bibinfo {author} {\bibfnamefont {Takahiro}\ \bibnamefont
  {Morimoto}}, \ and\ \bibinfo {author} {\bibfnamefont {Toshiya}\ \bibnamefont
  {Hikihara}},\ }\bibfield  {title} {\enquote {\bibinfo {title} {Quantum phase
  transitions beyond landau-ginzburg theory in one-dimensional space
  revisited},}\ }\href@noop {} {\bibfield  {journal} {\bibinfo  {journal}
  {Physical Review B}\ }\textbf {\bibinfo {volume} {99}},\ \bibinfo {pages}
  {205153} (\bibinfo {year} {2019})}\BibitemShut {NoStop}%
\bibitem [{\citenamefont {Huang}\ \emph {et~al.}(2019)\citenamefont {Huang},
  \citenamefont {Lu}, \citenamefont {You}, \citenamefont {Meng},\ and\
  \citenamefont {Xiang}}]{huang2019_1ddqcp}%
  \BibitemOpen
  \bibfield  {author} {\bibinfo {author} {\bibfnamefont {Rui-Zhen}\
  \bibnamefont {Huang}}, \bibinfo {author} {\bibfnamefont {Da-Chuan}\
  \bibnamefont {Lu}}, \bibinfo {author} {\bibfnamefont {Yi-Zhuang}\
  \bibnamefont {You}}, \bibinfo {author} {\bibfnamefont {Zi~Yang}\ \bibnamefont
  {Meng}}, \ and\ \bibinfo {author} {\bibfnamefont {Tao}\ \bibnamefont
  {Xiang}},\ }\bibfield  {title} {\enquote {\bibinfo {title} {Emergent symmetry
  and conserved current at a one-dimensional incarnation of deconfined quantum
  critical point},}\ }\href@noop {} {\bibfield  {journal} {\bibinfo  {journal}
  {Physical Review B}\ }\textbf {\bibinfo {volume} {100}},\ \bibinfo {pages}
  {125137} (\bibinfo {year} {2019})}\BibitemShut {NoStop}%
\bibitem [{\citenamefont {White}(1992)}]{white1992prl}%
  \BibitemOpen
  \bibfield  {author} {\bibinfo {author} {\bibfnamefont {Steven~R.}\
  \bibnamefont {White}},\ }\bibfield  {title} {\enquote {\bibinfo {title}
  {Density matrix formulation for quantum renormalization groups},}\ }\href
  {\doibase 10.1103/PhysRevLett.69.2863} {\bibfield  {journal} {\bibinfo
  {journal} {Phys. Rev. Lett.}\ }\textbf {\bibinfo {volume} {69}},\ \bibinfo
  {pages} {2863--2866} (\bibinfo {year} {1992})}\BibitemShut {NoStop}%
\bibitem [{\citenamefont {Schollw\"ock}(2005)}]{schollwock2005rmp}%
  \BibitemOpen
  \bibfield  {author} {\bibinfo {author} {\bibfnamefont {U.}~\bibnamefont
  {Schollw\"ock}},\ }\bibfield  {title} {\enquote {\bibinfo {title} {The
  density-matrix renormalization group},}\ }\href {\doibase
  10.1103/RevModPhys.77.259} {\bibfield  {journal} {\bibinfo  {journal} {Rev.
  Mod. Phys.}\ }\textbf {\bibinfo {volume} {77}},\ \bibinfo {pages} {259--315}
  (\bibinfo {year} {2005})}\BibitemShut {NoStop}%
\bibitem [{\citenamefont {Schollwöck}(2011)}]{SCHOLLWOCK201196}%
  \BibitemOpen
  \bibfield  {author} {\bibinfo {author} {\bibfnamefont {Ulrich}\ \bibnamefont
  {Schollwöck}},\ }\bibfield  {title} {\enquote {\bibinfo {title} {The
  density-matrix renormalization group in the age of matrix product states},}\
  }\href {\doibase https://doi.org/10.1016/j.aop.2010.09.012} {\bibfield
  {journal} {\bibinfo  {journal} {Annals of Physics}\ }\textbf {\bibinfo
  {volume} {326}},\ \bibinfo {pages} {96--192} (\bibinfo {year} {2011})},\
  \bibinfo {note} {january 2011 Special Issue}\BibitemShut {NoStop}%
\bibitem [{\citenamefont {Verstraete}\ \emph {et~al.}(2004)\citenamefont
  {Verstraete}, \citenamefont {Porras},\ and\ \citenamefont
  {Cirac}}]{verstraete2004prl}%
  \BibitemOpen
  \bibfield  {author} {\bibinfo {author} {\bibfnamefont {F.}~\bibnamefont
  {Verstraete}}, \bibinfo {author} {\bibfnamefont {D.}~\bibnamefont {Porras}},
  \ and\ \bibinfo {author} {\bibfnamefont {J.~I.}\ \bibnamefont {Cirac}},\
  }\bibfield  {title} {\enquote {\bibinfo {title} {Density matrix
  renormalization group and periodic boundary conditions: A quantum information
  perspective},}\ }\href {\doibase 10.1103/PhysRevLett.93.227205} {\bibfield
  {journal} {\bibinfo  {journal} {Phys. Rev. Lett.}\ }\textbf {\bibinfo
  {volume} {93}},\ \bibinfo {pages} {227205} (\bibinfo {year}
  {2004})}\BibitemShut {NoStop}%
\bibitem [{\citenamefont {Nigel}(1992)}]{Nigel1992}%
  \BibitemOpen
  \bibfield  {author} {\bibinfo {author} {\bibfnamefont {Goldenfeld}\
  \bibnamefont {Nigel}},\ }\href@noop {} {\emph {\bibinfo {title} {Lectures on
  Phase Transitions and the Renormalization Group}}},\ \bibinfo {edition}
  {1st}\ ed.\ (\bibinfo  {publisher} {CRC Press},\ \bibinfo {year}
  {1992})\BibitemShut {NoStop}%
\bibitem [{\citenamefont {Luo}\ \emph {et~al.}(2019)\citenamefont {Luo},
  \citenamefont {Zhao},\ and\ \citenamefont {Wang}}]{luo2019prb}%
  \BibitemOpen
  \bibfield  {author} {\bibinfo {author} {\bibfnamefont {Qiang}\ \bibnamefont
  {Luo}}, \bibinfo {author} {\bibfnamefont {Jize}\ \bibnamefont {Zhao}}, \ and\
  \bibinfo {author} {\bibfnamefont {Xiaoqun}\ \bibnamefont {Wang}},\ }\bibfield
   {title} {\enquote {\bibinfo {title} {Intrinsic jump character of first-order
  quantum phase transitions},}\ }\href {\doibase 10.1103/PhysRevB.100.121111}
  {\bibfield  {journal} {\bibinfo  {journal} {Phys. Rev. B}\ }\textbf {\bibinfo
  {volume} {100}},\ \bibinfo {pages} {121111} (\bibinfo {year}
  {2019})}\BibitemShut {NoStop}%
\bibitem [{\citenamefont {Luo}\ \emph {et~al.}(2023)\citenamefont {Luo},
  \citenamefont {Hu}, \citenamefont {Li}, \citenamefont {Zhao}, \citenamefont
  {Kee},\ and\ \citenamefont {Wang}}]{luo2023prb}%
  \BibitemOpen
  \bibfield  {author} {\bibinfo {author} {\bibfnamefont {Qiang}\ \bibnamefont
  {Luo}}, \bibinfo {author} {\bibfnamefont {Shijie}\ \bibnamefont {Hu}},
  \bibinfo {author} {\bibfnamefont {Jinbin}\ \bibnamefont {Li}}, \bibinfo
  {author} {\bibfnamefont {Jize}\ \bibnamefont {Zhao}}, \bibinfo {author}
  {\bibfnamefont {Hae-Young}\ \bibnamefont {Kee}}, \ and\ \bibinfo {author}
  {\bibfnamefont {Xiaoqun}\ \bibnamefont {Wang}},\ }\bibfield  {title}
  {\enquote {\bibinfo {title} {Spontaneous dimerization, spin-nematic order,
  and deconfined quantum critical point in a spin-1 kitaev chain with tunable
  single-ion anisotropy},}\ }\href {\doibase 10.1103/PhysRevB.107.245131}
  {\bibfield  {journal} {\bibinfo  {journal} {Phys. Rev. B}\ }\textbf {\bibinfo
  {volume} {107}},\ \bibinfo {pages} {245131} (\bibinfo {year}
  {2023})}\BibitemShut {NoStop}%
\bibitem [{\citenamefont {Binder}(1981)}]{Binder1981}%
  \BibitemOpen
  \bibfield  {author} {\bibinfo {author} {\bibfnamefont {K.}~\bibnamefont
  {Binder}},\ }\bibfield  {title} {\enquote {\bibinfo {title} {Finite size
  scaling analysis of ising model block distribution functions},}\ }\href
  {\doibase 10.1007/BF01293604} {\bibfield  {journal} {\bibinfo  {journal} {Z.
  Physik B - Condensed Matter}\ }\textbf {\bibinfo {volume} {43}},\ \bibinfo
  {pages} {119--140} (\bibinfo {year} {1981})}\BibitemShut {NoStop}%
\bibitem [{\citenamefont {Vollmayr}\ \emph {et~al.}(1993)\citenamefont
  {Vollmayr}, \citenamefont {Reger}, \citenamefont {Scheucher},\ and\
  \citenamefont {Binder}}]{Binder1993}%
  \BibitemOpen
  \bibfield  {author} {\bibinfo {author} {\bibfnamefont {Katharina}\
  \bibnamefont {Vollmayr}}, \bibinfo {author} {\bibfnamefont {Joseph~D.}\
  \bibnamefont {Reger}}, \bibinfo {author} {\bibfnamefont {Manfred}\
  \bibnamefont {Scheucher}}, \ and\ \bibinfo {author} {\bibfnamefont {Kurt}\
  \bibnamefont {Binder}},\ }\bibfield  {title} {\enquote {\bibinfo {title}
  {Finite size effects at thermally-driven first order phase transitions: A
  phenomenological theory of the order parameter distribution},}\ }\href
  {\doibase 10.1007/BF01316713} {\bibfield  {journal} {\bibinfo  {journal} {Z.
  Physik B - Condensed Matter}\ }\textbf {\bibinfo {volume} {91}},\ \bibinfo
  {pages} {113–125} (\bibinfo {year} {1993})}\BibitemShut {NoStop}%
\bibitem [{\citenamefont {Sandvik}(2010{\natexlab{c}})}]{sandvik2010aip}%
  \BibitemOpen
  \bibfield  {author} {\bibinfo {author} {\bibfnamefont {Anders~W.}\
  \bibnamefont {Sandvik}},\ }\bibfield  {title} {\enquote {\bibinfo {title}
  {Computational studies of quantum spin systems},}\ }\href {\doibase
  10.1063/1.3518900} {\bibfield  {journal} {\bibinfo  {journal} {AIP Conference
  Proceedings}\ }\textbf {\bibinfo {volume} {1297}},\ \bibinfo {pages}
  {135--338} (\bibinfo {year} {2010}{\natexlab{c}})}\BibitemShut {NoStop}%
\bibitem [{\citenamefont {Wang}\ \emph
  {et~al.}(2017{\natexlab{b}})\citenamefont {Wang}, \citenamefont {Qi},
  \citenamefont {Chen},\ and\ \citenamefont {Meng}}]{wang2017prb}%
  \BibitemOpen
  \bibfield  {author} {\bibinfo {author} {\bibfnamefont {Yan-Cheng}\
  \bibnamefont {Wang}}, \bibinfo {author} {\bibfnamefont {Yang}\ \bibnamefont
  {Qi}}, \bibinfo {author} {\bibfnamefont {Shu}\ \bibnamefont {Chen}}, \ and\
  \bibinfo {author} {\bibfnamefont {Zi~Yang}\ \bibnamefont {Meng}},\ }\bibfield
   {title} {\enquote {\bibinfo {title} {Caution on emergent continuous
  symmetry: A monte carlo investigation of the transverse-field frustrated
  ising model on the triangular and honeycomb lattices},}\ }\href {\doibase
  10.1103/PhysRevB.96.115160} {\bibfield  {journal} {\bibinfo  {journal} {Phys.
  Rev. B}\ }\textbf {\bibinfo {volume} {96}},\ \bibinfo {pages} {115160}
  (\bibinfo {year} {2017}{\natexlab{b}})}\BibitemShut {NoStop}%
\bibitem [{\citenamefont {Kuklov}\ \emph {et~al.}(2006)\citenamefont {Kuklov},
  \citenamefont {Prokof’Ev}, \citenamefont {Svistunov},\ and\ \citenamefont
  {Troyer}}]{kuklov2006deconfined}%
  \BibitemOpen
  \bibfield  {author} {\bibinfo {author} {\bibfnamefont {AB}~\bibnamefont
  {Kuklov}}, \bibinfo {author} {\bibfnamefont {NV}~\bibnamefont {Prokof’Ev}},
  \bibinfo {author} {\bibfnamefont {BV}~\bibnamefont {Svistunov}}, \ and\
  \bibinfo {author} {\bibfnamefont {Matthias}\ \bibnamefont {Troyer}},\
  }\bibfield  {title} {\enquote {\bibinfo {title} {Deconfined criticality,
  runaway flow in the two-component scalar electrodynamics and weak first-order
  superfluid-solid transitions},}\ }\href@noop {} {\bibfield  {journal}
  {\bibinfo  {journal} {Annals of Physics}\ }\textbf {\bibinfo {volume}
  {321}},\ \bibinfo {pages} {1602--1621} (\bibinfo {year} {2006})}\BibitemShut
  {NoStop}%
\bibitem [{\citenamefont {Sreejith}\ \emph
  {et~al.}(2019{\natexlab{a}})\citenamefont {Sreejith}, \citenamefont
  {Powell},\ and\ \citenamefont {Nahum}}]{nahum2019prl}%
  \BibitemOpen
  \bibfield  {author} {\bibinfo {author} {\bibfnamefont {G.~J.}\ \bibnamefont
  {Sreejith}}, \bibinfo {author} {\bibfnamefont {Stephen}\ \bibnamefont
  {Powell}}, \ and\ \bibinfo {author} {\bibfnamefont {Adam}\ \bibnamefont
  {Nahum}},\ }\bibfield  {title} {\enquote {\bibinfo {title} {Emergent so(5)
  symmetry at the columnar ordering transition in the classical cubic dimer
  model},}\ }\href {\doibase 10.1103/PhysRevLett.122.080601} {\bibfield
  {journal} {\bibinfo  {journal} {Phys. Rev. Lett.}\ }\textbf {\bibinfo
  {volume} {122}},\ \bibinfo {pages} {080601} (\bibinfo {year}
  {2019}{\natexlab{a}})}\BibitemShut {NoStop}%
\bibitem [{\citenamefont {Serna}\ and\ \citenamefont
  {Nahum}(2019{\natexlab{a}})}]{nahum2019prb}%
  \BibitemOpen
  \bibfield  {author} {\bibinfo {author} {\bibfnamefont {Pablo}\ \bibnamefont
  {Serna}}\ and\ \bibinfo {author} {\bibfnamefont {Adam}\ \bibnamefont
  {Nahum}},\ }\bibfield  {title} {\enquote {\bibinfo {title} {Emergence and
  spontaneous breaking of approximate $\mathrm{O}(4)$ symmetry at a weakly
  first-order deconfined phase transition},}\ }\href {\doibase
  10.1103/PhysRevB.99.195110} {\bibfield  {journal} {\bibinfo  {journal} {Phys.
  Rev. B}\ }\textbf {\bibinfo {volume} {99}},\ \bibinfo {pages} {195110}
  (\bibinfo {year} {2019}{\natexlab{a}})}\BibitemShut {NoStop}%
\bibitem [{\citenamefont {Sato}\ \emph {et~al.}(2017)\citenamefont {Sato},
  \citenamefont {Hohenadler},\ and\ \citenamefont {Assaad}}]{sato2017prl}%
  \BibitemOpen
  \bibfield  {author} {\bibinfo {author} {\bibfnamefont {Toshihiro}\
  \bibnamefont {Sato}}, \bibinfo {author} {\bibfnamefont {Martin}\ \bibnamefont
  {Hohenadler}}, \ and\ \bibinfo {author} {\bibfnamefont {Fakher~F.}\
  \bibnamefont {Assaad}},\ }\bibfield  {title} {\enquote {\bibinfo {title}
  {Dirac fermions with competing orders: Non-landau transition with emergent
  symmetry},}\ }\href {\doibase 10.1103/PhysRevLett.119.197203} {\bibfield
  {journal} {\bibinfo  {journal} {Phys. Rev. Lett.}\ }\textbf {\bibinfo
  {volume} {119}},\ \bibinfo {pages} {197203} (\bibinfo {year}
  {2017})}\BibitemShut {NoStop}%
\bibitem [{\citenamefont {Giamarchi}(2003)}]{giamarchi2003_book}%
  \BibitemOpen
  \bibfield  {author} {\bibinfo {author} {\bibfnamefont {Thierry}\ \bibnamefont
  {Giamarchi}},\ }\href@noop {} {\emph {\bibinfo {title} {Quantum physics in
  one dimension}}},\ Vol.\ \bibinfo {volume} {121}\ (\bibinfo  {publisher}
  {Clarendon press},\ \bibinfo {year} {2003})\BibitemShut {NoStop}%
\bibitem [{\citenamefont {Yang}\ and\ \citenamefont
  {Yang}(1966{\natexlab{a}})}]{yang1966aXXZ}%
  \BibitemOpen
  \bibfield  {author} {\bibinfo {author} {\bibfnamefont {Chen-Ning}\
  \bibnamefont {Yang}}\ and\ \bibinfo {author} {\bibfnamefont {Chen-Ping}\
  \bibnamefont {Yang}},\ }\bibfield  {title} {\enquote {\bibinfo {title}
  {One-dimensional chain of anisotropic spin-spin interactions. i. proof of
  bethe's hypothesis for ground state in a finite system},}\ }\href@noop {}
  {\bibfield  {journal} {\bibinfo  {journal} {Physical Review}\ }\textbf
  {\bibinfo {volume} {150}},\ \bibinfo {pages} {321} (\bibinfo {year}
  {1966}{\natexlab{a}})}\BibitemShut {NoStop}%
\bibitem [{\citenamefont {Yang}\ and\ \citenamefont
  {Yang}(1966{\natexlab{b}})}]{yang1966bXXZ}%
  \BibitemOpen
  \bibfield  {author} {\bibinfo {author} {\bibfnamefont {Chen-Ning}\
  \bibnamefont {Yang}}\ and\ \bibinfo {author} {\bibfnamefont {Chen-Ping}\
  \bibnamefont {Yang}},\ }\bibfield  {title} {\enquote {\bibinfo {title}
  {One-dimensional chain of anisotropic spin-spin interactions. ii. properties
  of the ground-state energy per lattice site for an infinite system},}\
  }\href@noop {} {\bibfield  {journal} {\bibinfo  {journal} {Physical Review}\
  }\textbf {\bibinfo {volume} {150}},\ \bibinfo {pages} {327} (\bibinfo {year}
  {1966}{\natexlab{b}})}\BibitemShut {NoStop}%
\bibitem [{\citenamefont {Yang}\ and\ \citenamefont
  {Yang}(1966{\natexlab{c}})}]{yang1966cXXZ}%
  \BibitemOpen
  \bibfield  {author} {\bibinfo {author} {\bibfnamefont {Chen-Ning}\
  \bibnamefont {Yang}}\ and\ \bibinfo {author} {\bibfnamefont {Chen-Ping}\
  \bibnamefont {Yang}},\ }\bibfield  {title} {\enquote {\bibinfo {title}
  {One-dimensional chain of anisotropic spin-spin interactions. iii.
  applications},}\ }\href@noop {} {\bibfield  {journal} {\bibinfo  {journal}
  {Physical Review}\ }\textbf {\bibinfo {volume} {151}},\ \bibinfo {pages}
  {258} (\bibinfo {year} {1966}{\natexlab{c}})}\BibitemShut {NoStop}%
\bibitem [{\citenamefont {Lu}\ \emph {et~al.}(2021)\citenamefont {Lu},
  \citenamefont {Xu},\ and\ \citenamefont {You}}]{lu2021prb}%
  \BibitemOpen
  \bibfield  {author} {\bibinfo {author} {\bibfnamefont {Da-Chuan}\
  \bibnamefont {Lu}}, \bibinfo {author} {\bibfnamefont {Cenke}\ \bibnamefont
  {Xu}}, \ and\ \bibinfo {author} {\bibfnamefont {Yi-Zhuang}\ \bibnamefont
  {You}},\ }\bibfield  {title} {\enquote {\bibinfo {title} {Self-duality
  protected multicriticality in deconfined quantum phase transitions},}\ }\href
  {\doibase 10.1103/PhysRevB.104.205142} {\bibfield  {journal} {\bibinfo
  {journal} {Phys. Rev. B}\ }\textbf {\bibinfo {volume} {104}},\ \bibinfo
  {pages} {205142} (\bibinfo {year} {2021})}\BibitemShut {NoStop}%
\bibitem [{\citenamefont {Zhao}\ \emph {et~al.}(2019)\citenamefont {Zhao},
  \citenamefont {Weinberg},\ and\ \citenamefont {Sandvik}}]{zhao2019symmetry}%
  \BibitemOpen
  \bibfield  {author} {\bibinfo {author} {\bibfnamefont {Bowen}\ \bibnamefont
  {Zhao}}, \bibinfo {author} {\bibfnamefont {Phillip}\ \bibnamefont
  {Weinberg}}, \ and\ \bibinfo {author} {\bibfnamefont {Anders~W}\ \bibnamefont
  {Sandvik}},\ }\bibfield  {title} {\enquote {\bibinfo {title}
  {Symmetry-enhanced discontinuous phase transition in a two-dimensional
  quantum magnet},}\ }\href@noop {} {\bibfield  {journal} {\bibinfo  {journal}
  {Nature Physics}\ }\textbf {\bibinfo {volume} {15}},\ \bibinfo {pages}
  {678--682} (\bibinfo {year} {2019})}\BibitemShut {NoStop}%
\bibitem [{\citenamefont {Sun}\ \emph {et~al.}(2021)\citenamefont {Sun},
  \citenamefont {Ma}, \citenamefont {Zhao}, \citenamefont {Sandvik},\ and\
  \citenamefont {Meng}}]{Sun_2021}%
  \BibitemOpen
  \bibfield  {author} {\bibinfo {author} {\bibfnamefont {Guangyu}\ \bibnamefont
  {Sun}}, \bibinfo {author} {\bibfnamefont {Nvsen}\ \bibnamefont {Ma}},
  \bibinfo {author} {\bibfnamefont {Bowen}\ \bibnamefont {Zhao}}, \bibinfo
  {author} {\bibfnamefont {Anders~W.}\ \bibnamefont {Sandvik}}, \ and\ \bibinfo
  {author} {\bibfnamefont {Zi~Yang}\ \bibnamefont {Meng}},\ }\bibfield  {title}
  {\enquote {\bibinfo {title} {Emergent o(4) symmetry at the phase transition
  from plaquette-singlet to antiferromagnetic order in quasi-two-dimensional
  quantum magnets*},}\ }\href {\doibase 10.1088/1674-1056/abf3b8} {\bibfield
  {journal} {\bibinfo  {journal} {Chinese Physics B}\ }\textbf {\bibinfo
  {volume} {30}},\ \bibinfo {pages} {067505} (\bibinfo {year}
  {2021})}\BibitemShut {NoStop}%
\bibitem [{\citenamefont {Sreejith}\ \emph
  {et~al.}(2019{\natexlab{b}})\citenamefont {Sreejith}, \citenamefont
  {Powell},\ and\ \citenamefont {Nahum}}]{sreejith2019prl}%
  \BibitemOpen
  \bibfield  {author} {\bibinfo {author} {\bibfnamefont {G.~J.}\ \bibnamefont
  {Sreejith}}, \bibinfo {author} {\bibfnamefont {Stephen}\ \bibnamefont
  {Powell}}, \ and\ \bibinfo {author} {\bibfnamefont {Adam}\ \bibnamefont
  {Nahum}},\ }\bibfield  {title} {\enquote {\bibinfo {title} {Emergent so(5)
  symmetry at the columnar ordering transition in the classical cubic dimer
  model},}\ }\href {\doibase 10.1103/PhysRevLett.122.080601} {\bibfield
  {journal} {\bibinfo  {journal} {Phys. Rev. Lett.}\ }\textbf {\bibinfo
  {volume} {122}},\ \bibinfo {pages} {080601} (\bibinfo {year}
  {2019}{\natexlab{b}})}\BibitemShut {NoStop}%
\bibitem [{\citenamefont {Serna}\ and\ \citenamefont
  {Nahum}(2019{\natexlab{b}})}]{serna2019prb}%
  \BibitemOpen
  \bibfield  {author} {\bibinfo {author} {\bibfnamefont {Pablo}\ \bibnamefont
  {Serna}}\ and\ \bibinfo {author} {\bibfnamefont {Adam}\ \bibnamefont
  {Nahum}},\ }\bibfield  {title} {\enquote {\bibinfo {title} {Emergence and
  spontaneous breaking of approximate $\mathrm{O}(4)$ symmetry at a weakly
  first-order deconfined phase transition},}\ }\href {\doibase
  10.1103/PhysRevB.99.195110} {\bibfield  {journal} {\bibinfo  {journal} {Phys.
  Rev. B}\ }\textbf {\bibinfo {volume} {99}},\ \bibinfo {pages} {195110}
  (\bibinfo {year} {2019}{\natexlab{b}})}\BibitemShut {NoStop}%
\bibitem [{\citenamefont {Wildeboer}\ \emph
  {et~al.}(2020{\natexlab{a}})\citenamefont {Wildeboer}, \citenamefont {Desai},
  \citenamefont {D'Emidio},\ and\ \citenamefont {Kaul}}]{wildeboer2020prb}%
  \BibitemOpen
  \bibfield  {author} {\bibinfo {author} {\bibfnamefont {Julia}\ \bibnamefont
  {Wildeboer}}, \bibinfo {author} {\bibfnamefont {Nisheeta}\ \bibnamefont
  {Desai}}, \bibinfo {author} {\bibfnamefont {Jonathan}\ \bibnamefont
  {D'Emidio}}, \ and\ \bibinfo {author} {\bibfnamefont {Ribhu~K.}\ \bibnamefont
  {Kaul}},\ }\bibfield  {title} {\enquote {\bibinfo {title} {First-order n\'eel
  to columnar valence bond solid transition in a model square-lattice $s=1$
  antiferromagnet},}\ }\href {\doibase 10.1103/PhysRevB.101.045111} {\bibfield
  {journal} {\bibinfo  {journal} {Phys. Rev. B}\ }\textbf {\bibinfo {volume}
  {101}},\ \bibinfo {pages} {045111} (\bibinfo {year}
  {2020}{\natexlab{a}})}\BibitemShut {NoStop}%
\bibitem [{\citenamefont {Wildeboer}\ \emph
  {et~al.}(2020{\natexlab{b}})\citenamefont {Wildeboer}, \citenamefont {Desai},
  \citenamefont {D'Emidio},\ and\ \citenamefont {Kaul}}]{wildeboer2020prb2}%
  \BibitemOpen
  \bibfield  {author} {\bibinfo {author} {\bibfnamefont {Julia}\ \bibnamefont
  {Wildeboer}}, \bibinfo {author} {\bibfnamefont {Nisheeta}\ \bibnamefont
  {Desai}}, \bibinfo {author} {\bibfnamefont {Jonathan}\ \bibnamefont
  {D'Emidio}}, \ and\ \bibinfo {author} {\bibfnamefont {Ribhu~K.}\ \bibnamefont
  {Kaul}},\ }\bibfield  {title} {\enquote {\bibinfo {title} {First-order n\'eel
  to columnar valence bond solid transition in a model square-lattice $s=1$
  antiferromagnet},}\ }\href {\doibase 10.1103/PhysRevB.101.045111} {\bibfield
  {journal} {\bibinfo  {journal} {Phys. Rev. B}\ }\textbf {\bibinfo {volume}
  {101}},\ \bibinfo {pages} {045111} (\bibinfo {year}
  {2020}{\natexlab{b}})}\BibitemShut {NoStop}%
\bibitem [{\citenamefont {Yu}\ \emph {et~al.}(2019)\citenamefont {Yu},
  \citenamefont {Roiban}, \citenamefont {Jian},\ and\ \citenamefont
  {Liu}}]{yu2019prb}%
  \BibitemOpen
  \bibfield  {author} {\bibinfo {author} {\bibfnamefont {Jiabin}\ \bibnamefont
  {Yu}}, \bibinfo {author} {\bibfnamefont {Radu}\ \bibnamefont {Roiban}},
  \bibinfo {author} {\bibfnamefont {Shao-Kai}\ \bibnamefont {Jian}}, \ and\
  \bibinfo {author} {\bibfnamefont {Chao-Xing}\ \bibnamefont {Liu}},\
  }\bibfield  {title} {\enquote {\bibinfo {title} {Finite-scale emergence of
  $2+1\mathrm{D}$ supersymmetry at first-order quantum phase transition},}\
  }\href {\doibase 10.1103/PhysRevB.100.075153} {\bibfield  {journal} {\bibinfo
   {journal} {Phys. Rev. B}\ }\textbf {\bibinfo {volume} {100}},\ \bibinfo
  {pages} {075153} (\bibinfo {year} {2019})}\BibitemShut {NoStop}%
\bibitem [{\citenamefont {Lee}\ \emph {et~al.}(2022)\citenamefont {Lee},
  \citenamefont {Ramette}, \citenamefont {Metlitski}, \citenamefont {Vuletic},
  \citenamefont {Ho},\ and\ \citenamefont {Choi}}]{lee2022landauforbidden}%
  \BibitemOpen
  \bibfield  {author} {\bibinfo {author} {\bibfnamefont {Jong~Yeon}\
  \bibnamefont {Lee}}, \bibinfo {author} {\bibfnamefont {Joshua}\ \bibnamefont
  {Ramette}}, \bibinfo {author} {\bibfnamefont {Max~A.}\ \bibnamefont
  {Metlitski}}, \bibinfo {author} {\bibfnamefont {Vladan}\ \bibnamefont
  {Vuletic}}, \bibinfo {author} {\bibfnamefont {Wen~Wei}\ \bibnamefont {Ho}}, \
  and\ \bibinfo {author} {\bibfnamefont {Soonwon}\ \bibnamefont {Choi}},\
  }\href@noop {} {\enquote {\bibinfo {title} {Landau-forbidden quantum
  criticality in rydberg quantum simulators},}\ } (\bibinfo {year} {2022}),\
  \Eprint {http://arxiv.org/abs/2207.08829} {arXiv:2207.08829
  [cond-mat.str-el]} \BibitemShut {NoStop}%
\bibitem [{\citenamefont {Chang}\ \emph {et~al.}(2019)\citenamefont {Chang},
  \citenamefont {Lin}, \citenamefont {Shao}, \citenamefont {Wang},\ and\
  \citenamefont {Yin}}]{chang2019topological}%
  \BibitemOpen
  \bibfield  {author} {\bibinfo {author} {\bibfnamefont {Chi-Ming}\
  \bibnamefont {Chang}}, \bibinfo {author} {\bibfnamefont {Ying-Hsuan}\
  \bibnamefont {Lin}}, \bibinfo {author} {\bibfnamefont {Shu-Heng}\
  \bibnamefont {Shao}}, \bibinfo {author} {\bibfnamefont {Yifan}\ \bibnamefont
  {Wang}}, \ and\ \bibinfo {author} {\bibfnamefont {Xi}~\bibnamefont {Yin}},\
  }\bibfield  {title} {\enquote {\bibinfo {title} {Topological defect lines and
  renormalization group flows in two dimensions},}\ }\href@noop {} {\bibfield
  {journal} {\bibinfo  {journal} {Journal of High Energy Physics}\ }\textbf
  {\bibinfo {volume} {2019}},\ \bibinfo {pages} {1--85} (\bibinfo {year}
  {2019})}\BibitemShut {NoStop}%
\bibitem [{\citenamefont {Thorngren}\ and\ \citenamefont
  {Wang}(2021)}]{thorngren2021fusion}%
  \BibitemOpen
  \bibfield  {author} {\bibinfo {author} {\bibfnamefont {Ryan}\ \bibnamefont
  {Thorngren}}\ and\ \bibinfo {author} {\bibfnamefont {Yifan}\ \bibnamefont
  {Wang}},\ }\bibfield  {title} {\enquote {\bibinfo {title} {Fusion category
  symmetry ii: categoriosities at c= 1 and beyond},}\ }\href@noop {} {\bibfield
   {journal} {\bibinfo  {journal} {arXiv preprint arXiv:2106.12577}\ }
  (\bibinfo {year} {2021})}\BibitemShut {NoStop}%
\bibitem [{\citenamefont {Fishman}\ \emph {et~al.}(2022)\citenamefont
  {Fishman}, \citenamefont {White},\ and\ \citenamefont
  {Stoudenmire}}]{itensor}%
  \BibitemOpen
  \bibfield  {author} {\bibinfo {author} {\bibfnamefont {Matthew}\ \bibnamefont
  {Fishman}}, \bibinfo {author} {\bibfnamefont {Steven~R.}\ \bibnamefont
  {White}}, \ and\ \bibinfo {author} {\bibfnamefont {E.~Miles}\ \bibnamefont
  {Stoudenmire}},\ }\bibfield  {title} {\enquote {\bibinfo {title} {{The
  ITensor Software Library for Tensor Network Calculations}},}\ }\href
  {\doibase 10.21468/SciPostPhysCodeb.4} {\bibfield  {journal} {\bibinfo
  {journal} {SciPost Phys. Codebases}\ ,\ \bibinfo {pages} {4}} (\bibinfo
  {year} {2022})}\BibitemShut {NoStop}%
\end{thebibliography}%

\onecolumngrid

\begin{appendix}

%\section*{Supplemental Material}

\renewcommand{\theequation}{A\arabic{equation}}
\setcounter{equation}{0}
\renewcommand{\thefigure}{A\arabic{figure}}
\setcounter{figure}{0}
\renewcommand{\thetable}{A\arabic{table}}
\setcounter{table}{0}

\section{EFFECTIVE THEORY FOR THE 1D SPIN CHAIN WITH SHORT-RANGE INTERACTION}
\label{sec:SM5}
In this appendix, we summary the basic effective continuum theory description of the generalized 1D spin-$1/2$ chain, JM model \cite{jiang2019_1ddqcp,roberts2019dqcp1d,mudry2019_1dqdcp,huang2019_1ddqcp},
\begin{equation}
\begin{aligned}
H=& \frac{1}{4} \sum_j \Big( -J_x \sigma_{j}^x \sigma_{j+1}^x
-J_z \sigma_{j}^z \sigma_{j+1}^z \Big)
+ \frac{1}{4} \sum_j \Big( +K_x \sigma_{j}^x \sigma_{j+2}^x
+K_z \sigma_{j}^z \sigma_{j+2}^z\Big)
\end{aligned}
\label{eq:BasicModel}
\end{equation}
where the spin operators have been represented by Pauli matrices 
$\bm{S}_j=\frac{1}{2}\bm{\sigma}_j=\frac{1}{2}(\sigma_j^x,\sigma_j^y,\sigma_j^z)$.
The effective field theory could be obtained from Bosonization approach based on the transformation \cite{giamarchi2003_book,jiang2019_1ddqcp,mudry2019_1dqdcp},
\begin{equation}
\begin{aligned}
\sigma_j^y \sim \frac{2}{\pi}\big( \theta_{j+1/2} -\theta_{j-1/2} \big),\quad
\sigma_j^z \sim  \cos(\phi_j),\quad
\sigma_j^x \sim  -\sin(\phi_j),\quad
[\theta_{j+1/2},\phi_{j^{\prime}}] 
=i\pi \Theta(j+1/2- j^{\prime}),
\end{aligned}
\label{eq:BosonizationDictionary1}
\end{equation}
where $\theta$ and $\phi$ is a pair of conjugate field, $\Theta(x)$ is a Heaviside step function.
Taking continuum limit, the effective action in the imaginary-time path integral formulation is given by \cite{jiang2019_1ddqcp},
\begin{equation}
\begin{aligned}
S[\phi,\theta] =&\int d\tau\,dx\ \Big[ \frac{i}{\pi} \partial_{\tau}\phi \partial_x\theta
+\frac{v}{2\pi} \Big( \frac{1}{g} (\partial_x\theta)^2 +g(\partial_x\phi)^2  \Big) \Big]
+ \int d\tau\,dx\ \Big[ \lambda_u \cos(4\theta) 
+\lambda_a \cos(2\phi) \Big].
\end{aligned}
\end{equation}
where $\tau$ is the imaginary time, $x$ is the spatial coordinate, $v$ is the velocity and $g$ is the Luttinger parameter.
$\lambda_u$ and $\lambda_a$ terms are the most relevant operators which have the largest scaling dimension and reflect the symmetries of the system.
They will drive the possible transition to the ordered phase, reflecting by the pinning of the $(\theta,\phi)$ fields.
In the continuum theory, the order parameters for the $z$-FM phase and VBS phase are represented by the continuous bosonic fields,
\begin{align*}
\Psi_{z\text{FM}} \sim  \cos(\phi),\qquad
\Psi_{\text{VBS}} \sim  \cos(2\theta).
\end{align*}
The $z$-FM state $\Psi_{z\text{FM}}$ is invariant under the lattice translational symmetry $T_x(\phi,\theta)\rightarrow(\phi,\theta+\pi/2)$ and $\mathbb{Z}_2^z$ spin rotation symmetry around $z$-axis $g_z(\phi,\theta)\rightarrow(-\phi,-\theta)$,
but breaks the $\mathbb{Z}_2^x$ spin rotation symmetry around $x$-axis $g_x(\phi,\theta)\rightarrow(-\phi+\pi,-\theta)$ and time-reversal symmetry $\mathcal{T}(\phi,\theta,i)\rightarrow(\phi+\pi,-\theta,-i)$.
The VBS state $\Psi_{\text{VBS}}$ is invariant under $g_x$, $g_z$ and $\mathcal{T}$, but breaks $T_x$.
The tree-level scalings of the $\cos$-operators are given by $\dim[\cos(2n\theta)] =n^2g$ and $\dim[\cos(m\phi)] =\frac{m^2}{4g}$.
The tree-level $\beta$-functions for the short-ranged $\lambda_u$- and $\lambda_a$-term are given by,
\begin{equation}
\begin{aligned}
\frac{d\lambda_u}{dl} 
=&\Big( 2- \dim[\cos(4\theta)] \Big) \lambda_u
=\Big( 2- 4g \Big) \lambda_u,\qquad
\frac{d\lambda_a}{dl} 
=&\Big( 2- \dim[\cos(2\phi)] \Big) \lambda_u
=\Big( 2- \frac{1}{g} \Big) \lambda_u.
\end{aligned}
\end{equation}
The scaling behaviors of the $z$FM correlation and VBS correlation are given by
\begin{align*}
\langle \Psi_{z\text{FM}}(r) \Psi_{z\text{FM}}(0)\rangle \sim \frac{1}{r^{1/2g}},\qquad
\langle \Psi_{\text{VBS}}(r) \Psi_{\text{VBS}}(0)\rangle \sim \frac{1}{r^{2g}}
\end{align*}
However, such continuum theory is not complete for the understanding of deconfined quantum critical point between $z$-FM and VBS order, 
and a dual theory is necessary to fully characterize the critical behavior \cite{jiang2019_1ddqcp}.

\paragraph{Dual Luttinger-like theory}
To describe the deconfined quantum phase transition between $z$-FM and VBS order, 
it is more sufficient to work in the duality formulation,
which is also represented by a Luttinger-like theory \cite{jiang2019_1ddqcp,roberts2019dqcp1d},
\begin{equation}
\begin{aligned}
S[\tilde{\phi},\tilde{\theta}] =&\int d\tau\,dx\ \Big[ \frac{i}{\pi} \partial_{\tau}\tilde{\phi} \partial_x\tilde{\theta}
+\frac{\tilde{v}}{2\pi} \Big( \frac{1}{\tilde{g}} (\partial_x\tilde{\theta})^2 +\tilde{g}(\partial_x\tilde{\phi})^2  \Big) \Big]
+ \int d\tau\,dx\ \Big[ 
\lambda \cos(2\tilde{\theta}) 
+\lambda^{\prime} \cos(4\tilde{\theta}) 
+\kappa \cos(4\tilde{\phi}) \Big]
\end{aligned}
\end{equation}
where $\tilde{\phi}$ and $\tilde{\theta}$ are a pair of conjugate fields in the dual theory, $\tilde{v}$ and $\tilde{g}$ are corresponding velocity and Luttinger parameter.
$\lambda,\lambda^{\prime}$ and $\kappa$ are several lowest-order short-range interactions.

The phase transition between $z$-FM and VBS order lies within the regime $\tilde{g}\in(1/2,2)$ such that $\lambda$-term is the single relevant operator which drives the phase transition happens when its sign alters.
The correlation length exponent at the critical point follows the scaling dimension of $\lambda \cos(2\tilde{\theta})$, that is \cite{jiang2019_1ddqcp},
\begin{align*}
\nu^{-1}=2- \mathrm{dim}[\cos(2\tilde{\theta})]
=2-\tilde{g},\qquad
\nu =\frac{1}{2-\tilde{g}}.
\end{align*}
In this dual Luttinger-like theory, the order parameters for the $z$-FM and VBS phase are decoded into a compatible form based on a single field $\tilde{\theta}$
\begin{align}
\Psi_{z\text{FM}} \sim  \sin(\tilde{\theta}),\qquad
\Psi_{\text{VBS}} \sim  \cos(\tilde{\theta}),
\label{SEQ:OPdualTheory}
\end{align}
which, of course, have the same scaling dimension $\mathrm{dim}[\Psi_{z\text{FM}}]=\mathrm{dim}[\Psi_{\text{VBS}}]=\tilde{g}/4$ and similar correlation behavior at the critical point,
\begin{align*}
\langle \Psi_{z\text{FM}}(r) \Psi_{z\text{FM}}(0)\rangle \sim 
\frac{1}{r^{2\mathrm{dim}\Psi_{z\text{FM}}}}
=\frac{1}{r^{\tilde{g}/2}},\qquad
\langle \Psi_{\text{VBS}}(r) \Psi_{\text{VBS}}(0)\rangle \sim 
\frac{1}{r^{2\mathrm{dim}\Psi_{\text{VBS}}}}
=\frac{1}{r^{\tilde{g}/2}}.
\end{align*}
The phase transition between $z$-FM  and VBS phase is tuned by $\lambda$,
while at the critical point $\lambda=0$ there exist an emergent O(2) symmetry corresponding to $\tilde{\theta}$ part.

\section{DETAILS OF RENORMALIZATION GROUP CALCULATIONS FOR THE LONG-RANGE SINE-GORDON MODEL}
\label{sec:SM6}

In this work, we consider the long-range interaction Hamiltonian,
\begin{equation}
H_{LRP} = \sum_{i}(-J_{z}S^{z}_{i}S^{z}_{i+1}+K_{x}S^{x}_{i}S^{x}_{i+2}+K_{z}S^{z}_{i}S^{z}_{i+2})
-\frac{J_{z}}{N(\alpha)}\sum_{i,j}\frac{S^{z}_{i}S^{z}_{j}}{|i-j|^{\alpha}},
\end{equation}
where $J$ is the interaction strength, parameter $\alpha$ tunes the power of long-range interactions. 
$N(\alpha)(= \frac{1}{N-1}\sum_{i\ne j}\frac{1}{|i-j|^{\alpha}}$) is the Kac factor to preserve the Hamiltonian extensive.
To connect this model with the known short-range model (\ref{eq:BasicModel}), we can separated the long-range term in the form,
\begin{align*}
-\frac{J_{z}}{N(\alpha)}\sum_{i,j}\frac{S^{z}_{i}S^{z}_{j}}{|i-j|^{\alpha}}
= -\frac{J_{z}}{N(\alpha)}\sum_{|i-j|=1} S^{z}_{i}S^{z}_{j}
-\frac{J_{z}}{N(\alpha)}\sum_{|i-j|>1} \frac{S^{z}_{i}S^{z}_{j}}{|i-j|^{\alpha}}
=-J_z^{\prime} \sum_{i} S^{z}_{i} S^{z}_{i+1}
-\frac{J_{z}}{N(\alpha)} \sum_{|i-j|>1} \frac{S^{z}_{i} S^{z}_{j}}{|i-j|^{\alpha}}.
\end{align*}
Here the first term is just the conventional nearest-neighbor coupling.
We consider the effective continuum theory for the long-range interactions,
\begin{align}
H_{LR}=&-\sum_{i,j} \frac{J_L}{|i-j|^{\alpha}} \sigma_{i}^z\sigma_{j}^z
=-\sum_{i,j} \frac{J_L}{|i-j|^{\alpha}} \cos\phi_i \cos\phi_j
\rightarrow -\int dxdx^{\prime} \frac{J_L}{|x-x^{\prime}|^{\alpha}} \cos\phi(x) \cos\phi(x^{\prime}) \nonumber	\\
\rightarrow&-\frac{1}{2}\int dxdr \frac{J_L}{|r|^{\alpha}} \Big(\cos\big[\phi(x+r)-\phi(x)\big]
+\cos\big[\phi(x+r)+\phi(x)\big]\Big) 
\label{eq:longrangeszsz}
\end{align}
Expand the first term for small $r$, we can obtain
\begin{align*}
&\frac{1}{2}\int dxdr \frac{J_L}{|r|^{\alpha}} \cos\big[\phi(x+r)-\phi(x)\big]
\approx \frac{1}{2}\int dxdr \frac{J_L}{|r|^{\alpha}} \Big( 1- \frac{1}{2} (r\nabla\phi)^2  \Big)
\end{align*}
which only normalize the Luttinger parameters.
In general, long-range correlated interaction part of the action is given by,
\begin{align*}
S_{LR}=\frac{1}{2}\int d\tau\int dxdr \frac{1}{|r|^{\alpha}} \Big(\lambda_+ \cos\big[\phi(x+r,\tau)+\phi(x,\tau)\big]
+\lambda_- \cos\big[\phi(x+r,\tau)-\phi(x,\tau)\big]\Big)
\end{align*}
with the bare interaction strength $\lambda_+<0$ and $\lambda_-<0$.

For the effective Bosonization theory, we follow the conventional tree-level RG analysis of the sine-Gordon model \cite{giamarchi2003_book}.
Separating the field into slow and fast modes ($\phi=\phi_<+\phi_>$ and $\phi=\theta_<+\theta_>$) and integrating over the fast mode $(\phi_{>},\theta_{>})$, the partition function can be expanded in the form,
\begin{equation*}
Z=\int D\phi D\theta\, e^{-S_0-S_1}
=\int D\phi_< D\theta_<\, \int D\phi_> D\theta_>\, e^{-S_{0,<} -S_{0,>} -S_1}
=\int D\phi_< D\theta_<\, e^{-S_{0,<}}\, \sum_{n=0}^{\infty} \frac{1}{n!} \big\langle (-S_1)^n\big\rangle_>
\end{equation*}
where the integral of fast mode gives the average,
\begin{align*}
\langle \cdots \rangle_>\equiv \int D\phi_> D\theta_>\, e^{-S_{0,>}} \big(\cdots\big).
\end{align*}
The effective action under the renormalization is,
\begin{align}
S_{eff}= S_{0,<} +\big\langle S_1\big\rangle_{>} 
-\frac{1}{2} \big\langle S_1^2\big\rangle_{>,c}.
\end{align}
The tree-level scalings of the operators can be obtained from the first order term $\big\langle S_1\big\rangle_{>}$.
We consider the tree-level scaling for the long-range correlated terms,
\begin{align*}
S_{\sigma}=&\frac{1}{2} \lambda_{\sigma} 
\int d\tau\int dxdr \frac{1}{|r|^{\alpha}} 
\cos\big[\phi(x+r,\tau) +\sigma \phi(x,\tau)\big]
\equiv \frac{1}{2} \lambda_{\sigma} 
\int d\tau\int dxdr \frac{1}{|r|^{\alpha}} 
\cos\big[\Delta_{r}^{\sigma} \phi \big]	\\
=&\frac{1}{2} \lambda_{\sigma} 
\int d\tau\int dxdr \frac{1}{|r|^{\alpha}} 
\cos\big[\Delta_{r}^{\sigma} \phi_< +\Delta_{r}^{\sigma} \phi_>\big]
\end{align*}
Integrating out the fast mode, the lowest order correction is
\begin{align*}
\langle S_{\sigma}\rangle_>
=&\frac{1}{2} \lambda_{\sigma} 
\int d\tau\int dxdr \frac{1}{|r|^{\alpha}} 
\cos\big[\Delta_{r}^{\sigma} \phi_< \big] 
\langle \cos\big[\Delta_{r}^{\sigma} \phi_>\big]\rangle_>
\end{align*}
Here, the renormalization gives the contribution,
\begin{align*}
&\langle \cos\big[\Delta_{r}^{\sigma} \phi_>\big]\rangle_>
=\exp\Big( -\frac{1}{2} \big\langle [\Delta_{r}^{\sigma} \phi_>]^2 \big\rangle_{>} \Big)
=\exp\Big( -\frac{1}{2} \big\langle \big[\phi(x+r,\tau) +\sigma \phi(x,\tau) \big]^2 \big\rangle_{>} \Big)	\\
=&\exp\Big( -\frac{1}{2} \big\langle \big[\phi(r,0) +\sigma \phi(0) \big]^2 \big\rangle_{>} \Big)
\end{align*}
where the correlation function of $\phi(r)$ field can be calculated out directly,
\begin{align*}
&\frac{1}{2} \big\langle \big[\phi(r,0)+\sigma \phi(0) \big]^2 \big\rangle_{>}
=\frac{1}{2} \int_> \frac{d\omega dq}{(2\pi)^2}  
\big(2 +2\sigma \cos(|qr|) \big) 
\frac{v}{g} \frac{\pi}{v^2q^2 +\omega^2}
=\frac{1}{2g} \Big(1 +\sigma \cos(|\Lambda r|)\Big) \frac{d\Lambda}{\Lambda}
\end{align*}
The lowest order correction is
\begin{align*}\langle S_{\sigma}\rangle_>
=&\frac{1}{2} \lambda_{\sigma} 
\int d\tau\int dxdr \frac{1}{|r|^{\alpha}} 
\cos\big[\Delta_{r}^{\sigma} \phi_< \big] 
\exp\Big( -\frac{1}{2g} \Big(1 +\sigma \cos(|\Lambda r|)\Big) \frac{d\Lambda}{\Lambda}  \Big).
\end{align*}
Up to tree-level, the effective action under renormalization is $S_{\sigma,eff} =S_{\sigma,<}+\langle S_{\sigma}\rangle_>$.
Under the rescaling transformation, 
\begin{align*}
\tau \rightarrow e^{dl} \tau,\qquad x\rightarrow e^{dl} x,\qquad
r\rightarrow e^{dl} r,\qquad
\Lambda\rightarrow e^{-dl} \Lambda,
\end{align*}
the effective action becomes,
\begin{align*}&S_{\sigma,eff}
\rightarrow \frac{1}{2} \lambda_{\sigma} e^{(3-\alpha) dl}
\int d\tau\int dxdr \frac{1}{|r|^{\alpha}} 
\cos\big[\Delta_{r}^{\sigma} \phi \big] 
\Big( 1-\frac{1}{2g}dl -\frac{\sigma}{2g} \cos(|\Lambda r|) dl \Big)	\\
=& \frac{1}{2} \lambda_{\sigma}
\int d\tau\int dxdr \frac{1}{|r|^{\alpha}} 
\cos\big[\Delta_{r}^{\sigma} \phi \big] 
\Big( 1+ \big( 3-\alpha -\frac{1}{2g}\big) dl -\frac{\sigma}{2g} \cos(|\Lambda r|) dl  \Big)
\end{align*}
and the RG functional equations for $\lambda_{\sigma}$ are
\begin{align*}
\frac{d\lambda_{\sigma}(r)}{dl} 
=\Big( 3-\alpha -\frac{1+ \sigma\cos(|\Lambda r|)}{2g} \Big) \lambda_{\sigma}(r).
\end{align*}
Here, the effective coupling has non-trivial dependence on the momentum cutoff $\Lambda$.
In the lattice formulation, the coordinates of the system are represented by $r=r_n=na$ ($n=0,1,\cdots,N-1$), where $a$ is the lattice constant and the total lattice size is $L=Na$.
The corresponding discrete set of momentum is $k=k_m=m\frac{\pi }{L}=m\frac{\pi }{Na}$ with
$-\frac{N}{2}+1,\cdots,\frac{N}{2}$.
In the infrared (long-length) limit, the momentum will flow to the shortest momentum scale $\sim \pi/L$.
For smaller $r$, the oscillation factor $\cos(|\Lambda r|)$ becomes nearly unity, and we can approximately obtain 
\begin{align*}
r<r_c:\quad
\frac{d\lambda_{\sigma}(r)}{dl} 
=\Big( 3-\alpha -\frac{1+ \sigma}{2g} \Big) \lambda_{\sigma}(r).
\end{align*}
For larger $r$, the scaling dimension of $\lambda_-$ is bigger than $\lambda_+$.
$\lambda_-$ is more relevant than $\lambda_+$ in general.
There exists a critical power $\alpha_c$ below which $\lambda_{\sigma}$ becomes most relevant, dominating the physical behavior of the system.

\paragraph{long-range interaction in the dual theory}
We now transform to the effect of long-range interaction in the dual theory.
From the representation of the order parameter in Eq.~(\ref{SEQ:OPdualTheory}), the long-range interaction in the continuous dual theory is given by,
\begin{align*}
&S_{LR}=\int d\tau\int dxdr \frac{\tilde{\lambda}}{|r|^{\alpha}} \sin\big[\tilde{\theta}(x+r,\tau)\big] 
\sin\big[\tilde{\theta}(x,\tau)\big]   \\
=&\frac{1}{2}\int d\tau\int dxdr \frac{1}{|r|^{\alpha}} \Big(\tilde{\lambda}_- \cos\big[\tilde{\theta}(x+r,\tau)-\tilde{\theta}(x,\tau)\big]
-\tilde{\lambda}_+ \cos\big[\tilde{\theta}(x+r,\tau)+\tilde{\theta}(x,\tau)\big]
\Big).
\end{align*}
The renormalization of $\tilde{\lambda}_{\pm}$ takes same form as $\lambda_{\pm}$, only taking the substitution $g\rightarrow 1/\tilde{g}$,
\begin{align*}
\frac{d\tilde{\lambda}_{\sigma}(r)}{dl} 
=\Big( 3-\alpha -\tilde{g} \frac{1+ \sigma\cos(|\Lambda r|)}{2} \Big) \tilde{\lambda}_{\sigma}(r).
\end{align*}
Similarly as before, long-range $\tilde{\lambda}_-$ interaction will dominate the system when the power smaller than some critical value, $\alpha<\alpha_c$. 
The scaling dimension of $\tilde{\lambda}_-$ is approximately given by
\begin{align*}
\dim[\Tilde{\lambda}_-] =3-\alpha -\tilde{g}\frac{1-\delta}{2}.
\end{align*}
with some constant $\delta$.
Compare the scaling dimension the long-range $\tilde{\lambda}_-$ and short-range interaction $\lambda$ at the tree-level,
we can obtain the critical power $\alpha_c$,
\begin{align*}
3-\alpha-\tilde{g}\frac{1-\delta}{2}
=2- \tilde{g},
\quad\rightarrow\quad
\alpha_c= 1+\frac{\tilde{g}}{2}(\delta+1).
\end{align*}
The true critical power could lie between $1+\frac{\tilde{g}}{2}<\alpha_c<1+\tilde{g}$.

\section{ADDITIONAL DATA FOR $z{\rm FM}$ BINDER RATIO AND THE RATIO OF SQUARED ORDER PARAMETERS}
\label{sec:appC}

In this section, we provide additional results of the $z{\rm FM}$ Binder ratio $U_{z\rm FM}$ and the ratio of squared order parameters $R_{2}$ for other $\alpha$ values. 

\begin{figure*}[tb]
\includegraphics[width=0.9\linewidth]{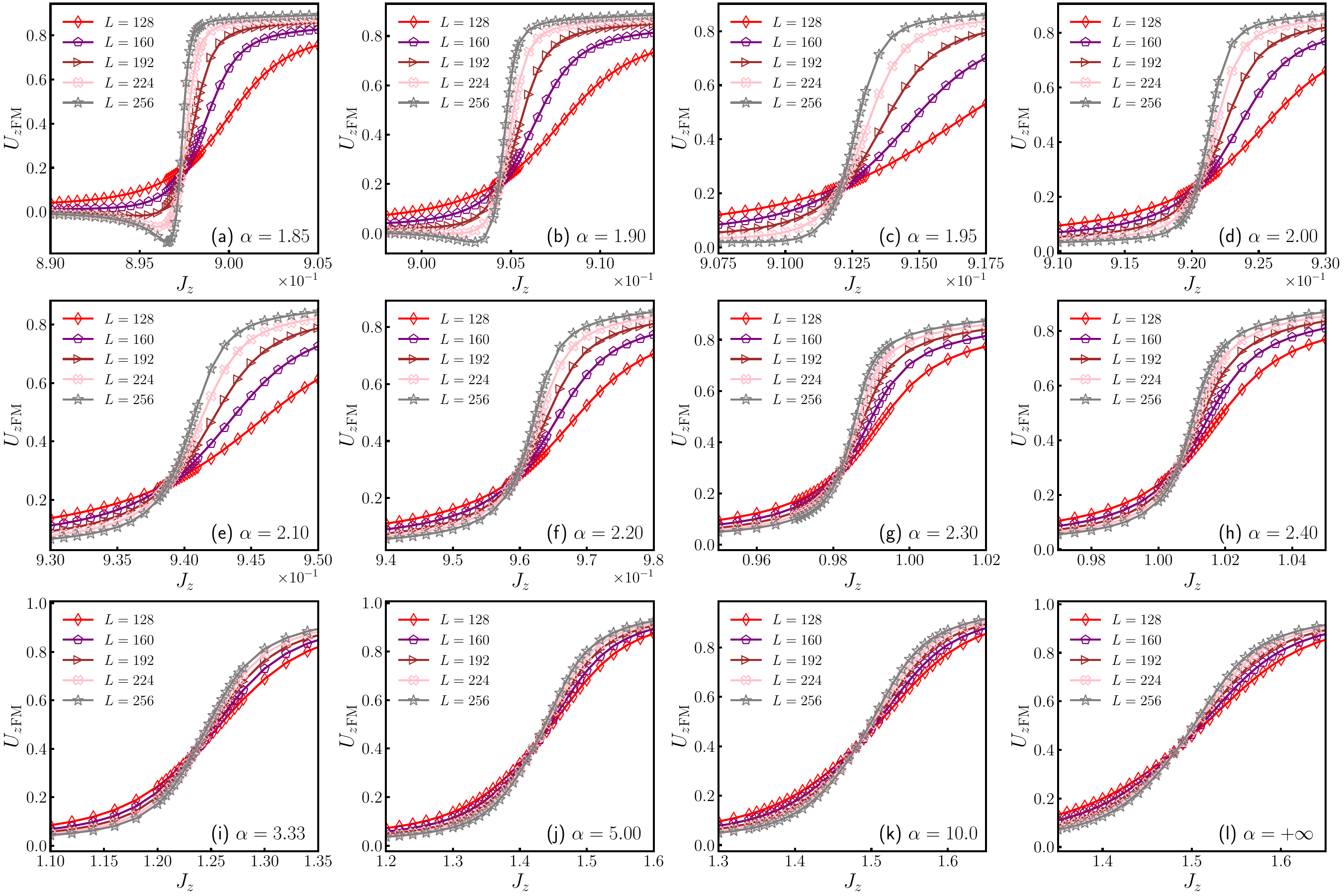}
\caption{The Binder ratio of the $z{\rm FM}$ order as a function of $J_{z}$ for (a) $\alpha=1.85$, (b) $\alpha=1.90$, (c) $\alpha=1.95$, (d) $\alpha=2.00$, (e) $\alpha=2.10$, (f) $\alpha=2.20$, (g) $\alpha=2.30$, (h) $\alpha=2.40$, (i) $\alpha=3.33$, (j) $\alpha=5.00$, (k) $\alpha=10.0$, and (l) $\alpha=+\infty$.
}
\label{fig:Binder_all}
\end{figure*}

\begin{figure*}[tb]
\includegraphics[width=0.9\linewidth]{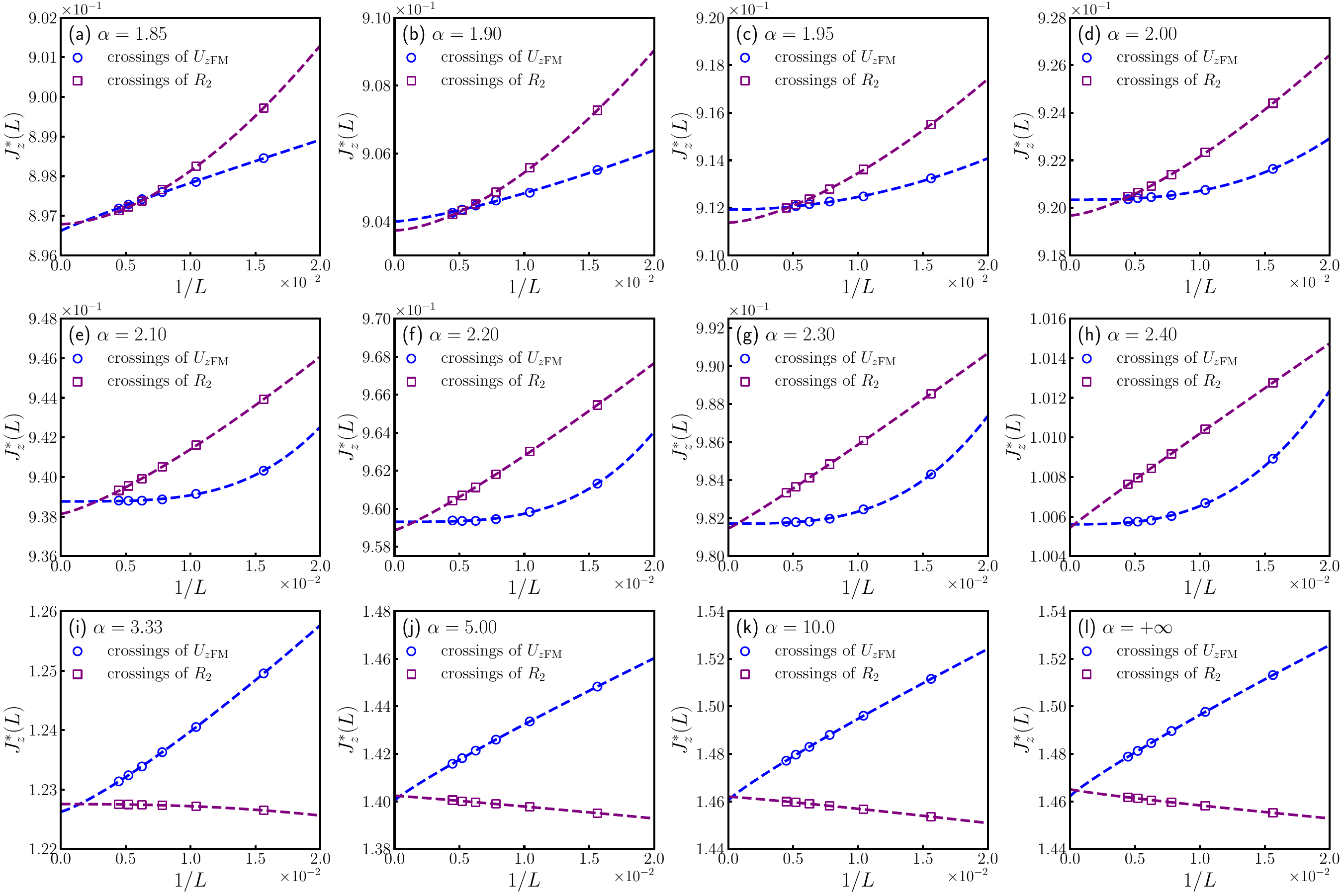}
\caption{The crossing locations $J_{z}^{*}(L)$ of $U_{z\rm FM}(L)$ [$R_{2}(L)$] and $U_{z\rm FM}(L+32)$ [$R_{2}(L+32)$] are shown versus $1/L$ for (a) $\alpha=1.85$, (b) $\alpha=1.90$, (c) $\alpha=1.95$, (d) $\alpha=2.00$, (e) $\alpha=2.10$, (f) $\alpha=2.20$, (g) $\alpha=2.30$, (h) $\alpha=2.40$, (i) $\alpha=3.33$, (j) $\alpha=5.00$, (k) $\alpha=10.0$, and (l) $\alpha=+\infty$. The curves are least-squares fits according to $J_{z}^{*}(L) = J_{z}^{\rm c} + aL^{-b}$. The critical points obtained respectively from $U_{z\rm FM}$ and $R_{2}$ are consistent with each other within numerical accuracy. 
}
\label{fig:cp_all}
\end{figure*}

In Fig.~\ref{fig:Binder_all}, we first present $U_{z\rm FM}$ as a function of $J_{z}$ for various system sizes at other representative $\alpha$ values. The distinct behaviours of $U_{z\rm FM}$ for $\alpha>\alpha_{\rm c}$ or $\alpha<\alpha_{\rm c}$ indicate a fundamentally change of the transition nature as explained in the main text. An extrapolation of the crossing points of $U_{z\rm FM}$, according to the relation $J_{z}^{*}(L)=J_{z}^{\rm c}+aL^{-b}$ where $J_{z}^{*}(L)$ is the crossing point of $U_{z\rm FM}(L)$ and $U_{z\rm FM}(L+32)$, is also performed to determine the precise boundary between the ordered phases (see Fig.~\ref{fig:cp_all}). The obtained critical points are then used to complete the ground-state phase diagram displayed in the main text (see Fig.~\ref{fig:phase_diagram}). 

\begin{figure*}[tb]
\includegraphics[width=0.9\linewidth]{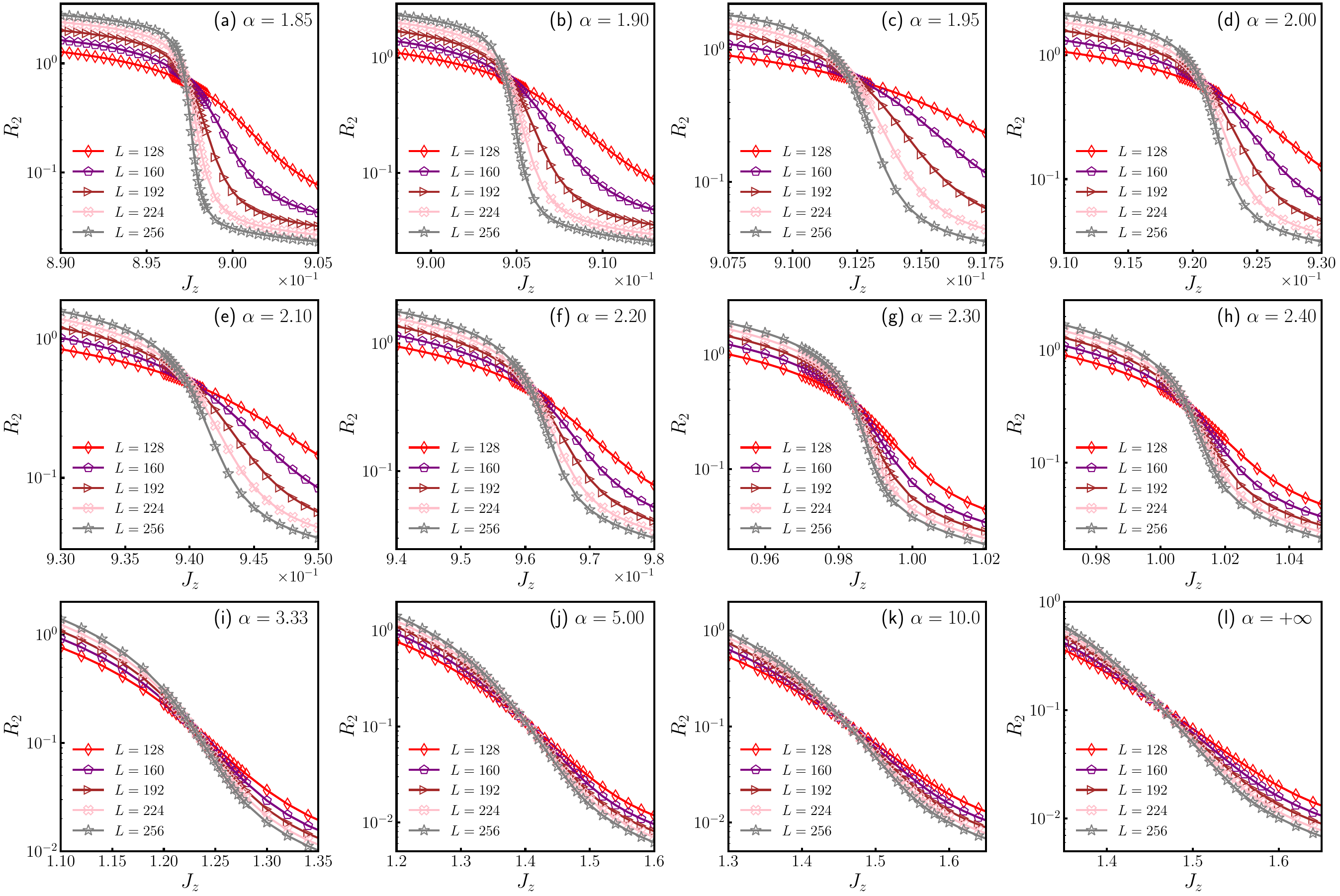}
\caption{The ratio of the squared order parameters $R_{2}$ as a function of $J_{z}$ for (a) $\alpha=1.85$, (b) $\alpha=1.90$, (c) $\alpha=1.95$, (d) $\alpha=2.00$, (e) $\alpha=2.10$, (f) $\alpha=2.20$, (g) $\alpha=2.30$, (h) $\alpha=2.40$, (i) $\alpha=3.33$, (j) $\alpha=5.00$, (k) $\alpha=10.0$, and (l) $\alpha=+\infty$.
}
\label{fig:r2_all}
\end{figure*}

On the other hand, the ratio of squared order parameters $R_{2}$ versus $J_{z}$ is also analyzed in Fig.~\ref{fig:r2_all} for other $\alpha$ values. It is clear that all the curves of different $L$ intersect almost at a single point, which means that $R_{2}$ becomes universal at the critical point. The result can be a supportive evidence for the O(2)$\times$O(2) symmetry appeared along the whole transition line (the dashed line in Fig.~\ref{fig:phase_diagram}). Furthermore, a similar extrapolation of the $R_{2}$ crossing points is also exhibited in Fig.~\ref{fig:cp_all}, from which we can see that the extrapolated critical points are consistent with the ones extracted from $U_{z\rm FM}$ quite well.

%---------------------------------------------------------------------
%---------------------------------------------------------------------
%---------------------------------------------------------------------

%\bibliography{SM}
\end{appendix}

\end{document}